\documentclass[a4paper,11pt]{article}
\usepackage{a4wide}
\usepackage{amsmath}
\usepackage{cite}
\usepackage[normalem]{ulem}
\usepackage{amssymb}
\usepackage{epsf}
\usepackage{amsfonts}
\usepackage{graphicx}
\usepackage{braket}
\usepackage{ragged2e}
\usepackage{appendix}
\numberwithin{equation}{section}
\usepackage{bbm}
\usepackage{xcolor}
\usepackage{subfigure}
\usepackage[export]{adjustbox}
\usepackage{siunitx,pslatex}
\usepackage{booktabs}
\usepackage{hyperref}
\usepackage{ulem}
\usepackage{caption}
\hypersetup{
    colorlinks,
    linkcolor={green!50!black},
    citecolor={magenta},
    urlcolor={blue!80!black}
}

\begin{document}
	
	\title{\bf Classical and quantum chaos of closed strings on a charged confining holographic background}
	\author{ \textbf{Bhaskar Shukla}\thanks{bhasker\_shukla@nitrkl.ac.in},~ \textbf{Owais Riyaz}\thanks{ovaisreyaz1@gmail.com},~ \textbf{Subhash Mahapatra}\thanks{mahapatrasub@nitrkl.ac.in}
		\\\\\textit{{\small Department of Physics and Astronomy, National Institute of Technology Rourkela, Rourkela - 769008, India}}\\
	}
    \date{}
\maketitle
\begin{abstract}
We discuss the classical and quantum chaos of closed strings on a recently constructed charged confining holographic background. The confining background corresponds to the charged soliton, which is a solution of minimal $d=5$ gauged supergravity. The solution has a compact spacelike direction with a Wilson line on a circle and asymptotes to $AdS_5$ with a planar boundary. For the classical case, we analyze the chaos using the power spectrum, Poincar\'{e} sections, and Lyapunov exponents, finding that both energy and charge play constructive effects on enhancing the chaotic nature of the system. We similarly analyze quantum chaos using the distribution of the spectrum's level-spacing and out-of-time-ordered correlators and thoroughly investigate the effects of charge and energy. A gradual transition from a chaotic to an integrable regime is obtained as the energy and charge increase from lower to higher values, with charge playing a subdominant role. 
\end{abstract}

\section{Introduction}\label{sec:intro}
Chaos theory has emerged as a cornerstone of contemporary scientific inquiry, influencing diverse disciplines such as meteorology, astronomy, ecology, economics, etc~\cite{Shen2021Jan, Fang2015Dec, Hastings1993Nov, Kelsey1988Mar}. It elucidates the intricate dynamics of complex systems, unveiling the hidden patterns that govern their ostensibly unpredictable behavior. From modeling atmospheric turbulence~\cite{Xin2001May} to understanding population fluctuations~\cite{Hassell1991Sep}, chaos theory reveals a fundamental, yet often elusive, order within disorder. It is within this framework that we discern not mere randomness but an inherent structure embedded in complexity, offering profound insights into the physics of natural phenomena. Chaos theory has applications in the quantum domain as well. Quantum chromodynamics (QCD) is one such example. QCD is the theory of strong interactions that exhibit a rich phase structure. It is characterized by confinement and chiral symmetry breaking at low temperatures and chemical potential~\cite{Roberts:1994dr}. However, as temperature and chemical potential increase, the system transitions to a state where chiral symmetry is restored, accompanied by deconfinement~\cite{Fang:2015ytf, Fischer:2011mz, Bazavov:2007zz}. Whether QCD is chaotic or not in both confined and deconfined phases is an important question. A substantial body of research has been conducted in this area ~\cite{Beisert:2010jr, Pullirsch:1998ke, Pullirsch:1998wp, Markum:1997fk, Bittner:2000nu, Bittner:2004ff}, leading to consistent advancements in our understanding of QCD physics from the chaos perspective. 

In the confined and near-deconfined region, QCD becomes difficult to probe due to its strongly coupled nature, making the standard perturbative techniques ineffective. This is where the AdS/CFT correspondence~\cite{Maldacena:1997re}, also known as the gauge/gravity duality, a powerful tool from string theory, becomes invaluable. The AdS/CFT correspondence offers a dual framework by mapping a strongly coupled quantum field theory, such as QCD, to a more manageable weakly coupled gravitational theory in a higher-dimensional spacetime~\cite{Witten:1998qj, Gubser:1998bc}. One of the foundational goals of this duality is to explore non-perturbative, strongly coupled gauge theories, especially QCD. Over time, many essential features of QCD have been successfully reproduced using this framework, and in certain cases, novel and profound insights into the strongly coupled regime of QCD have emerged. For further details, we refer to several excellent reviews on this subject~\cite{Casalderrey-Solana:2011dxg, Gubser:2009md, Rougemont:2023gfz}.

Various works have analyzed chaos in QCD observables in different holographic settings. The study of chaos in QCD chiral condensates in $\mathcal{N}=2$ supersymmetric QCD with $SU(N_c)$ gauge group at large $N_c$ and 't Hooft coupling was done in ~\cite{Hashimoto:2016wme} and the $\mathcal{N}=2$ theory was found to be chaotic. Similarly, a chaos-based mapping of the holographic QCD phase diagram in linear sigma models was obtained in~\cite{Akutagawa:2018yoe}. Using a hanging open string approach, the chaos analysis in the quark-antiquark pair in the QCD deconfined phase was carried out in~\cite{Hashimoto:2018fkb}. Their work was extended to analyze the anisotropic effects of the magnetic field and chemical potential on chaotic properties of QCD in top-down and bottom-up holographic settings in~\cite{Colangelo:2021kmn, Shukla:2023pbp, Colangelo:2020tpr, Shukladeconf, Losacco:2022lnz}. Additionally, tools like out-of-time-ordered correlator (OTOC), Pole-skipping, and Krylov complexity have been suggested as order parameters for the deconfinement transition~\cite{Anegawa:2024wov, Baishya:2023ojl, Li:2024yma}. One must use a confining background geometry to study chaos in the dual confined phase. This is usually done by taking the AdS soliton background, having a cigar-shaped geometry, in the gravity side~\cite{Horowitz:1998ha, Witten:1998zw}. The geometry proves instrumental in investigating confinement within holographic gauge theories, as it encapsulates key characteristics such as the mass gap and confinement, phenomena commonly observed in low-temperature QCD. Both open and closed strings have been utilized to analyze QCD chaos in the confined phase by making use of AdS soliton backgrounds~\cite{Akutagawa:2019awh, PandoZayas:2010xpn, Basu:2012ae, Ishii:2016rlk}. Open strings, whose endpoints represent quark-antiquark pairs, offer insight into quark confinement and meson dynamics, while closed strings, which correspond to glueballs in the dual QCD theory, allow for the exploration of the gluonic sector~\cite{Sonnenschein:2015zaa, Pons:2004dk, Sakai:2004cn, Ghodrati:2021ozc}. For other related works on chaos in open and closed strings, see \cite{Basu:2016zkr, Basu:2011di, Panigrahi:2016zny, Barik:2017opb, Banerjee:2018twd,  Xie:2022yef, Ma:2014aha, Ma:2022tvs, Penin:2024rqb, Dutta:2023yhx, Rigatos:2020hlq} for a necessarily biased selection.

In a series of earlier works~\cite{Basu:2011dg, Basu:2013uva}, the dynamics of closed strings in an uncharged AdS soliton background were examined using classical chaos diagnostics, such as Poincaré sections and Lyapunov exponents, as well as quantum chaos measures, including the analysis of energy level-spacing distributions. It was found that the bosonic strings on an AdS soliton background are nonintegrable and exhibit chaos. The uncharged AdS soliton is simply a double analytic continuation of the Schwarzschild AdS black hole solution and can be thought of as the ground state of the $\mathcal{N}=4$ supersymmetric Yang-Mills theory. This is interesting, considering that the AdS soliton corresponds to a confined phase in the dual field theory side, and the nonintegrable nature of the closed string dynamics in the AdS soliton background accordingly provides valuable information about the chaotic behavior of the glueball spectrum of the dual confined phase. 

Recently, new deformed AdS soliton solutions have been constructed with a Wilson line on the circle $S^1$ with aperiodic boundary condition for fermions~\cite{Anabalon:2021tua, Chatzis:2024kdu, Giliberti:2024eii}. Adding a Wilson line around the circle gives rise to non-trivial holonomy for the bulk gauge field, i.e., the deformed AdS solitons are characterized by an additional charge parameter. These charged AdS solitons are solutions of gauged supergravity in five dimensions, which can be further embedded in ten and eleven-dimensional string theory. Essentially, the supergravity background is dual to $\mathcal{N}=(1,0)$ six-dimensional superconformal field theory (SCFT$_6$). The SCFT$_6$ theory flows to a four-dimensional non-conformal theory after compactification and then to a gapped three-dimensional quantum field theory (QFT) by deformation with the vacuum expectation value~\cite{Anabalon:2021tua, Chatzis:2024kdu, Giliberti:2024eii}. The gravity background allows the confinement phenomenon to be examined by studying fundamental strings in bulk, representing quark-antiquark pairs in the dual field theory. In particular, it was shown that the quark-antiquark potential, obtained by minimizing the string's Nambu-Goto action, exhibits three distinct regimes: a Coulomb-like interaction at short distances, a linear confining potential at intermediate separations, and a screening effect at large distances where the potential saturates due to flux tube breaking. The linear rise in the potential, a hallmark of the confined phase in QCD, reflects the increasing energy required to separate quarks, embodying the principle of color confinement. These regimes and behavior qualitatively align with QCD at low temperature confined phase.  The top-down model of \cite{Anabalon:2021tua, Chatzis:2024kdu, Giliberti:2024eii}, therefore, provides a robust gravity background for exploring non-perturbative effects in QCD holographically.

The charged deformed AdS soliton background is interesting, considering that charge densities play a central role in QCD physics. The charge densities (or their conjugate chemical potentials) are important parameters that introduce nontrivial complexities in the structure of the QCD confined phase and modify its properties significantly \cite{Akemann:2004dr, Aoki:2007ka, Tanji:2016dka, Dudal:2017max, Bohra:2019ebj}. Thus, investigating the effect of charge is a necessary step ahead to have a comprehensive analysis of chaos in the confined phase. It may not only be important in the discussion of the overall integrability (or loss) of the system but may also provide important information about the chaotic nature of the glueball spectrum at finite charge density. Hence, the prime aim of this paper is to take this step, add a charge to our analysis, and build upon the above studies to investigate chaotic features of closed strings in the QCD confined phase. We find that the charge indeed modifies the string dynamics nontrivially and leaves significant imprints on both classical and quantum chaotic observables. In particular, the overall effect of the charge is similar to that of energy, albeit subdominant, which can derive the system into chaotic form in the classical domain and into integrable form in the quantum domain.

The paper is structured as follows. In the next section~\ref{sec:susy}, we briefly discuss the charged AdS soliton background and set up the relevant equations of motion for the closed string. The classical chaos of closed string is analyzed in section~\ref{sec:susyclassicalchaos} using power spectrum, Poincar\'{e} sections, and Lyapunov exponents. Subsequently, quantum chaos is studied in section~\ref{sec:susyquantumchaos} using level-spacing distribution, Dyson-Mehta statistics, and OTOC. Finally, we conclude the paper in section~\ref{sec:conclusion} with a summary of key findings and further discussions.

\section{Charged AdS solution and closed string dynamics}\label{sec:susy}
Let us first briefly discuss the charged AdS soliton background of \cite{Anabalon:2021tua, Chatzis:2024kdu} to set up the stage. The Einstein-Maxwell-AdS system in five dimensions is used here, which can be embedded in higher-dimensional string theory ~\cite{Cvetic:1999xp, Schwarz:1983qr, Howe:1983sra, Gunaydin:1984fk, Pernici:1985ju, Gunaydin:1984qu}. Considering the bosonic part in the action,
\begin{equation}
   \mathcal{S}=\frac{1}{16\pi G}\int d^5x \sqrt{-g}\left(R+\frac{12}{\ell^2}-\frac{3}{4}\mathcal{F}_{\mu\nu}\mathcal{F}^{\mu\nu}\right)+\frac{1}{16\pi G} \int \mathcal{F}\wedge\mathcal{F}\wedge\mathcal{A}\,,
\end{equation}
where $\ell$ is the AdS radius and the second integral is the Chern-Simons term. Extremizing the action, the equations of motion reads,
\begin{eqnarray}
    && d \star \mathcal{F} + \mathcal{F}\wedge\mathcal{F}=0\,, \nonumber \\
    && R_{\mu\nu}-\frac{1}{2}g_{\mu\nu} R -\frac{3}{2}\left[\mathcal{F}_{\mu\rho}\mathcal{F}_\nu^\rho-\frac{1}{6}g_{\mu\nu}\mathcal{F}_{\rho\sigma}\mathcal{F}^{\rho\sigma}\right]-\frac{6}{\ell^2}g_{\mu\nu}=0.
\end{eqnarray}
After restricting the solutions that satisfy $\mathcal{F}\wedge\mathcal{F}=0$, so that the  Chern-Simons term does not play any part,  the charged AdS soliton is obtained by considering the electrically charged black hole solution with a flat boundary and applying a double Wick rotation on it. The following solitonic solution is found,
\begin{eqnarray}
    & & ds_5^2=\frac{r^2}{\ell^2}\left(-dt^2+dx_1^2+dx_2^2\right)+\frac{\ell^2 dr^2}{r^2 f(r)}+\frac{r^2}{\ell^2}f(r) d\phi^2, \nonumber \\ 
   & & f(r)=1-\frac{\mu \ell^2}{r^4}-\frac{q^2 \ell^2}{r^6}, \nonumber \\
   & &  \mathcal{A}=q\left(\frac{1}{r^2}-\frac{1}{r_0^2}\right)d\phi,\quad \mathcal{F}=d\mathcal{A}=-\frac{2q}{r^3} dr\wedge d\phi,
\end{eqnarray}
where $r_0$ is the largest positive root of $f(r)$ and $\phi$ is the spatial boundary coordinate compactified on a circle. The remaining boundary
coordinates $t$ and $x_i$ are non-compact. The size of the $\phi$ coordinate shrinks to zero at $r=r_0$, thereby smoothly cutting the infrared region of AdS. In order to avoid conical singularity in the plane $(r,\phi)$ and have a smooth solution at $r=r_0$, the periodicity of $\phi$ is fixed to
\begin{equation}
 L_\phi = \frac{4 \pi \ell^2}{r_{0}^{2}f'(r_0)}\,.
\end{equation}
Similarly, one can find the reduced mass $\mu$ from $f(r_0)=0$
\begin{equation}
    \mu = \frac{(r_0^6 - q^2 \ell^2)}{\ell^2 r_0^2}\,,
\end{equation}
which puts a constraint on the largest allowed value of $q$ for a fixed $r_0$. The infrared cutoff of the above solution at $r=r_0$ dynamically generates a mass scale, much like in real QCD~\cite{Witten:1998zw}. The theory is, therefore, confining and has a mass gap.\\

Next, we consider the Polyakov action:
\begin{equation}
    \mathcal{S}_p = -\frac{1}{2\pi l_{s}^2} \int d\tau d\sigma \sqrt{-\gamma}\gamma^{ab}G_{\mu\nu}\partial_a X^\mu \partial_b X^\nu,
\end{equation}
where $X^\mu$ are the string coordinates, $G_{\mu\nu}$ is the fixed background spacetime metric, and $\gamma_{ab}$ is the induced worldsheet metric, with indices $a$, $b$ representing the coordinates on the string worldsheet $(\tau,\sigma)$. Following \cite{PandoZayas:2010xpn,Basu:2011dg}, we work in the conformal gauge $\gamma_{ab}=\eta_{ab}$ and consider the following Ans\"atze for the  closed string embedding:
\begin{eqnarray}
    t=t(\tau), \qquad \phi=\phi(\tau), \qquad r=r(\tau), \nonumber \\
    x_1=x(\tau)\cos{(\phi(\sigma))}, \qquad x_2=x(\tau)\sin{(\phi(\sigma))},
\end{eqnarray}
where $\phi(\sigma)=\alpha \sigma$, with $\alpha \in \mathbb{Z}$ being the winding number of the string. The location of the string is at a certain value of $r$, wrapped around $(x_1, x_2) -$ directions for a circle of radius $x$. The string is thus allowed to propagate the $r$ direction and change the radius $x$.

After substituting the embedding Ans\"atze in the Polyakov action, we obtain the following effective Lagrangian for the motion:
\begin{equation}
    \label{eq:lagrangian}
    \mathcal{L}=\frac{r^2}{\ell^2}\left(-\dot{t}^2+f(r)\dot{\phi}^2+\dot{x}^2-x^2\alpha^2\right)+\frac{\ell^2}{r^2 f(r)}\dot r^2,
\end{equation}
where the dot represents a derivative with respect to $\tau$. The coordinates $t$ and $\phi$ are cyclic with corresponding conserved momenta $E$ and $k$ are,
\begin{eqnarray}
    &&p_t = -\frac{2r^2 \dot{t}}{\ell^2} \equiv E, \nonumber \\
    &&p_\phi = \frac{2 f(r) r^2 \dot{\phi}}{\ell^2} \equiv k.
\end{eqnarray}
These conserved momenta can be used to replace $\dot{t}$ and $\dot{\phi}$ from the Lagrangian, making the overall dynamics effectively lower-dimensional. Similarly, the conjugate momenta corresponding to free coordinates $r$ and $x$ are:
\begin{eqnarray}
   && p_r = \frac{2\ell^2 \dot{r}}{f(r) r^2}, \nonumber \\
    && p_x = \frac{2 r^2 \dot{x}}{\ell^2}.
\end{eqnarray}
Using the above equation, we can find the effective Hamiltonian. This is given by,
\begin{equation}
    \label{eq:hamil}
    \mathcal{H} = \frac{1}{4 l^2 f(r) r^2}\left[k^2 \ell^4 + f(r)^2 p_r^2 r^4 +f(r) \left(-E^2 \ell^4 + \ell^4 p_x^2 +4\alpha^2 r^4 x^2\right)\right].
\end{equation}
The corresponding Hamilton's equations of motion are
\begin{eqnarray}\label{eq:susyhamiltoneqns}
\dot{r}&=&\frac{r^2 f(r) p_r}{2 \ell^2}, \\
\dot{p_r}&=&\frac{k^2 \ell^2 f'(r)}{4 r^2 f(r)^2}-\frac{r^2 p_r^2 f'(r)}{4 \ell^2}+\frac{k^2 \ell^2}{2 r^3 f(r)}-\frac{r f(r) p_r^2}{2 \ell^2}+\frac{\ell^2 p_x^2}{2 r^3}-\frac{\ell^2 E^2}{2 r^3}-\frac{2 \alpha ^2 r x^2}{\ell^2}, \\
\dot{x}&=&\frac{\ell^2 p_x}{2 r^2}, \\
\dot{p_x}&=&-\frac{2 \alpha ^2 r^2 x}{\ell^2}.
\end{eqnarray}
The Virasoro constraint equations are given by,
\begin{eqnarray}
    &&G_{\mu\nu}\left(\partial_\tau X^\mu \partial_\tau X^\nu + \partial_\sigma X^\mu \partial_\sigma X^\nu\right) = 0, \\
    &&G_{\mu\nu}\partial_\tau X^\mu \partial_\sigma X^\nu = 0.
\end{eqnarray}
The first equation corresponds to $\mathcal{H}=0$, and the second equation is trivially satisfied for the Ans\"atze of our embedding.

\section{Classical chaos analysis}\label{sec:susyclassicalchaos}
\begin{figure}[htbp!]
	\centering
	\subfigure[$r(\tau)$ profile ($E=0.25$)]{\label{fig:timeseries-susy-low-energy}\includegraphics[width=0.32\linewidth]{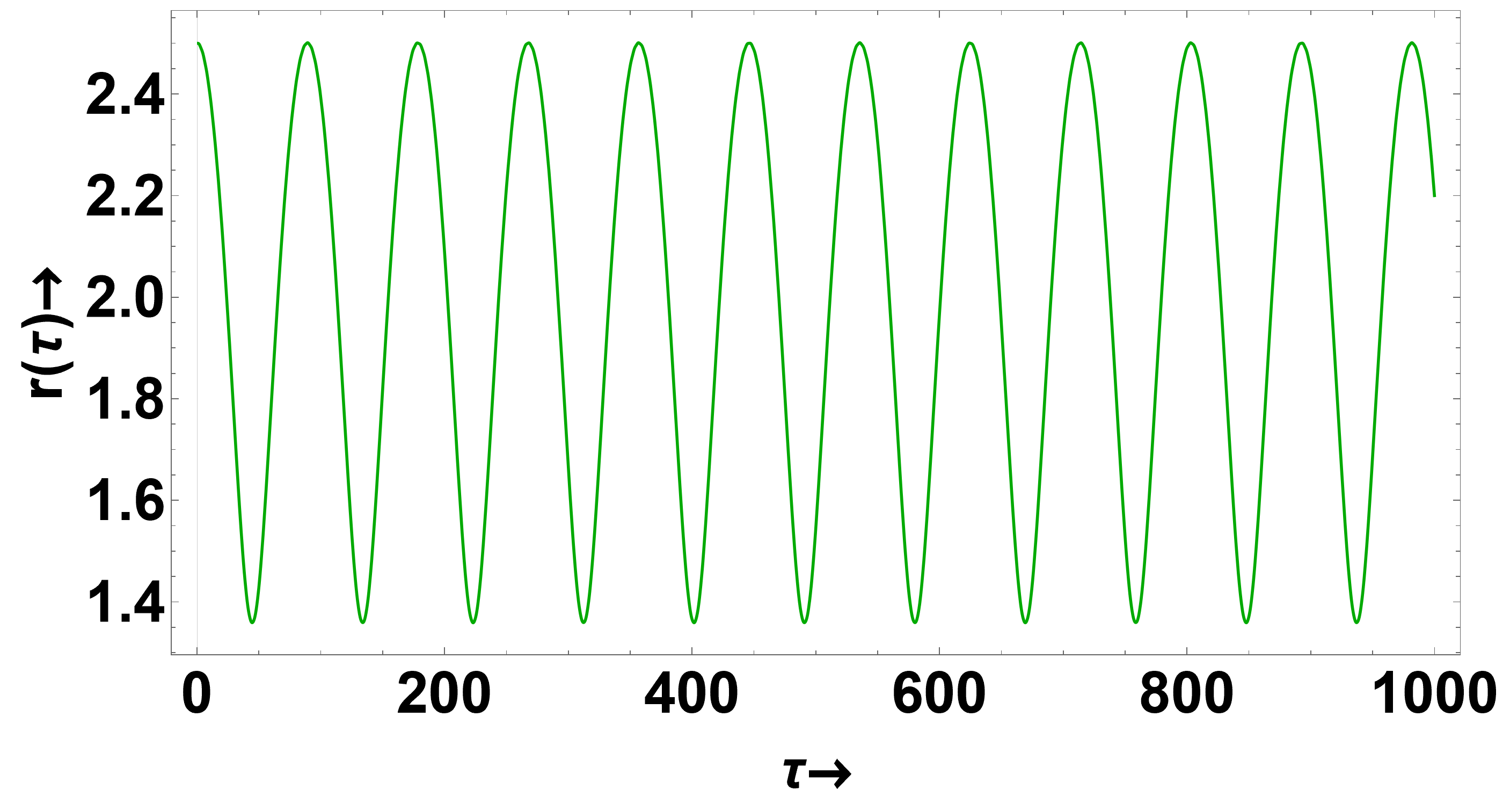}}
    \subfigure[$x(\tau)$ profile ($E=0.25$)]{\label{fig:radiusprofile-susy-low-energy}\includegraphics[width=0.32\linewidth]{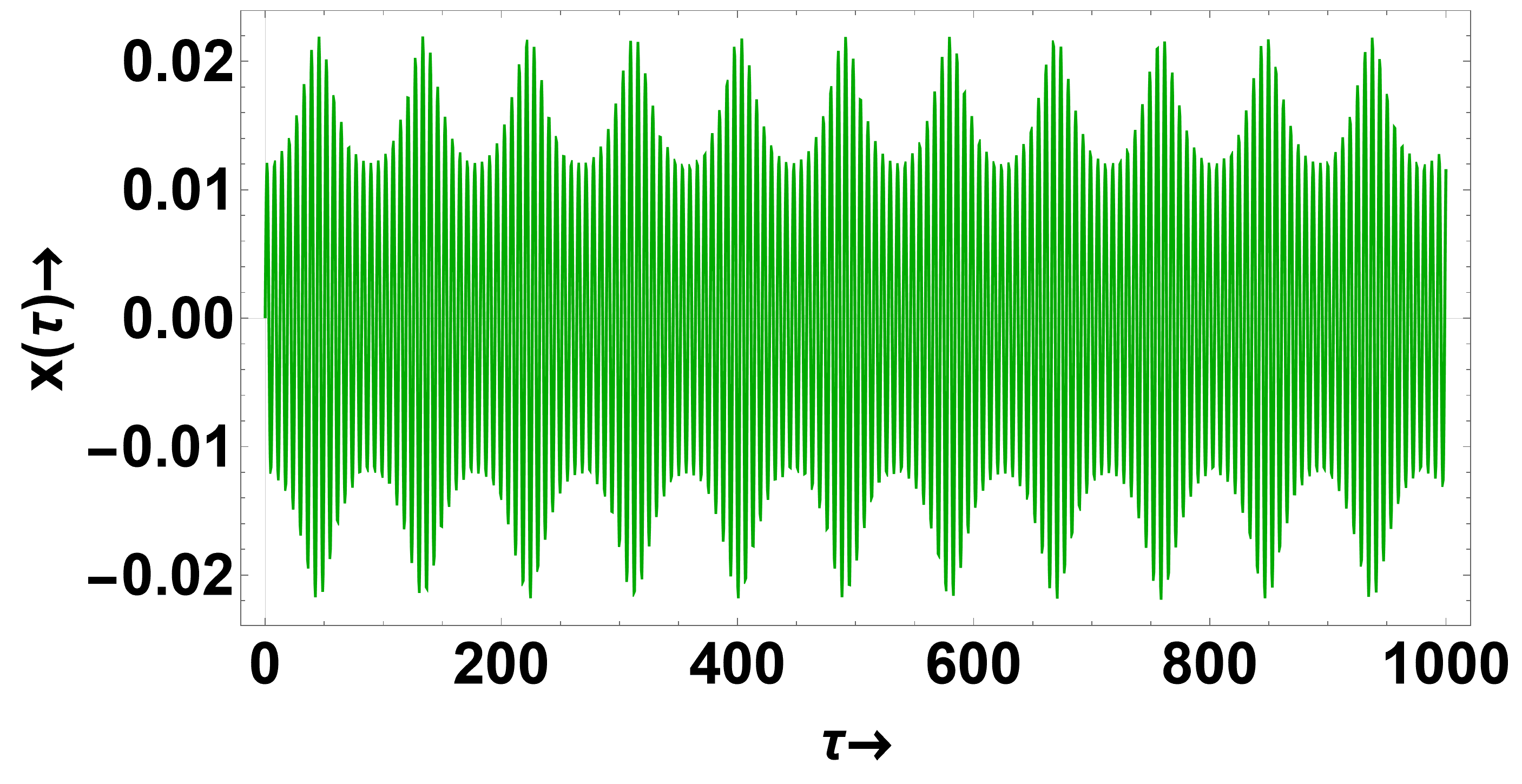}}
    \subfigure[Power spectrum ($E=0.25$)]{\label{fig:powerspectrum-susy-low-energy}\includegraphics[width=0.32\linewidth]{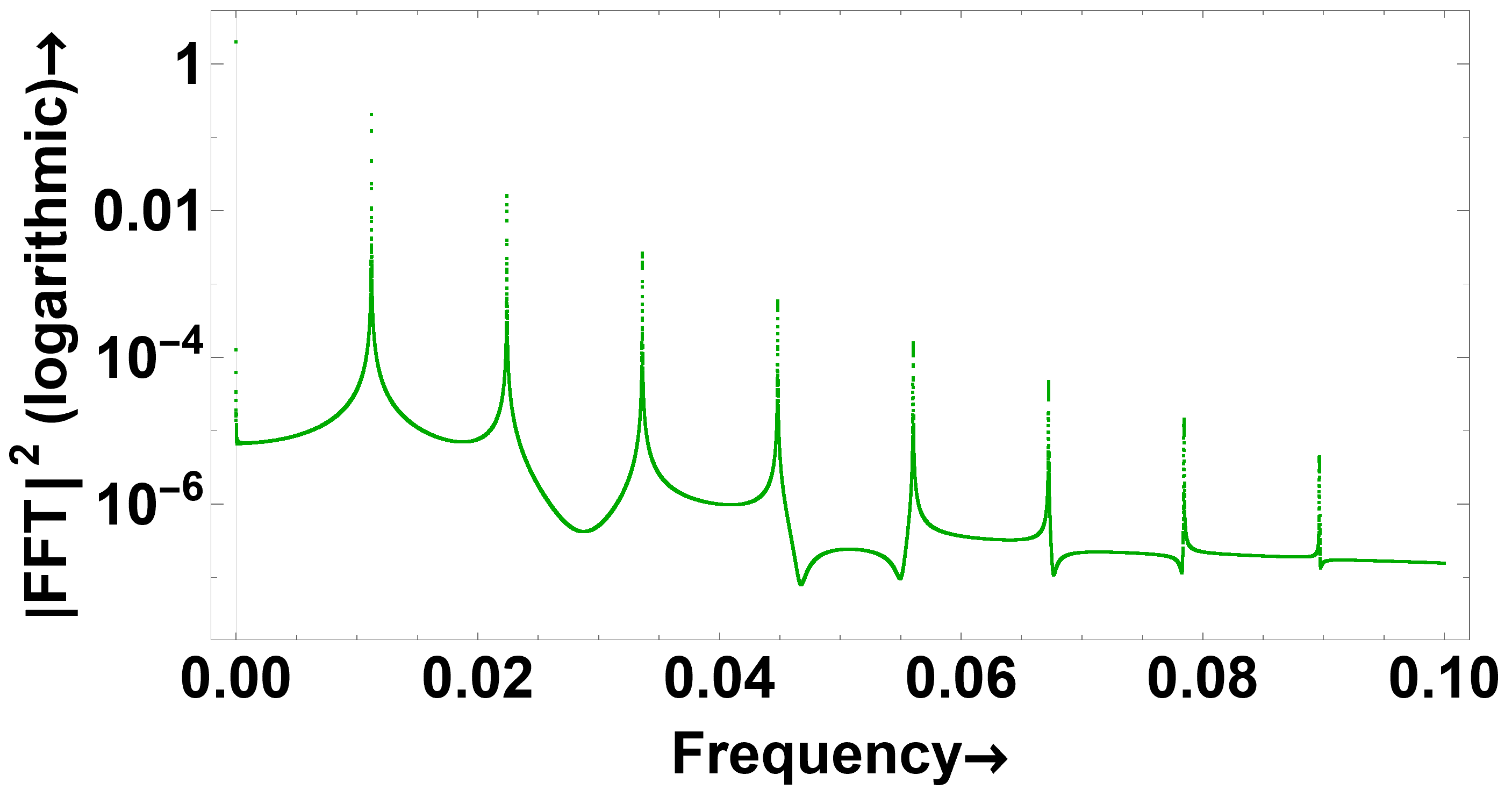}}\\
    \subfigure[$r(\tau)$ profile ($E=0.50$)]{\label{fig:timeseries-susy-medium-energy}\includegraphics[width=0.32\linewidth]{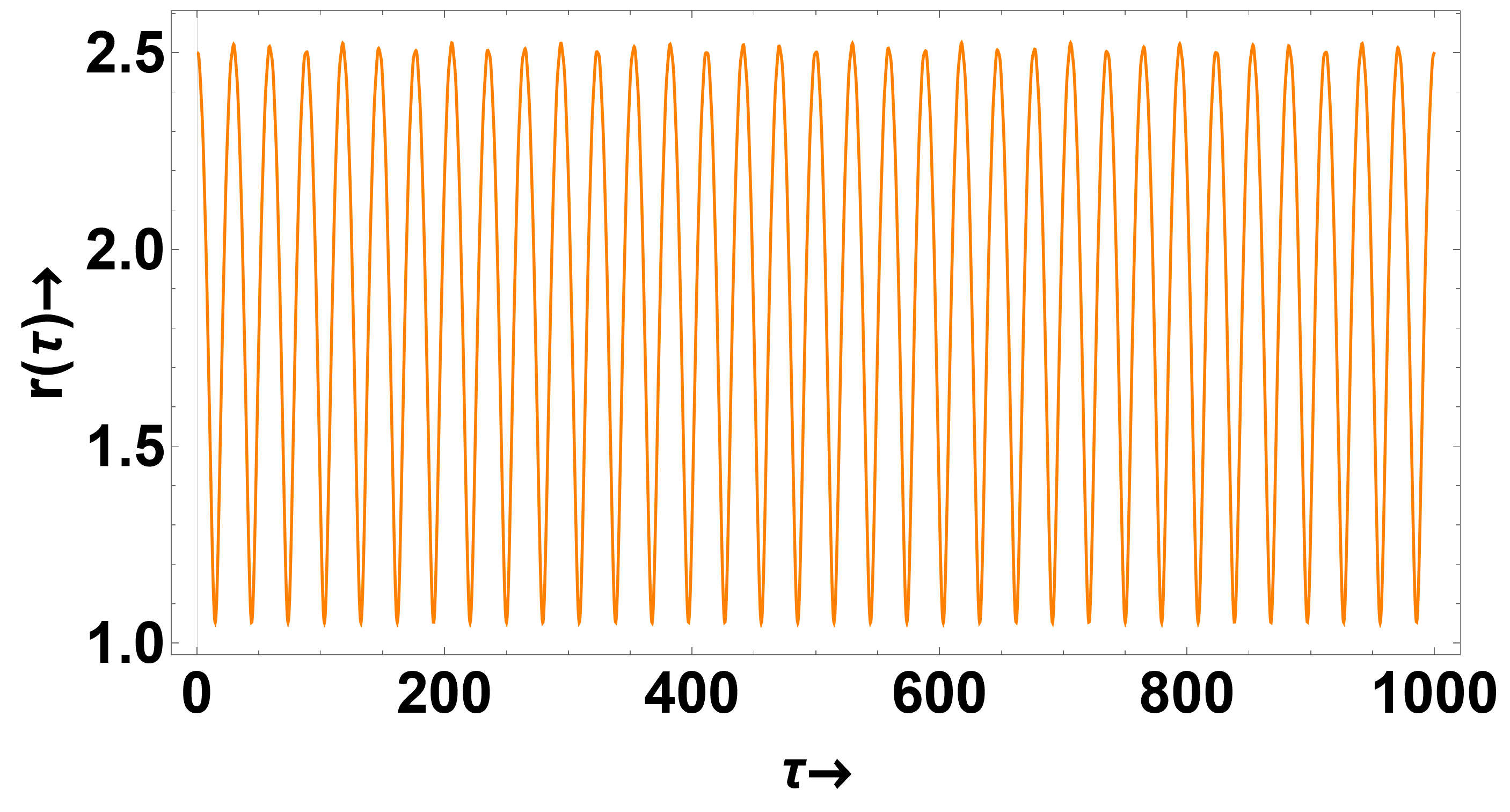}}
    \subfigure[$x(\tau)$ profile ($E=0.50$)]{\label{fig:radiusprofile-susy-medium-energy}\includegraphics[width=0.32\linewidth]{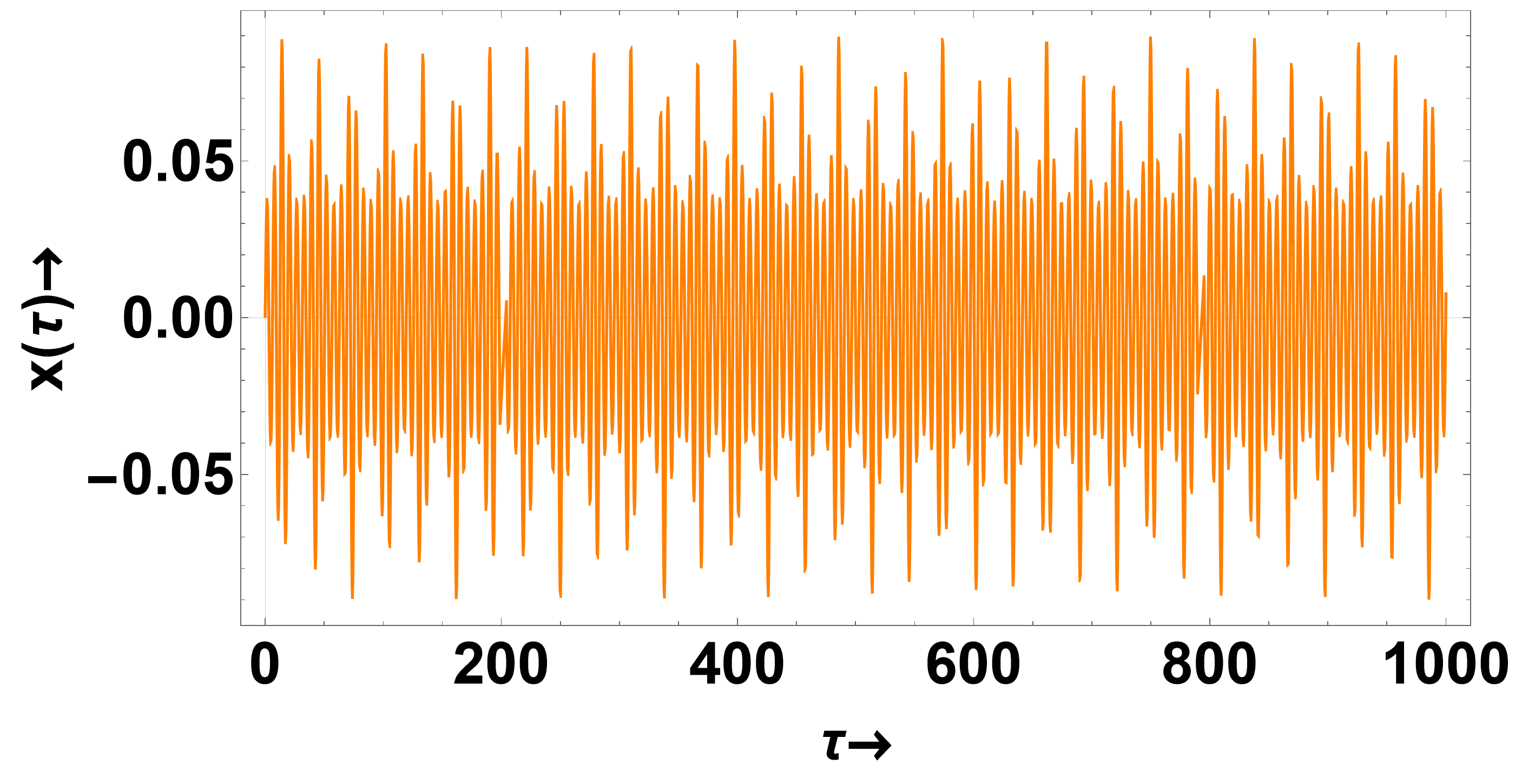}}
    \subfigure[Power spectrum ($E=0.50$)]{\label{fig:powerspectrum-susy-medium-energy}\includegraphics[width=0.32\linewidth]{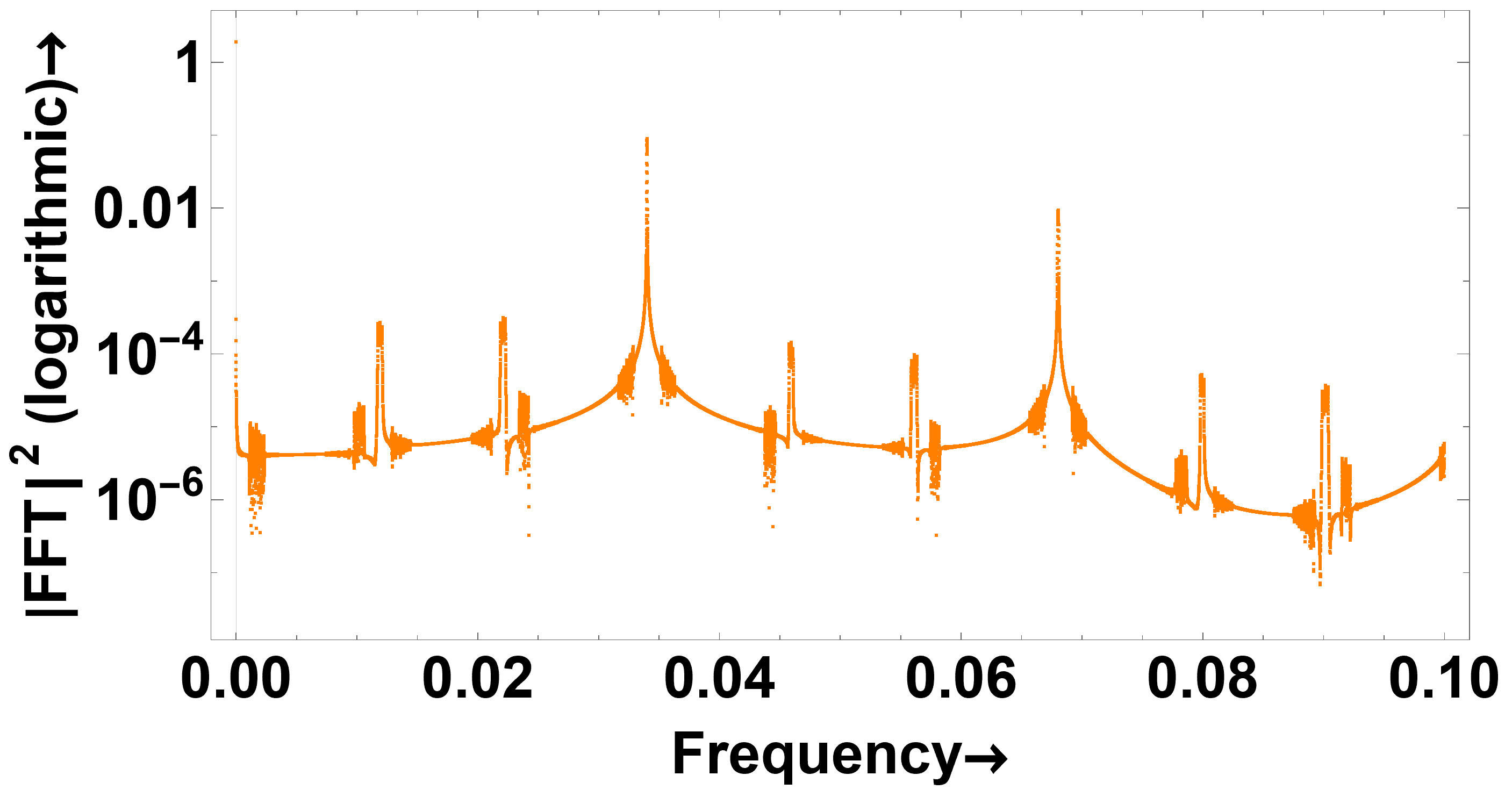}}\\
    \subfigure[$r(\tau)$ profile ($E=2.00$)]{\label{fig:timeseries-susy-high-energy}\includegraphics[width=0.32\linewidth]{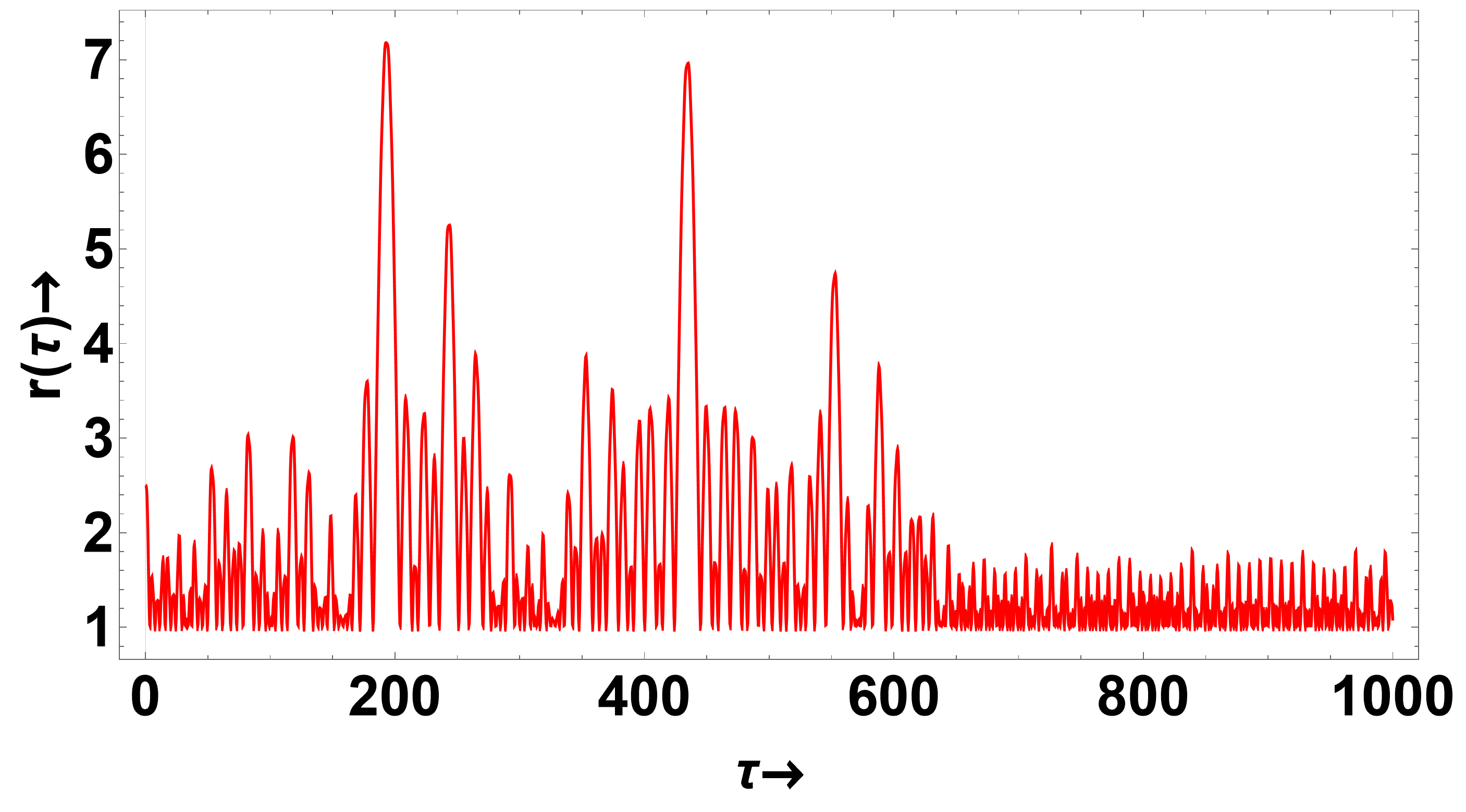}}
    \subfigure[$x(\tau)$ profile ($E=2.00$)]{\label{fig:radiusprofile-susy-high-energy}\includegraphics[width=0.32\linewidth]{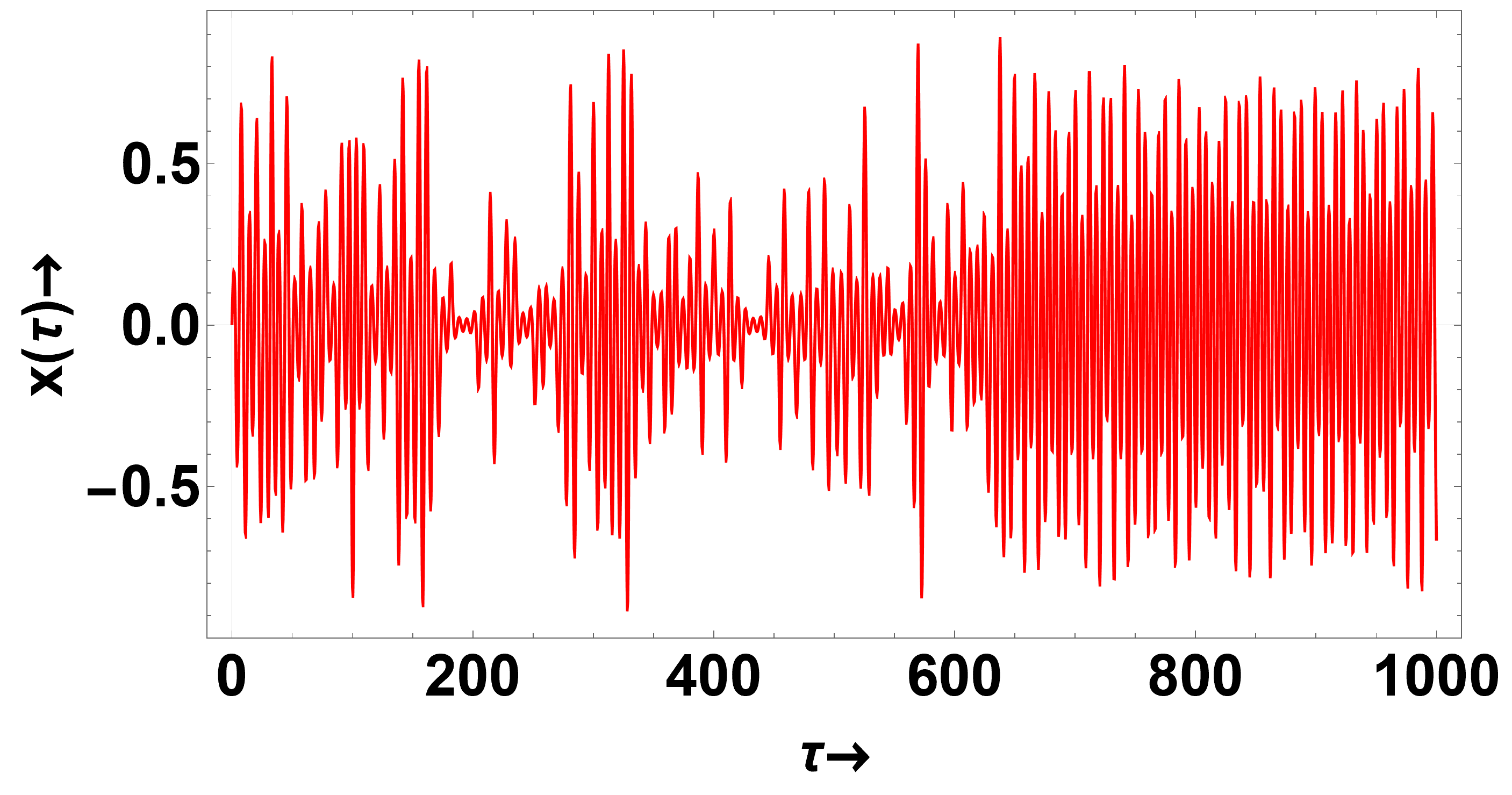}}
    \subfigure[Power spectrum ($E=2.00$)]{\label{fig:powerspectrum-susy-high-energy}\includegraphics[width=0.32\linewidth]{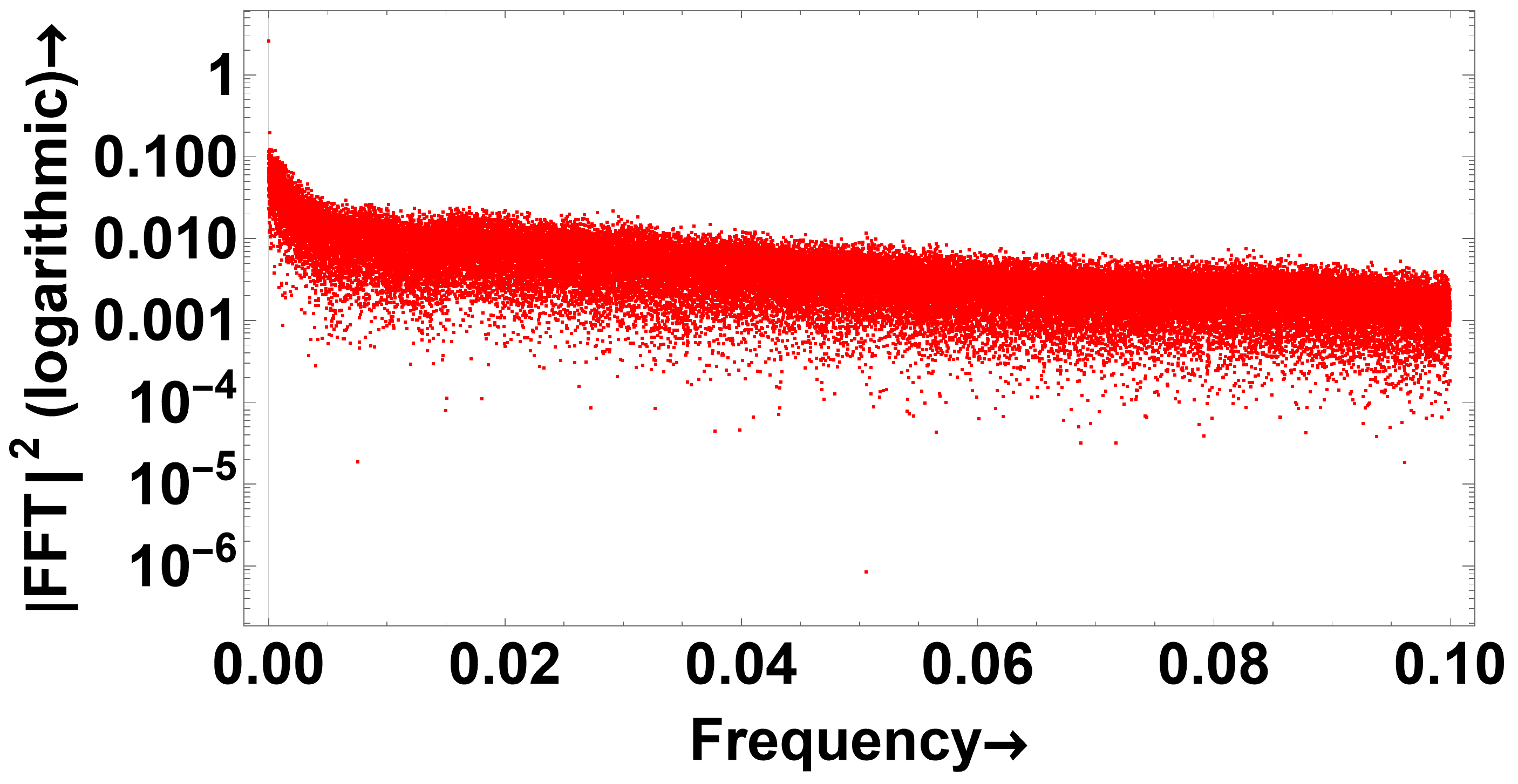}}\\
	\caption{\label{fig:supersymmetric-ads-soliton-timepower-energy}Coordinate profile $r(\tau)$, radius profile $x(\tau)$, and power spectrum plots obtained for different values of $E$. The parameter values are $\alpha=1$, $\ell=1$, $k=0.2$, $r_0=1$, and $q=0.2$. The initial conditions are taken as $r(0)=2.5$, $p_r(0)=0$, and $x(0)=0$. The initial condition $p_x(0)$ is determined using the constrain $\mathcal{H}=0$.}
\end{figure}

The solution for Hamilton's equations of motion gives us the coordinates profile, as shown in Fig.~\ref{fig:supersymmetric-ads-soliton-timepower-energy}, along with its power spectrum (FFT). The plots are for a fixed charge of $q=0.2$. We observe that the motion profiles of $r(\tau)$ and $x(\tau)$ are quasi-periodic at low energy (Fig.~\ref{fig:timeseries-susy-low-energy}), and the corresponding power spectrum has well-defined peaks (Fig.~\ref{fig:powerspectrum-susy-low-energy}). As we increase the energy from $E=0.25$ to $E=2.0$, the motion becomes aperiodic (Fig.~\ref{fig:timeseries-susy-high-energy}), and the corresponding power spectrum becomes noisy (Fig.~\ref{fig:powerspectrum-susy-high-energy}), a hallmark of chaotic systems. These observations are similar to results obtained in \cite{Basu:2011dg}.

It can be inferred from our equations of motion that an exact solution, for $k = 0$, at the tip of our geometry, can be given as 
\begin{eqnarray}
& & r(\tau) = 0\,, \\
& & x(\tau) = A \sin(\tau + \chi)\,,
\end{eqnarray}
where $A$ and $\chi$ are constants of integration. No solution exists for a constant $r(\tau)$ when $k \neq 0$. However, we can construct approximate quasi-periodic solutions for small values of $x(\tau)$ and $p_x(\tau)$. In the equation for $\dot{p}_x$, we can substitute $x(\tau)^2$ with a time-averaged value, allowing the motion in the $r$-direction to approximately follow the behavior of an anharmonic oscillator dependent on a single variable, which is theoretically solvable.

Another approximation can be made by treating $r(\tau)$ as a slowly varying field to obtain a quasi-periodic and semi-analytical expression for $x(\tau)$. Under this treatment, we obtain the following functional form for $x(\tau)$:
\begin{equation}
x(\tau) \approx  A \exp(-a r(\tau)^2) \sin(\tau + \chi).
\end{equation}
To explore the validity of our quasi-periodic solution [Fig. \ref{fig:radiusprofile-susy-low-energy}], we fit our solution for $x(\tau)$ with the numerical data points of $r(\tau)$ under a small $x$ regime and obtain the following values: 
\[
A = -0.024, \quad a = -0.24, \quad \chi = 2.
\]
An overlap of our semi-analytical solution with the numerical radius profile $x(\tau)$ justifies our statement that the motion in $r$ behaves like an anharmonic oscillator for small $x$. As we move further from this limit, the two-scale analysis loses validity, and the effect of the non-linear coupling term grows significant. In other words, the interaction between the oscillators strengthens as the string's energy increases. This non-linearity causes the fluctuations in the $x$ and $r$ coordinates to affect each other's motion, leading to aperiodic behavior. Ultimately, the system transitions to a fully chaotic state, as evident in [Fig. \ref{fig:radiusprofile-susy-high-energy}]. 

\begin{figure}[htbp!]
	\centering
	\subfigure[$r(\tau)$ profile ($q=0.0$)]{\label{fig:timeseries-susy-low-charge}\includegraphics[width=0.32\linewidth]{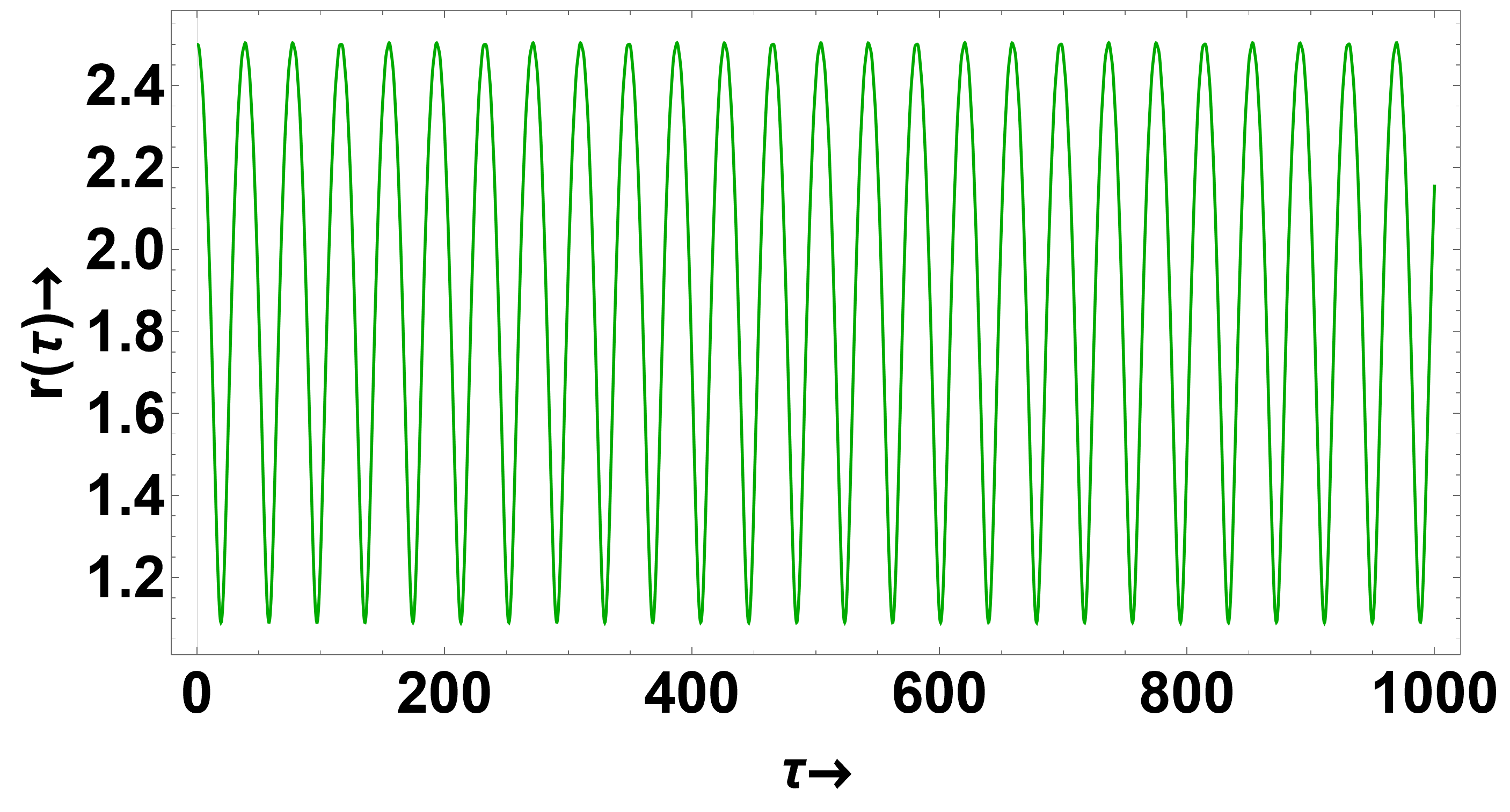}}
    \subfigure[$x(\tau)$ profile ($q=0.0$)]{\label{fig:radiusprofile-susy-low-charge}\includegraphics[width=0.32\linewidth]{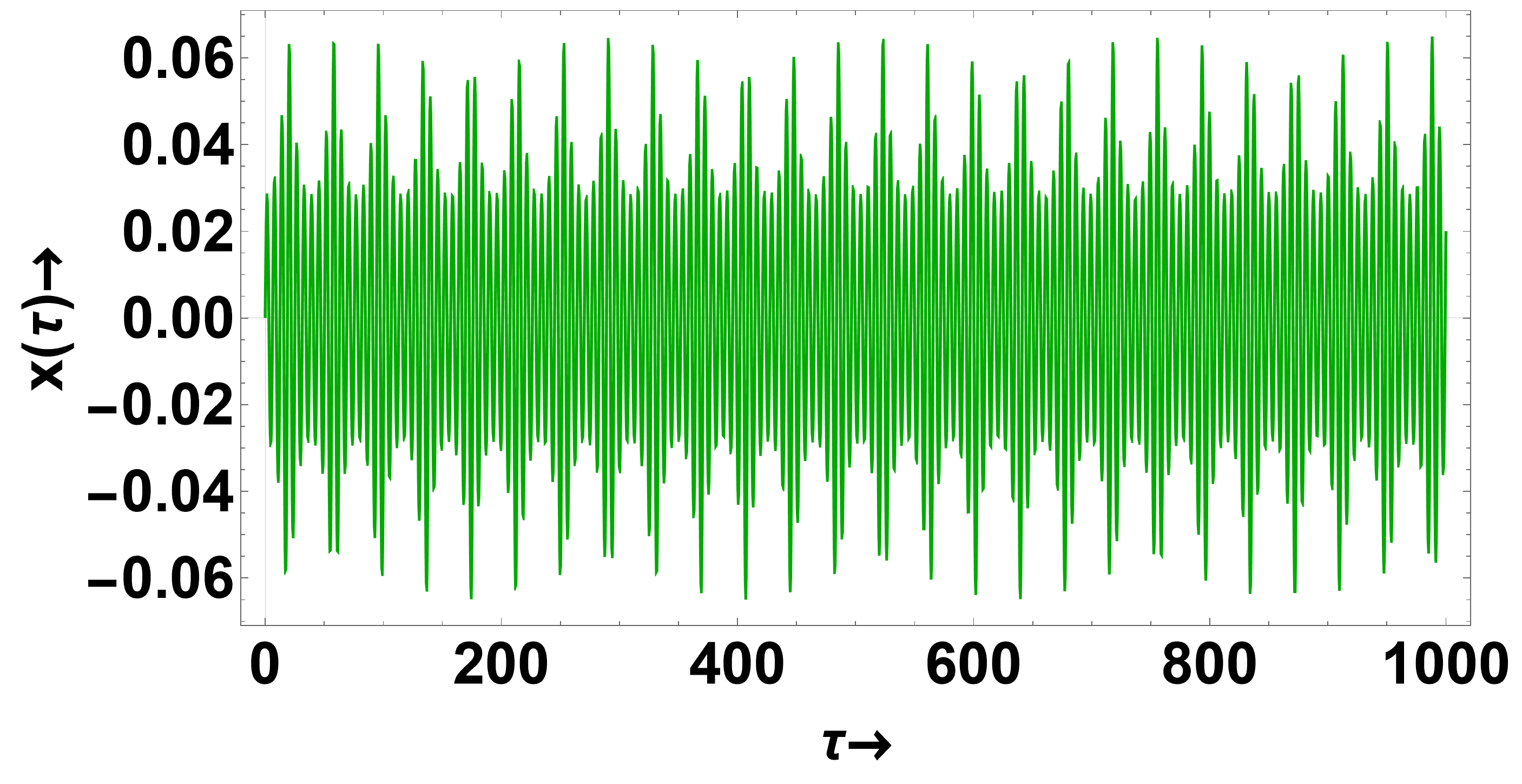}}
    \subfigure[Power spectrum ($q=0.0$)]{\label{fig:powerspectrum-susy-low-charge}\includegraphics[width=0.32\linewidth]{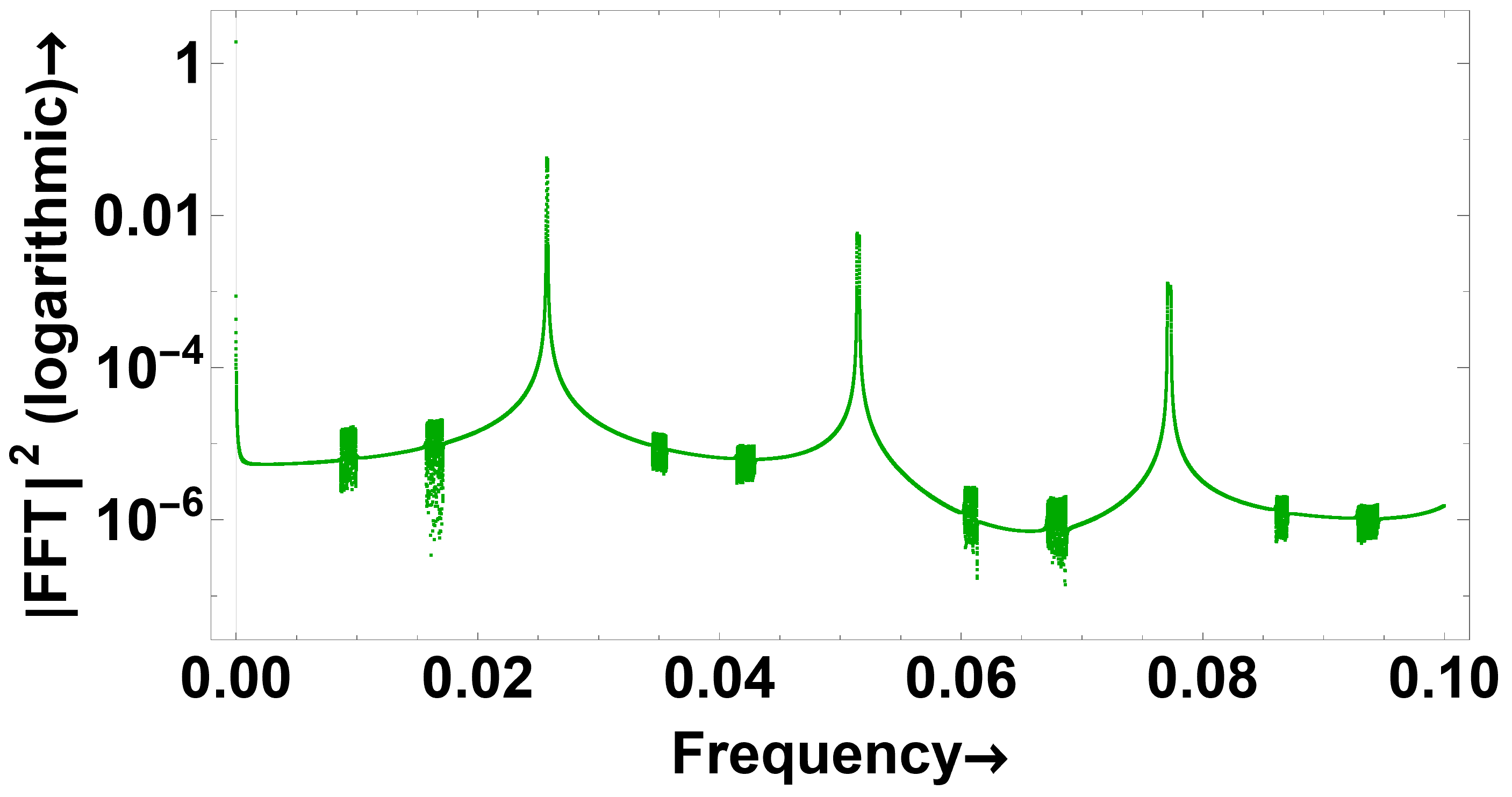}}\\
    \subfigure[$r(\tau)$ profile ($q=0.5$)]{\label{fig:timeseries-susy-medium-charge}\includegraphics[width=0.32\linewidth]{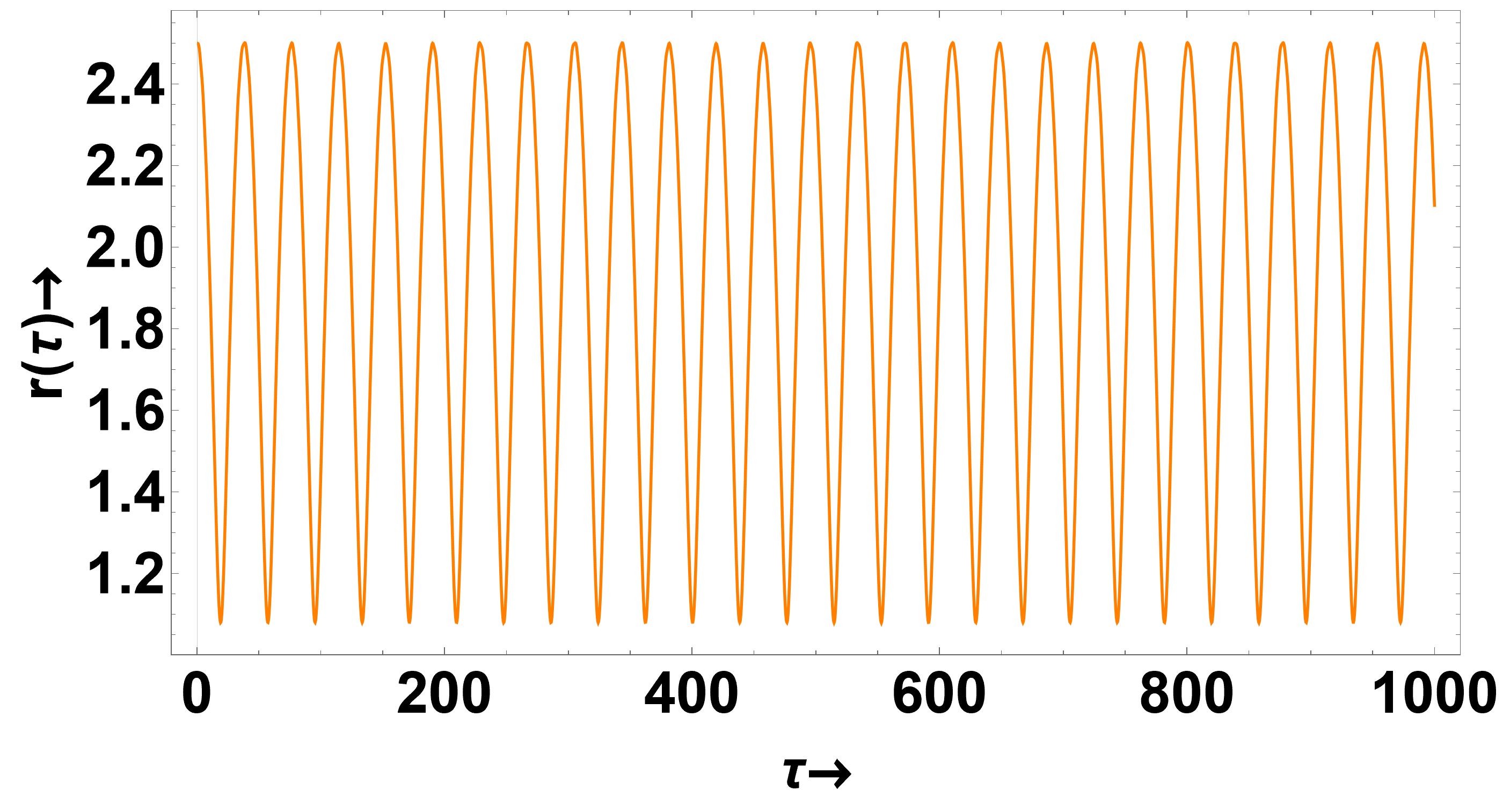}}
    \subfigure[$x(\tau)$ profile ($q=0.5$)]{\label{fig:radiusprofile-susy-medium-charge}\includegraphics[width=0.32\linewidth]{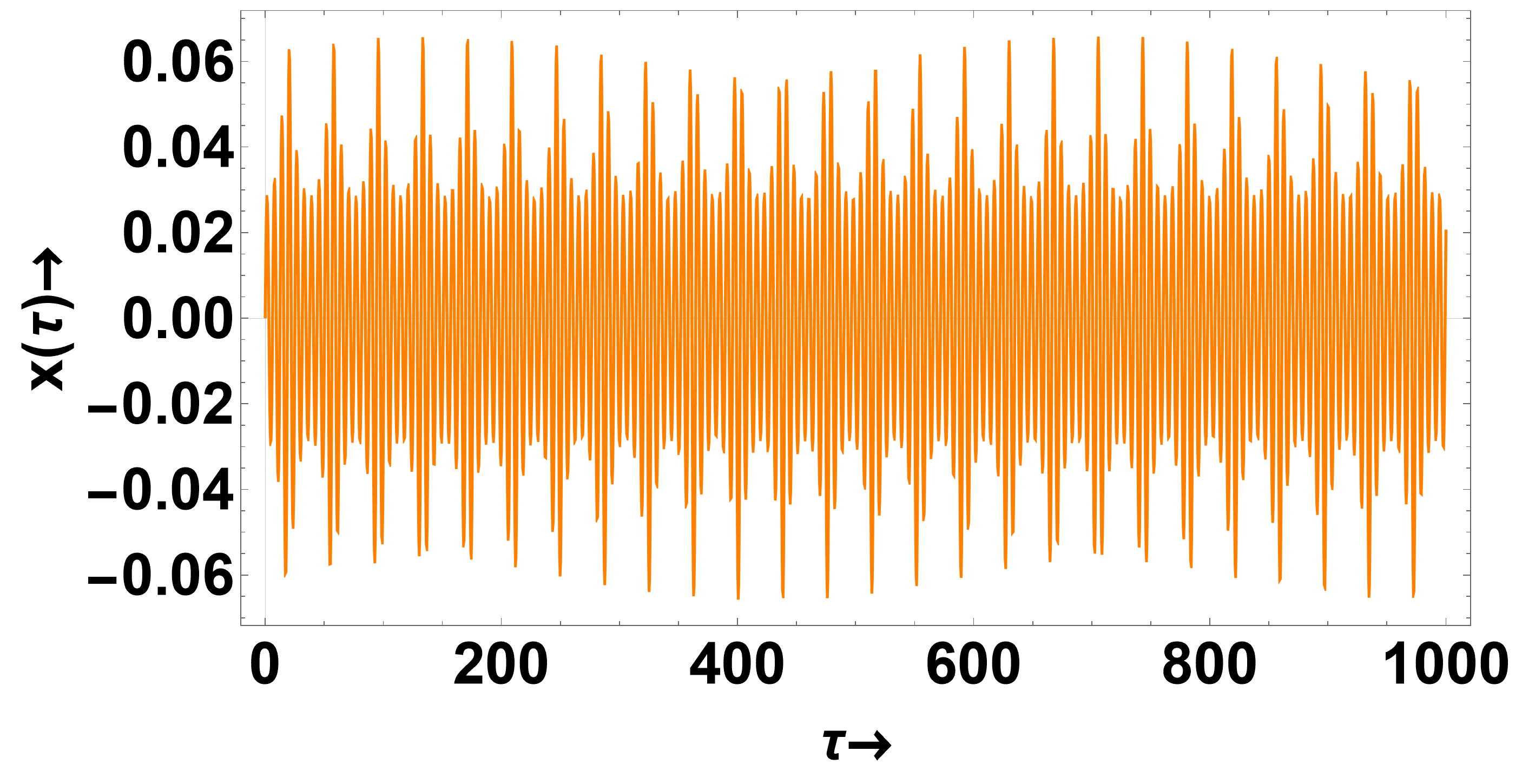}}
    \subfigure[Power spectrum ($q=0.5$)]{\label{fig:powerspectrum-susy-medium-charge}\includegraphics[width=0.32\linewidth]{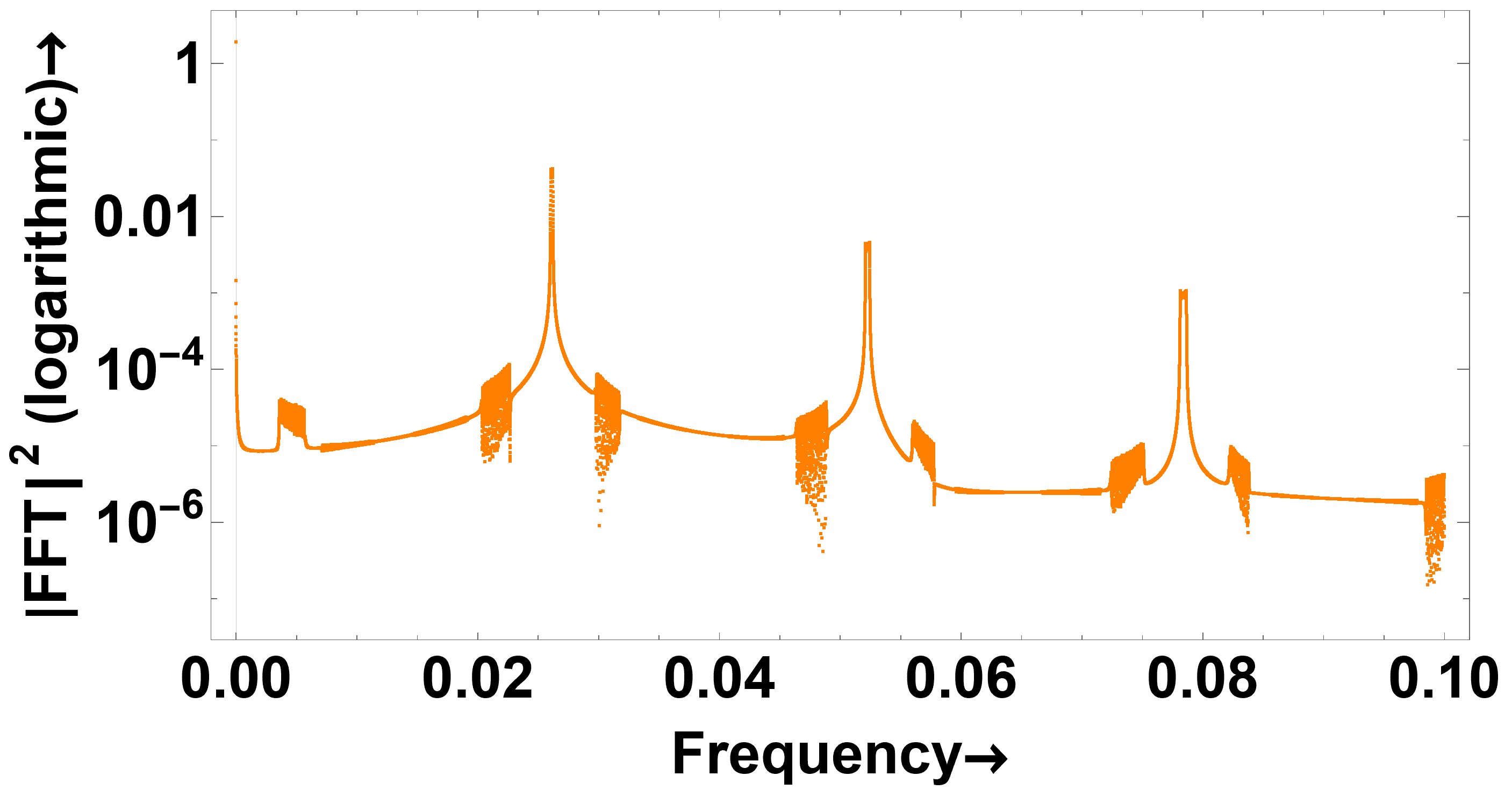}}\\
    \subfigure[$r(\tau)$ profile ($q=0.9$)]{\label{fig:timeseries-susy-high-charge}\includegraphics[width=0.32\linewidth]{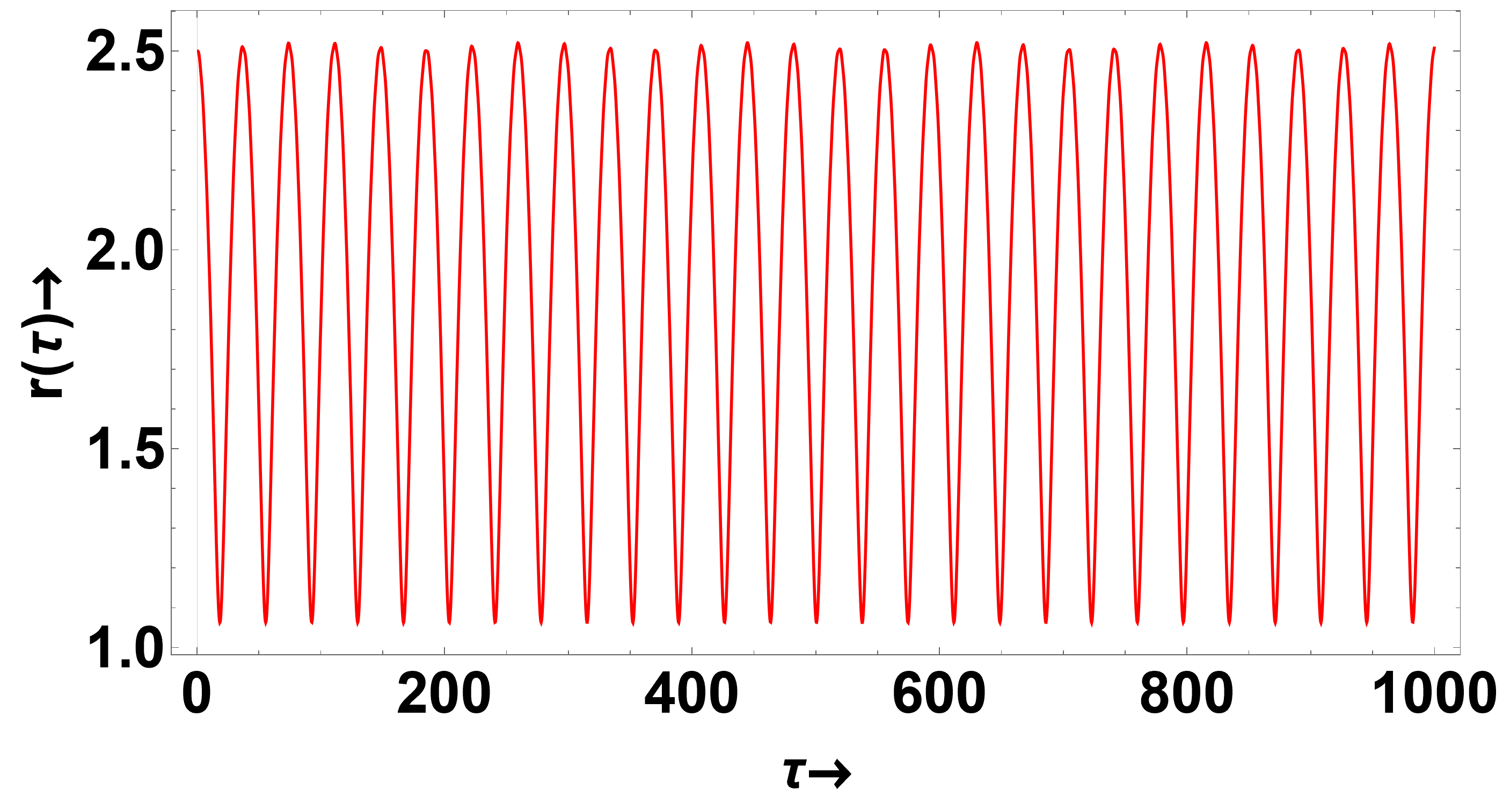}}
    \subfigure[$x(\tau)$ profile ($q=0.9$)]{\label{fig:radiusprofile-susy-high-charge}\includegraphics[width=0.32\linewidth]{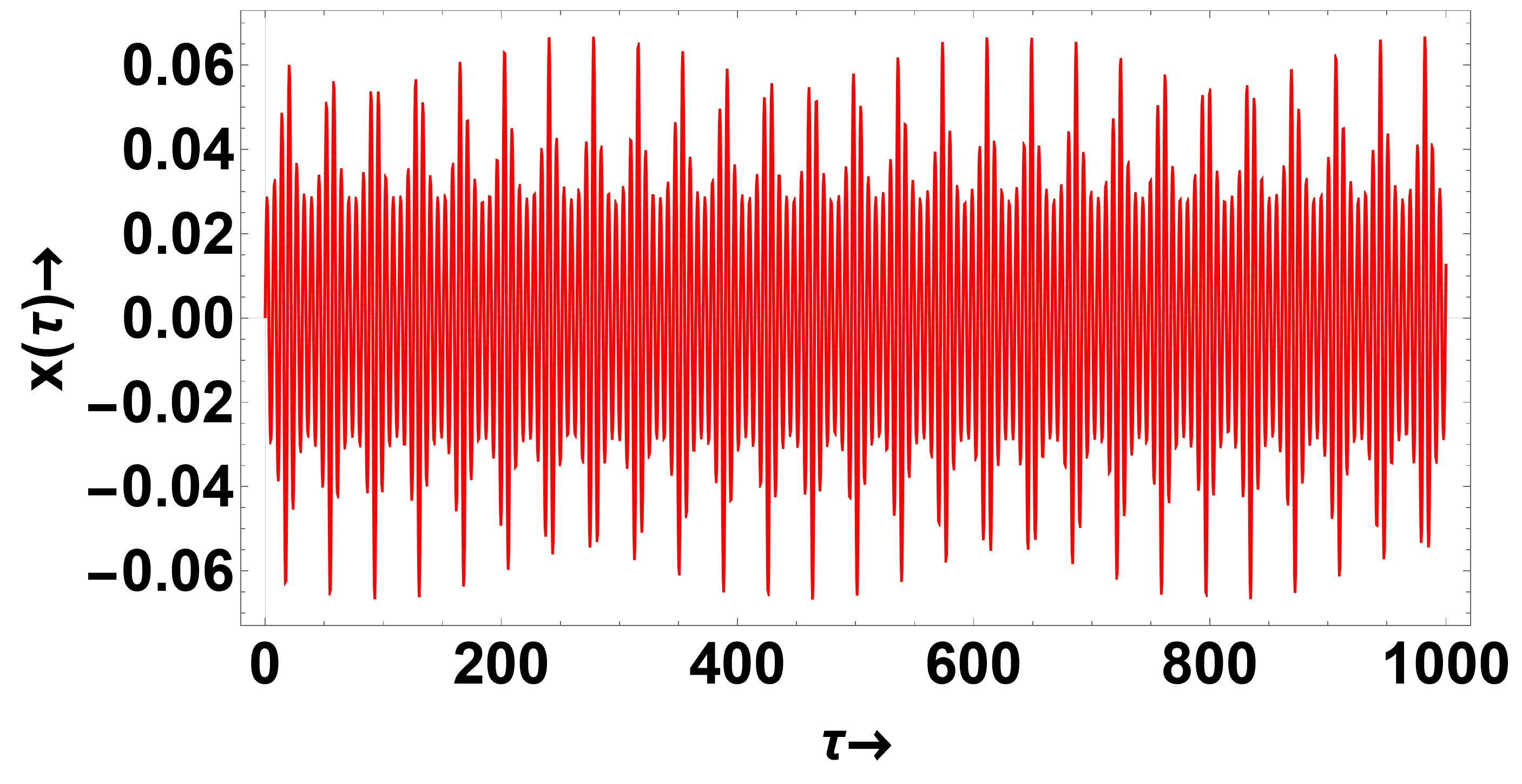}}
    \subfigure[Power spectrum ($q=0.9$)]{\label{fig:powerspectrum-susy-high-charge}\includegraphics[width=0.32\linewidth]{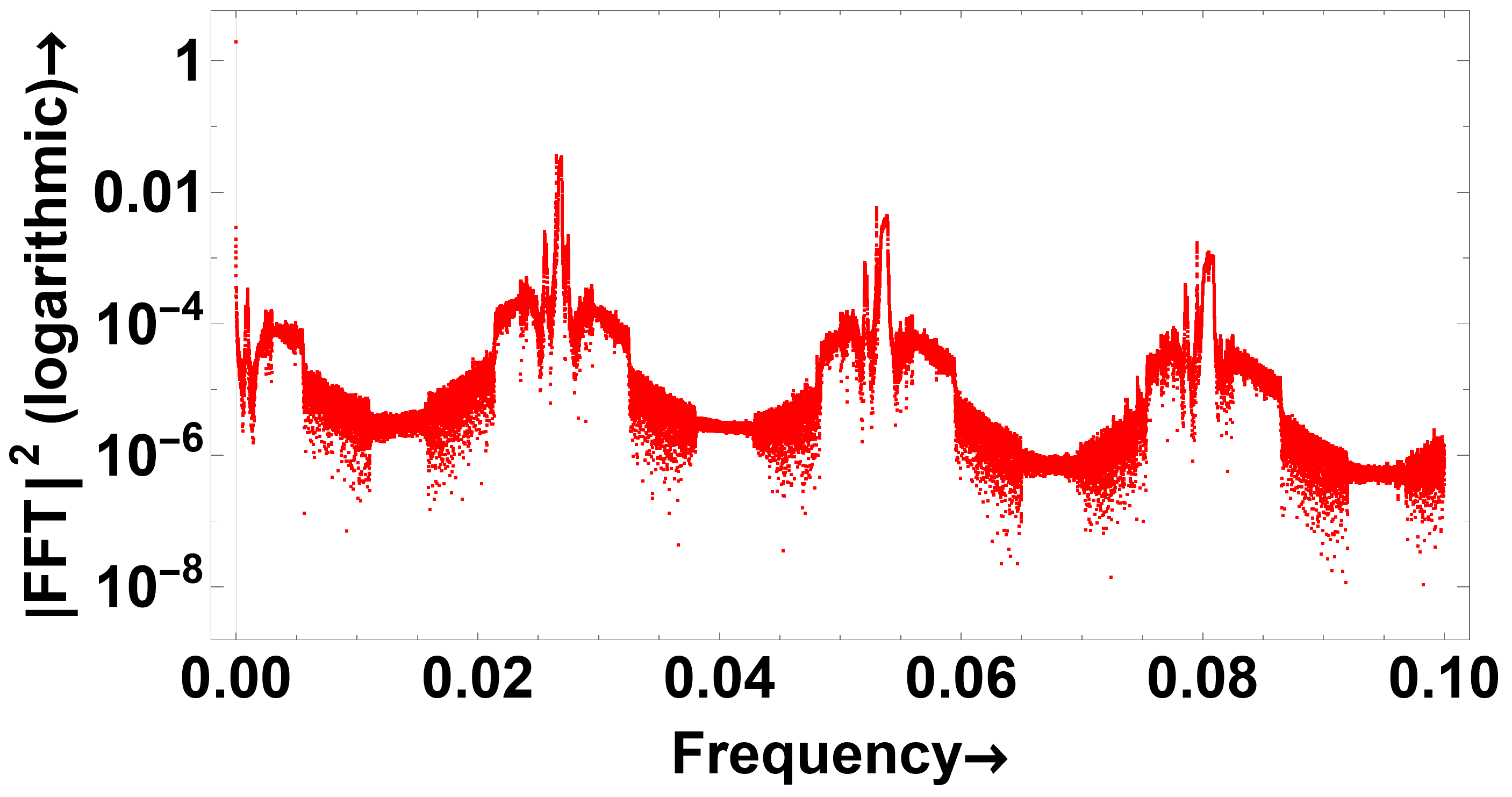}}\\
	\caption{\label{fig:supersymmetric-ads-soliton-timepower-charge}Coordinate profile $r(\tau)$, radius profile $x(\tau)$, and power spectrum plots obtained for small and large values of charge $q$. The parameter values are $\alpha=1$, $\ell=1$, $k=0.2$, $r_0=1$, and $E=0.4$. The initial conditions are taken as $r(0)=2.5$, $p_r(0)=0$, and $x(0)=0$. The initial condition $p_x(0)$ is determined using the constrain $\mathcal{H}=0$.}
\end{figure}

It is also instructive to analyze $r(\tau)$ profile and power spectrum for different charges. The results are shown in Fig.~\ref{fig:supersymmetric-ads-soliton-timepower-charge}. Here, we have fixed the energy to $E=0.4$ and varied the charge from $q=0$ to $q=0.9$. We observe substantial differences in the motion profile and power spectrum as we sequentially vary the charge. In particular, for a low charge, the motion is quasi-periodic with the corresponding power spectrum exhibiting well-defined peaks, whereas for a higher value of charge, the motion becomes aperiodic, and the power spectrum shows peaks getting diffused, effectively turning noisy. One could also see mixed behavior for the intermediate value of $q$. These results indicate that similar to $E$, increasing values of $q$ have a constructive effect in inducing chaos in the closed string dynamics. This result is further explored and confirmed from the Poincar\'{e} sections and Lyapunov exponents analysis in the subsequent sub-sections.

\subsection{Poincar\'{e} section}\label{sec:supersymmetricpoincare}
The Kolmogorov-Arnold-Moser (KAM) theorem~\cite{Gutzwiller1990} provides profound insight into the stability of Hamiltonian systems under small perturbations. It establishes that, for sufficiently small perturbations, most of the invariant tori of an integrable system remain intact while others break down, giving rise to chaotic trajectories. The theorem thus reveals a complex structure in the phase space of the perturbed system, where regions of regular motion, corresponding to the surviving tori, coexist with chaotic zones arising from the destruction of others. As the perturbation grows, the chaotic regions progressively expand, encroaching upon and destabilizing the ordered behavior typical of integrable systems. This delicate balance between stability and chaos underpins many physical phenomena, demonstrating how order can persist even as a system approaches chaotic dynamics.

\begin{figure}[htbp!]
	\centering
	\subfigure[$E=0.2$]{\label{fig:supersymmetric-ads-soliton-poincare-E0pt2}\includegraphics[width=0.35\linewidth, height=0.35\linewidth]{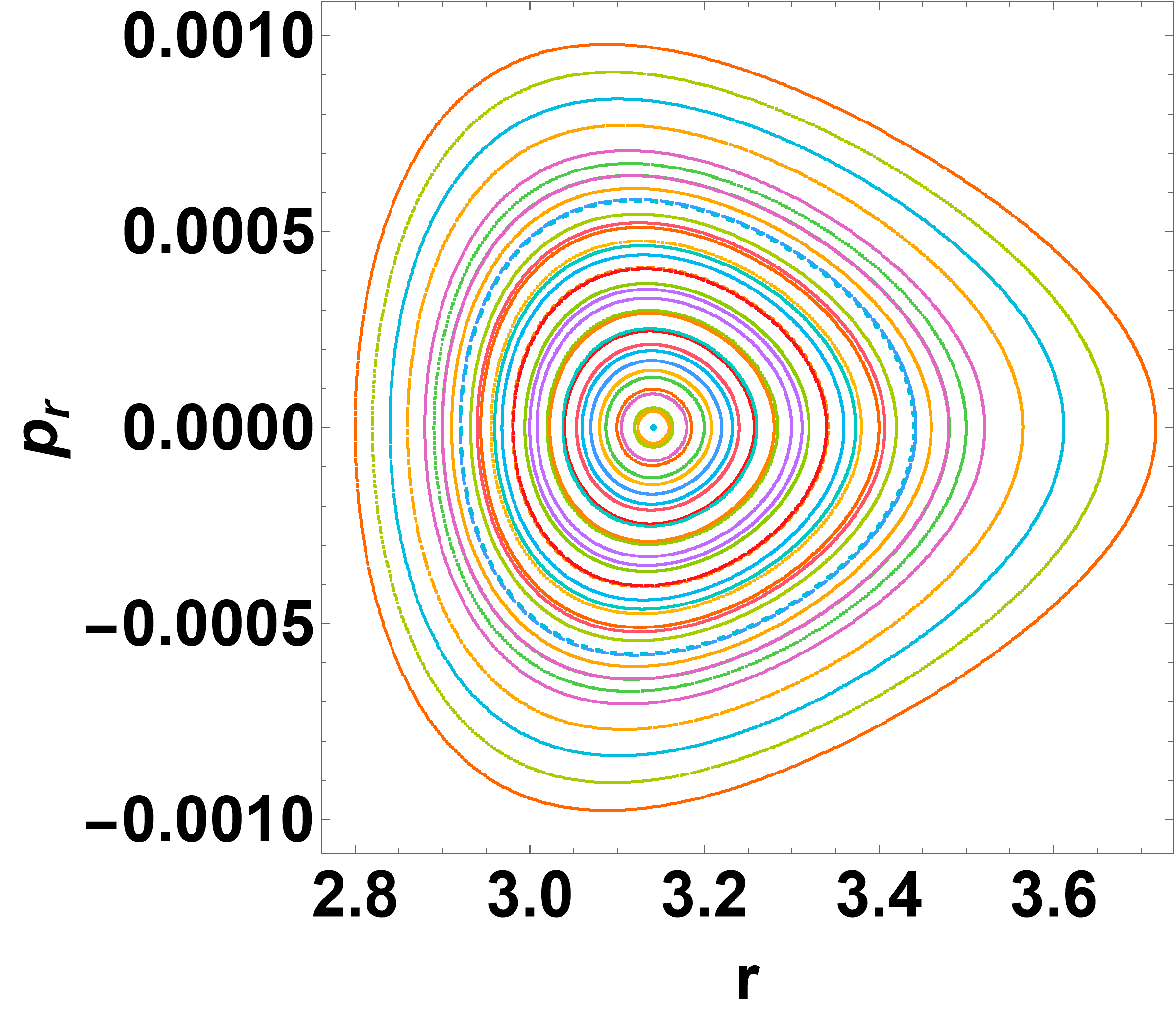}}\hspace{1cm}
    \subfigure[$E=0.4$]{\label{fig:supersymmetric-ads-soliton-poincare-E0pt4}\includegraphics[width=0.35\linewidth, height=0.35\linewidth]{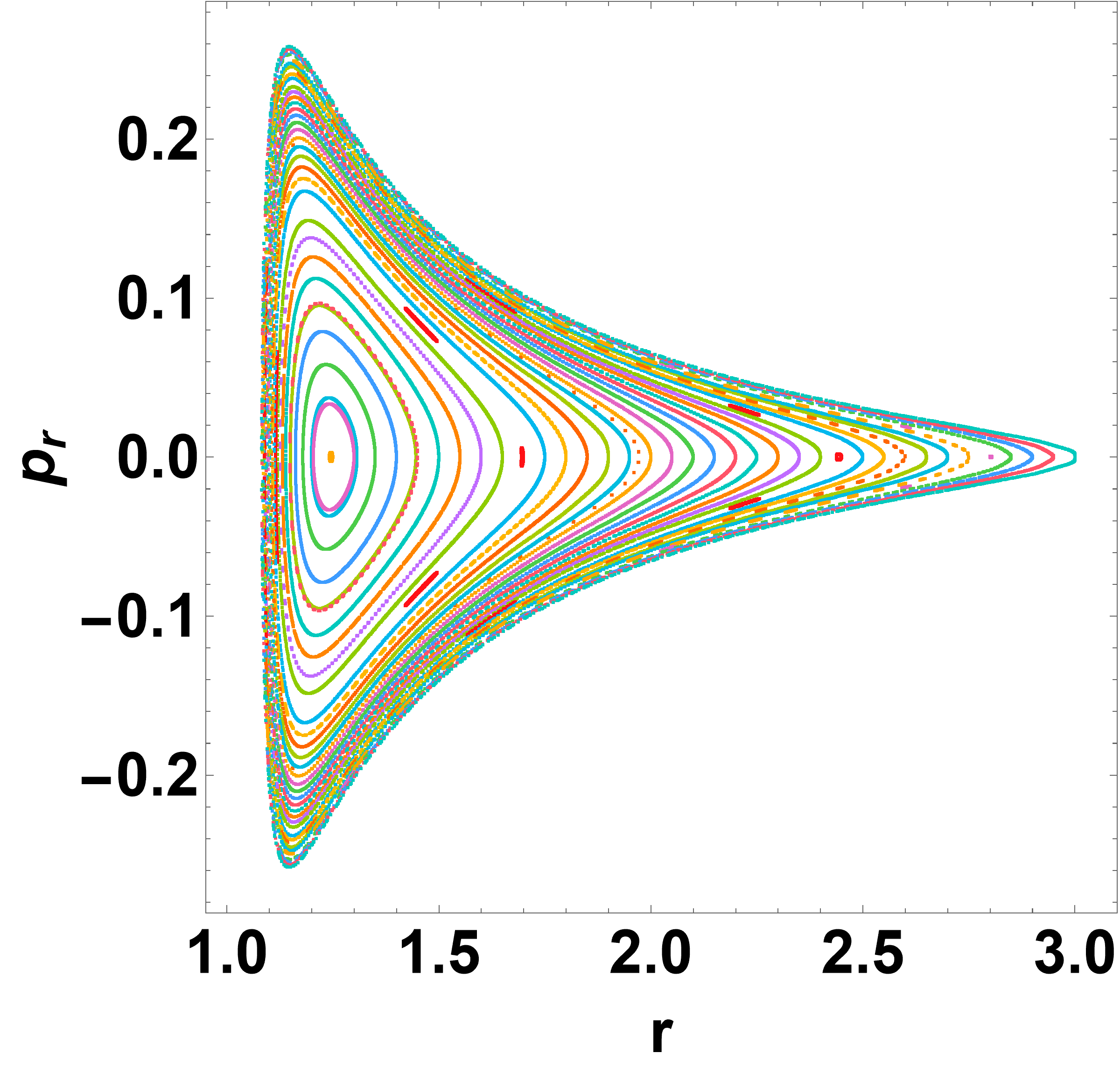}}\\
    \subfigure[$E=0.6$]{\label{fig:supersymmetric-ads-soliton-poincare-E0pt6}\includegraphics[width=0.35\linewidth, height=0.35\linewidth]{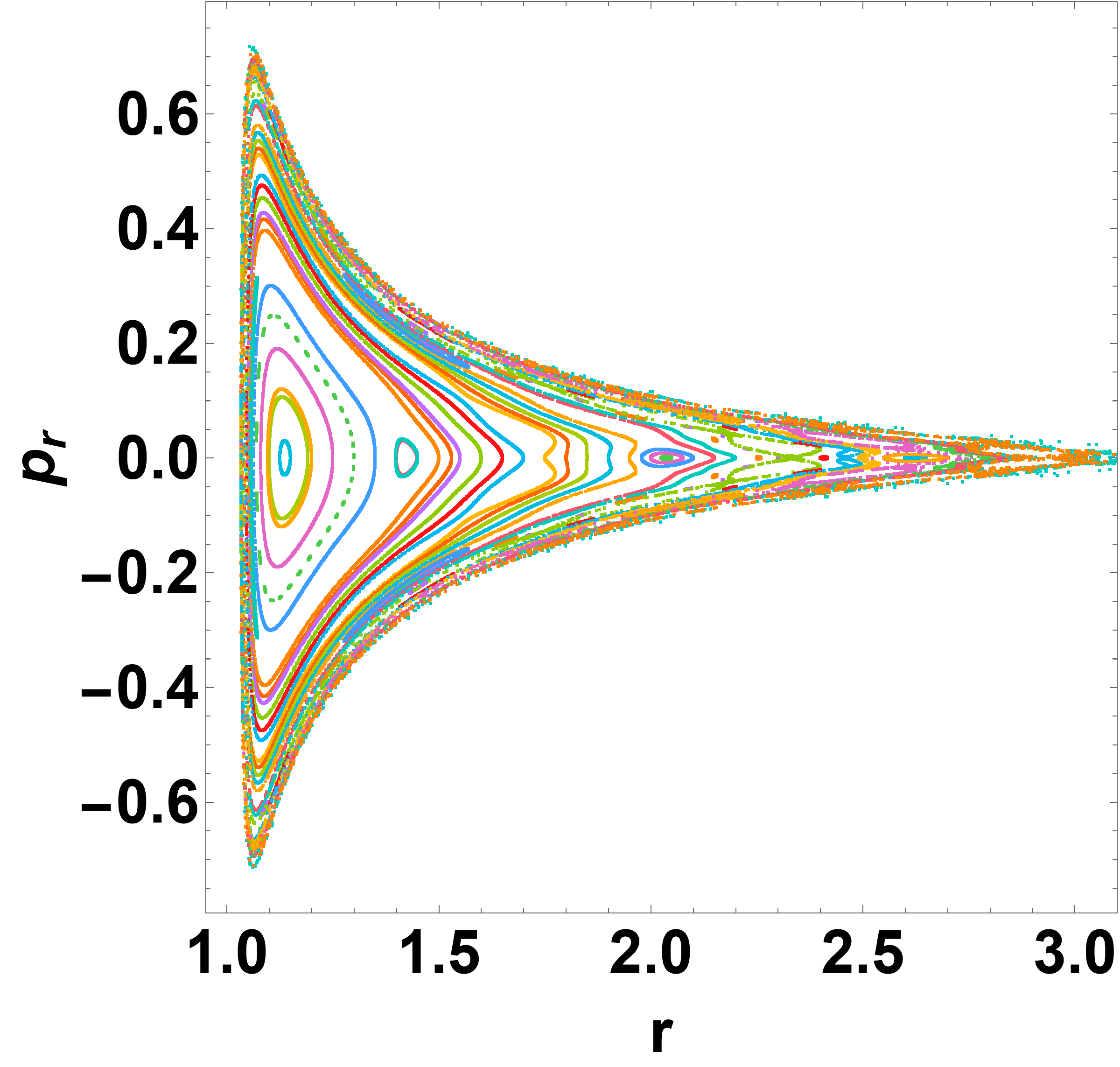}}\hspace{1cm}
    \subfigure[$E=0.8$]{\label{fig:supersymmetric-ads-soliton-poincare-E0pt8}\includegraphics[width=0.35\linewidth, height=0.35\linewidth]{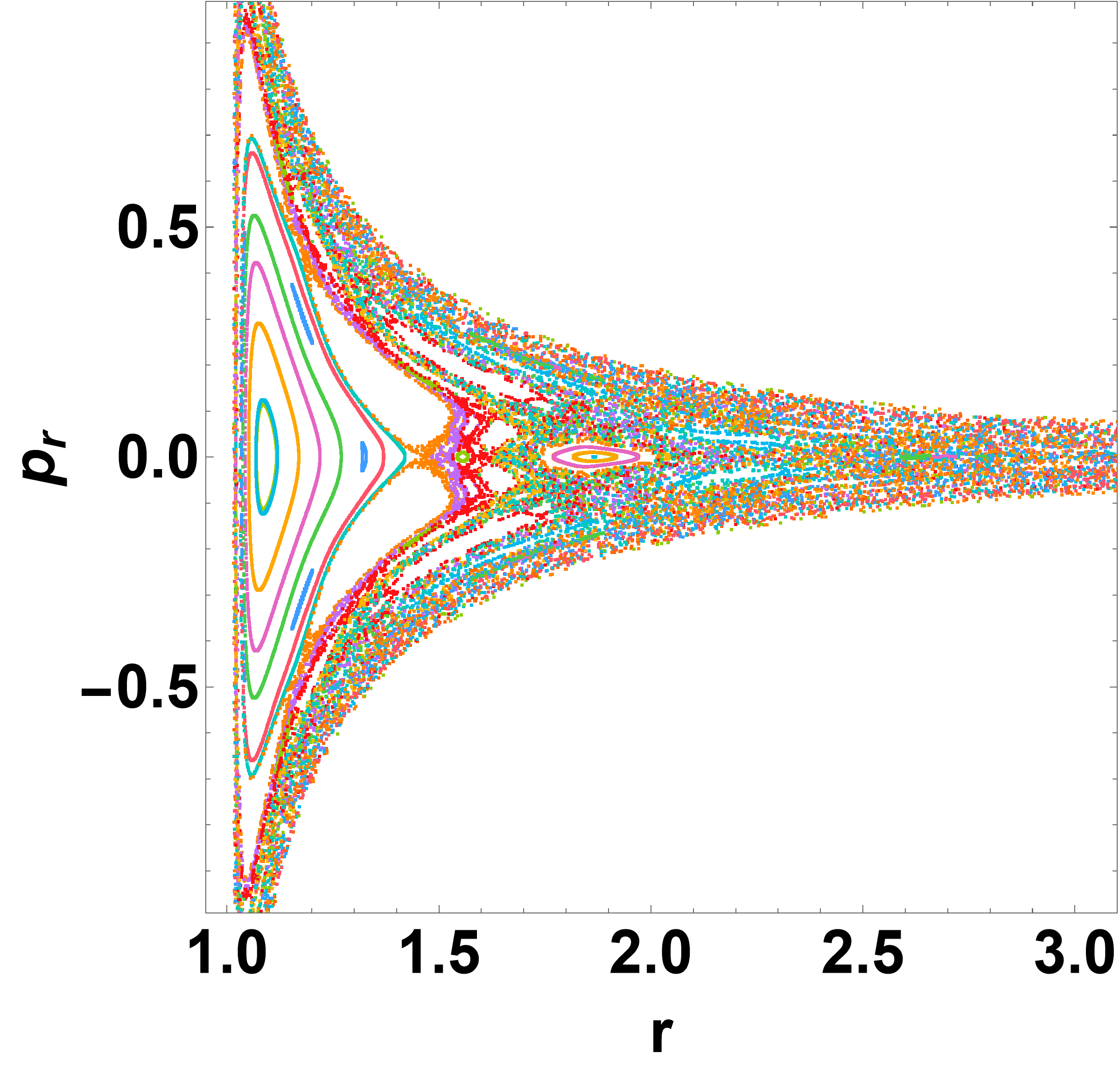}}\\
    \subfigure[$E=1.0$]{\label{fig:supersymmetric-ads-soliton-poincare-E1pt0}\includegraphics[width=0.35\linewidth, height=0.35\linewidth]{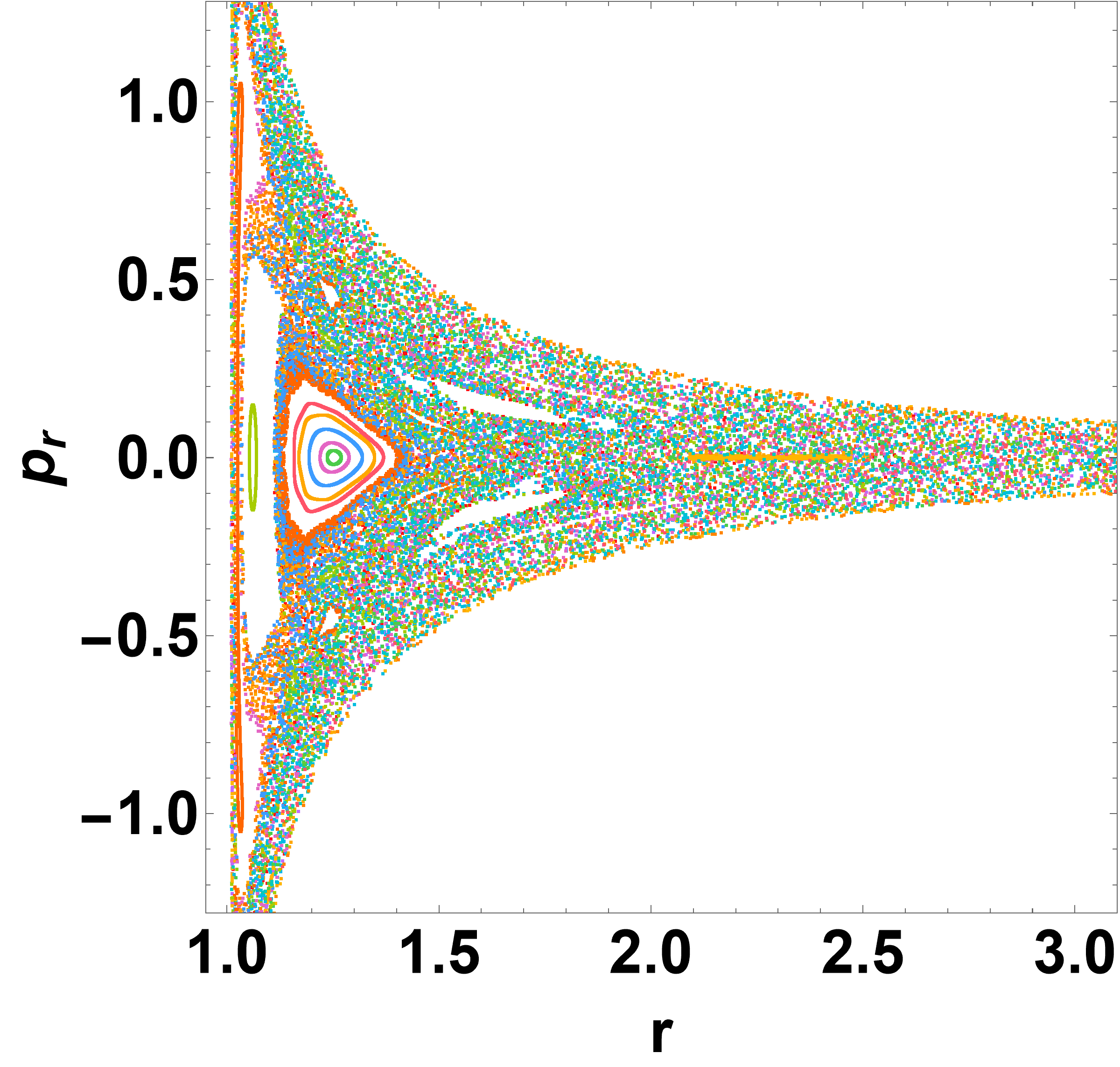}}\hspace{1cm}
    \subfigure[$E=1.2$]{\label{fig:supersymmetric-ads-soliton-poincare-E1pt2}\includegraphics[width=0.35\linewidth, height=0.35\linewidth]{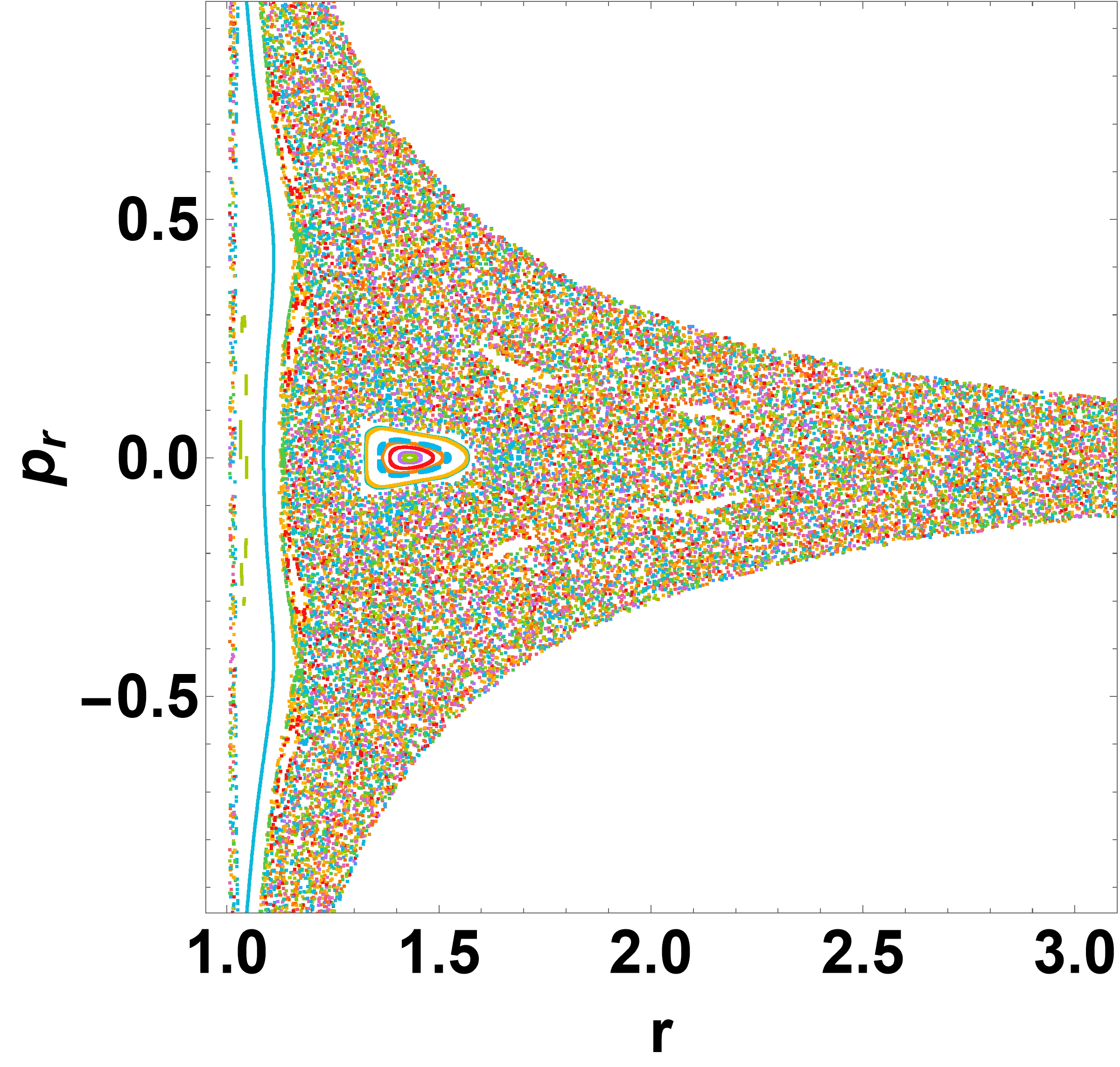}}
	\caption{\label{fig:supersymmetric-ads-soliton-poincare-E}Poincar\'{e} sections for different values of energy. Here charge $q=0.1$ is fixed and $\alpha=1$, $\ell=1$, $k=0.2$, and $r_0=1$ are used. }
\end{figure}

\begin{figure}[htbp!]
	\centering
	\begin{tabular}{c c c}
		\textbf{Charge} & \textbf{$E=0.6$} & \textbf{$E=0.9$} \\
		\textbf{$q=0.0$} & \includegraphics[scale=0.26,valign=c]{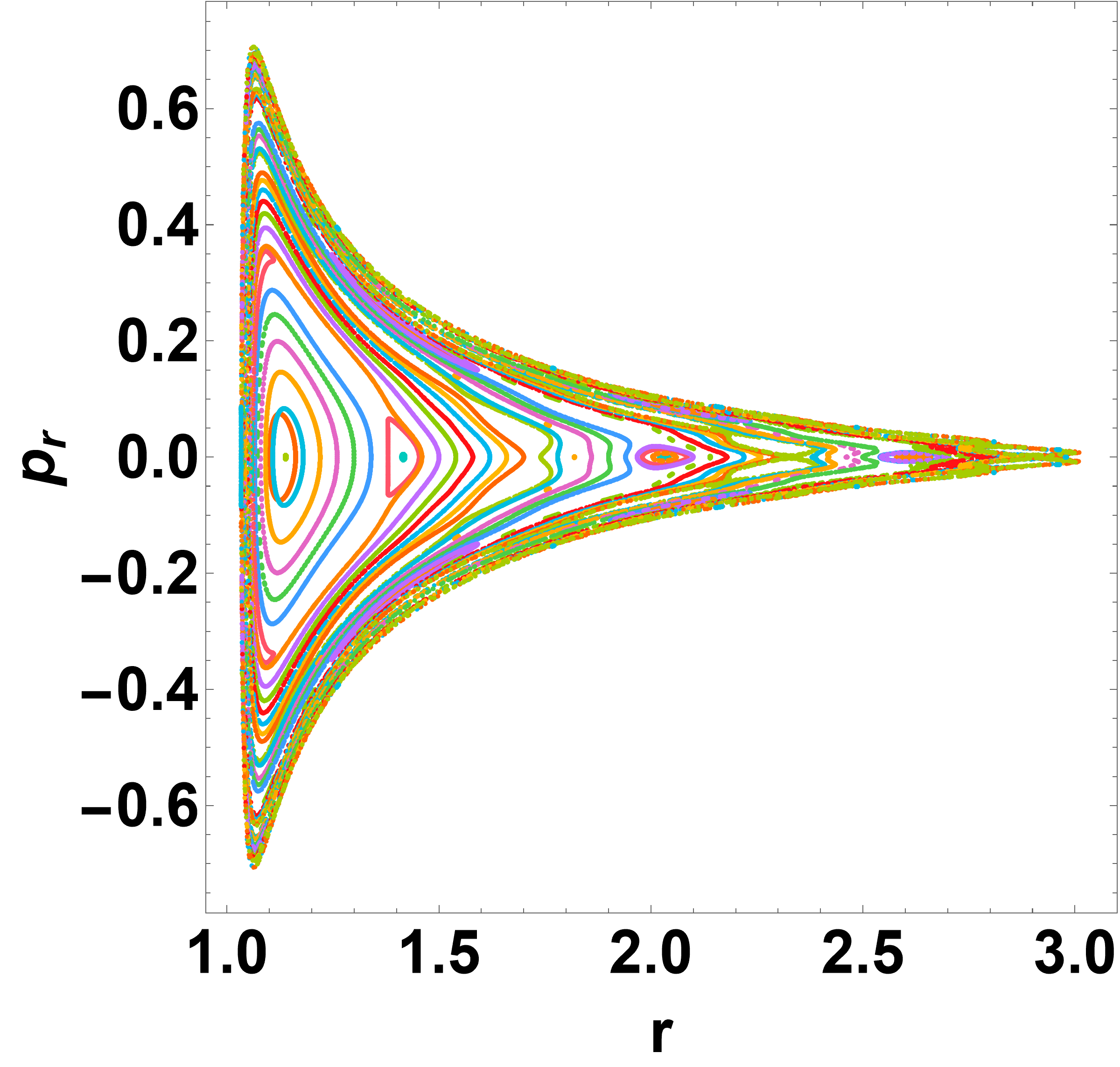} & \includegraphics[scale=0.26,valign=c]{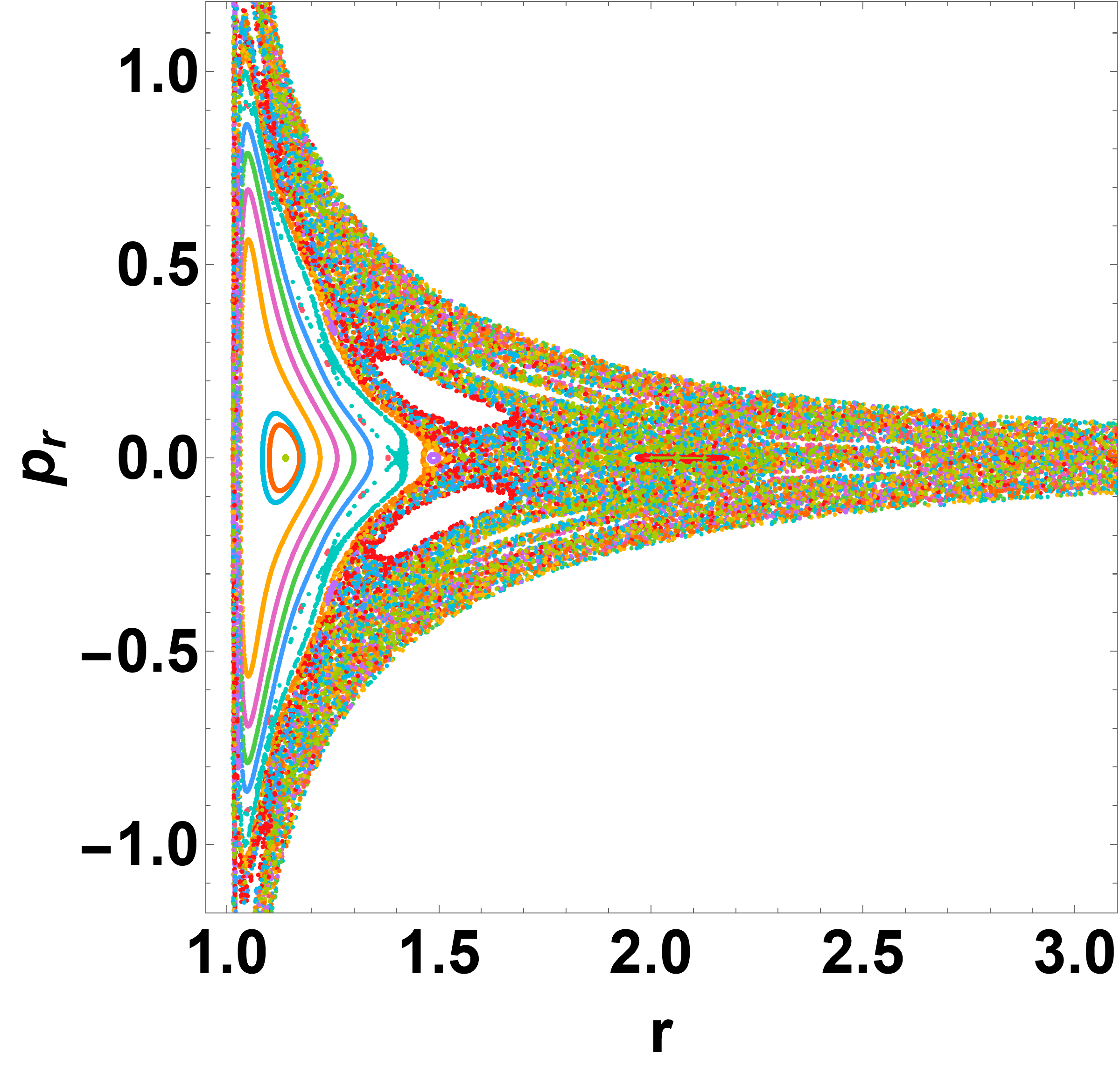} \\
		\textbf{$q=0.3$} & \includegraphics[scale=0.26,valign=c]{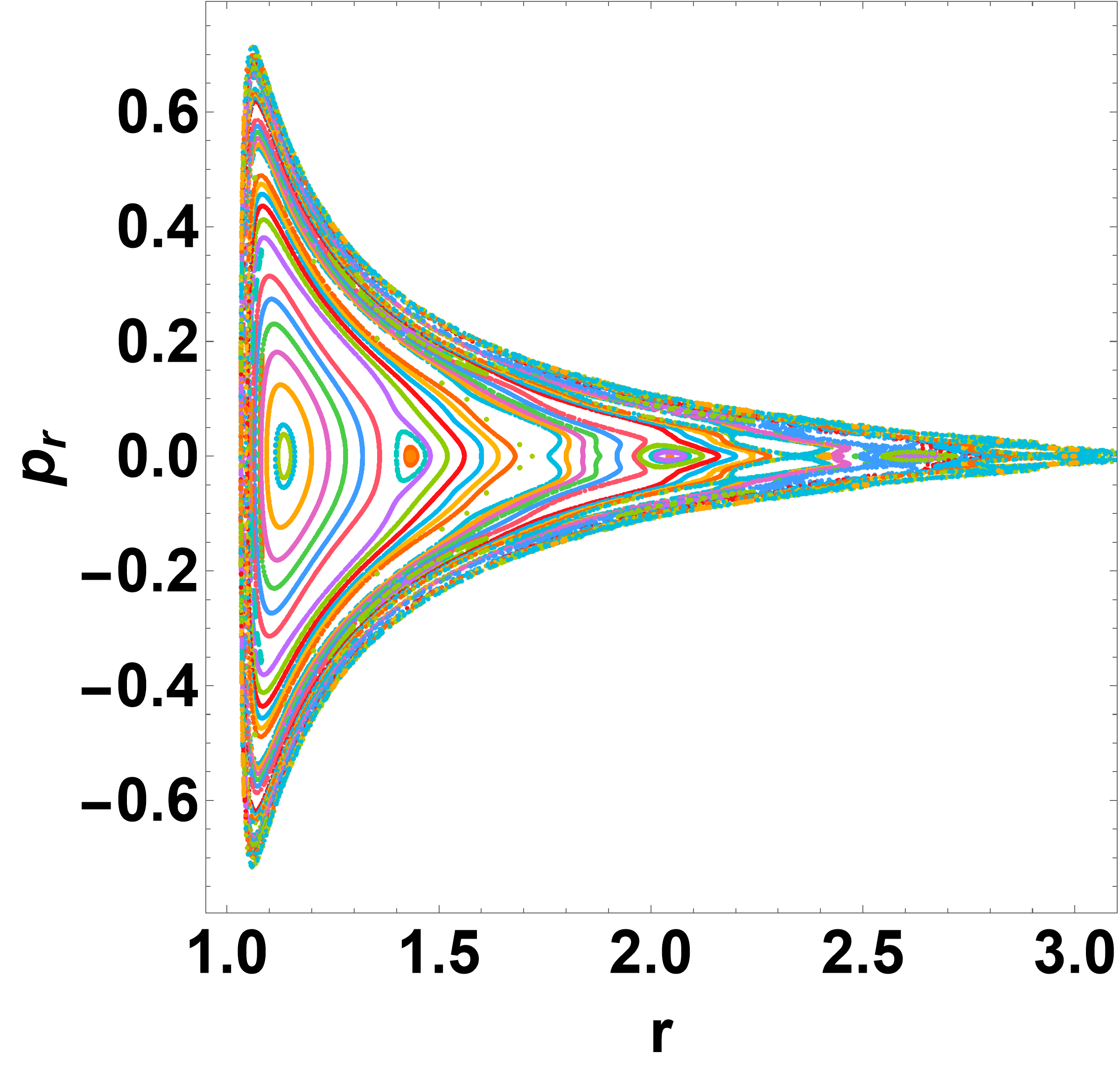} & \includegraphics[scale=0.26,valign=c]{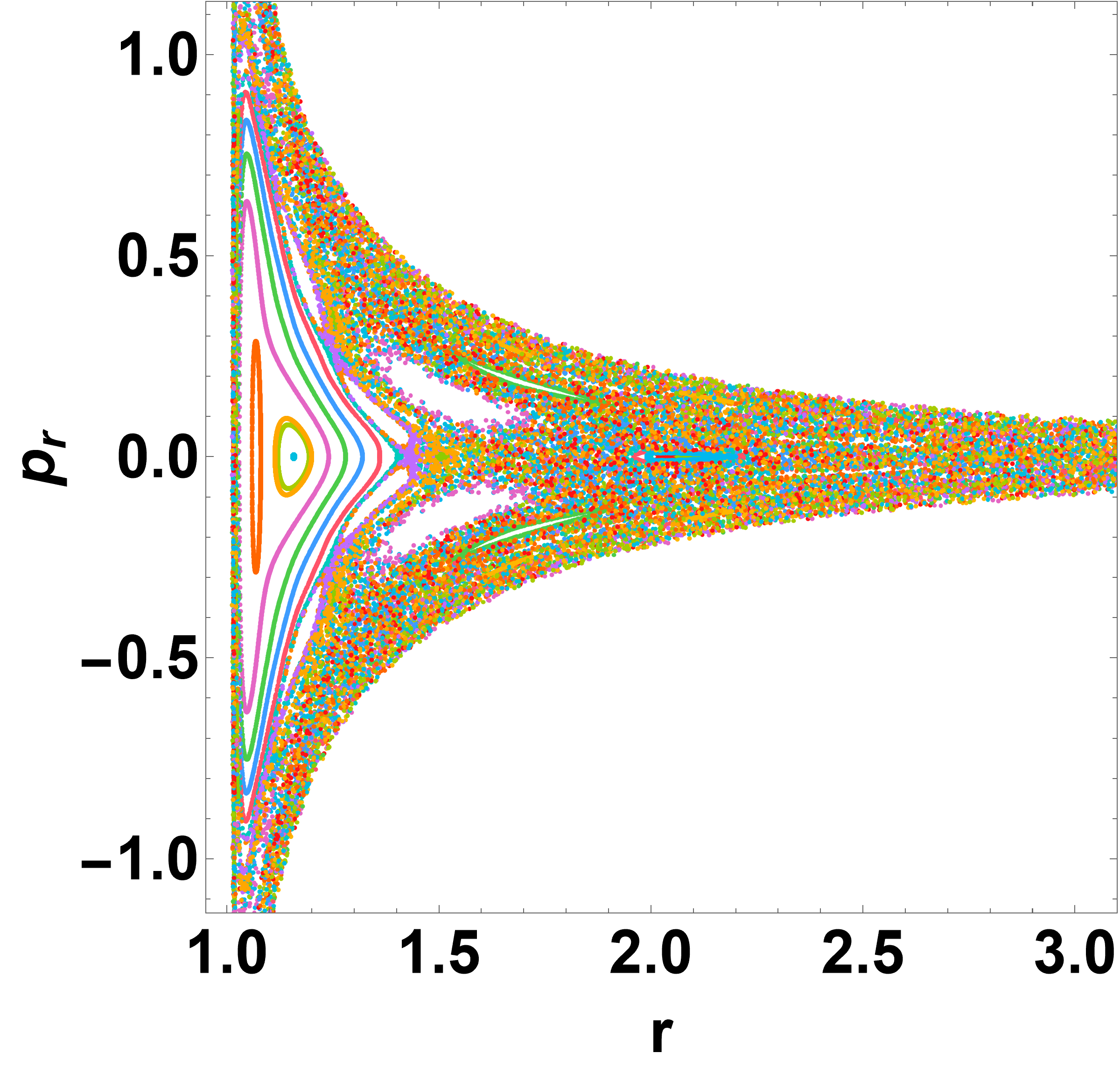} \\
		\textbf{$q=0.6$} & \includegraphics[scale=0.26,valign=c]{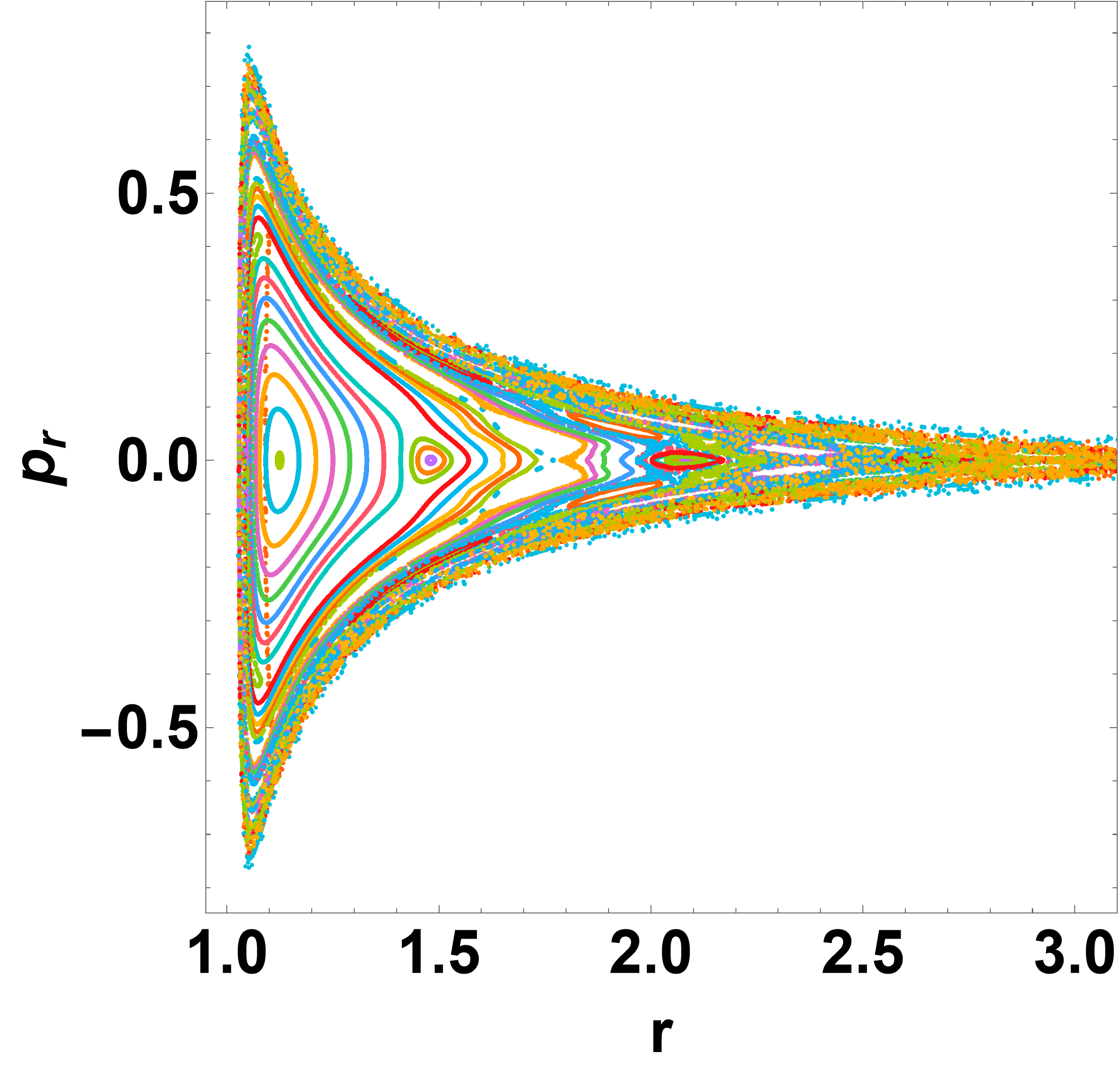} & \includegraphics[scale=0.26,valign=c]{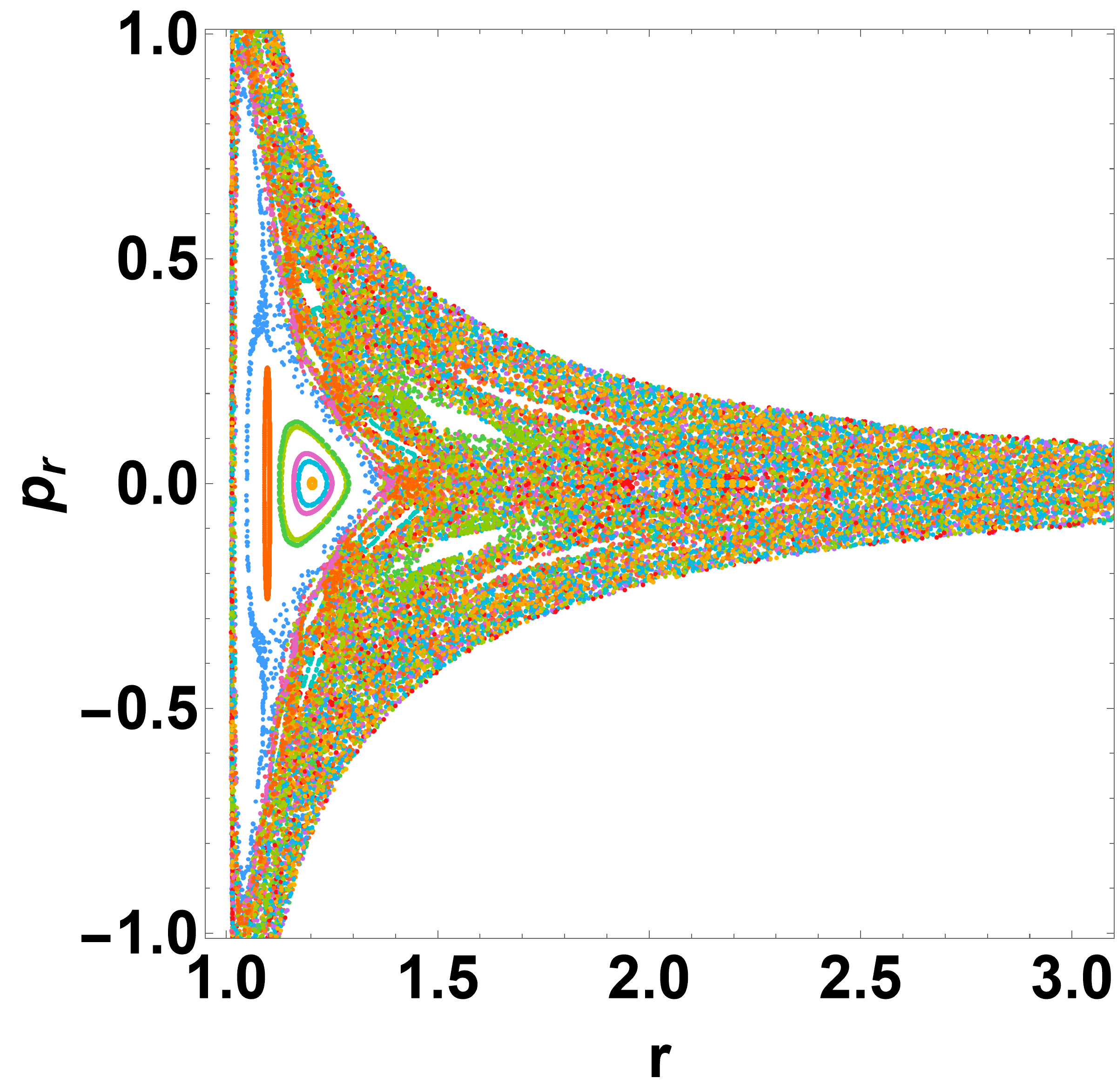} \\
        \textbf{$q=0.9$} & \includegraphics[scale=0.26,valign=c]{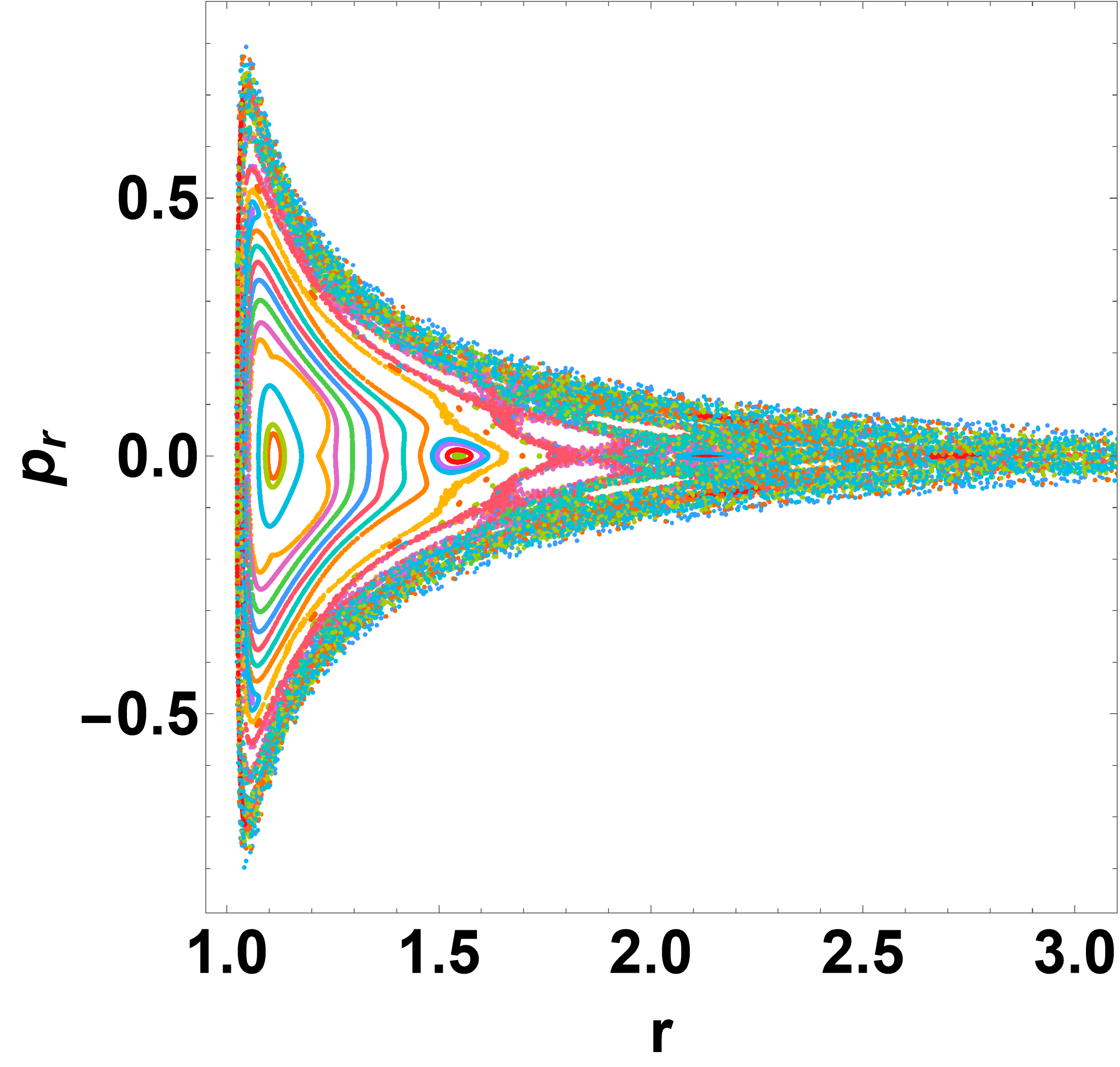} & \includegraphics[scale=0.26,valign=c]{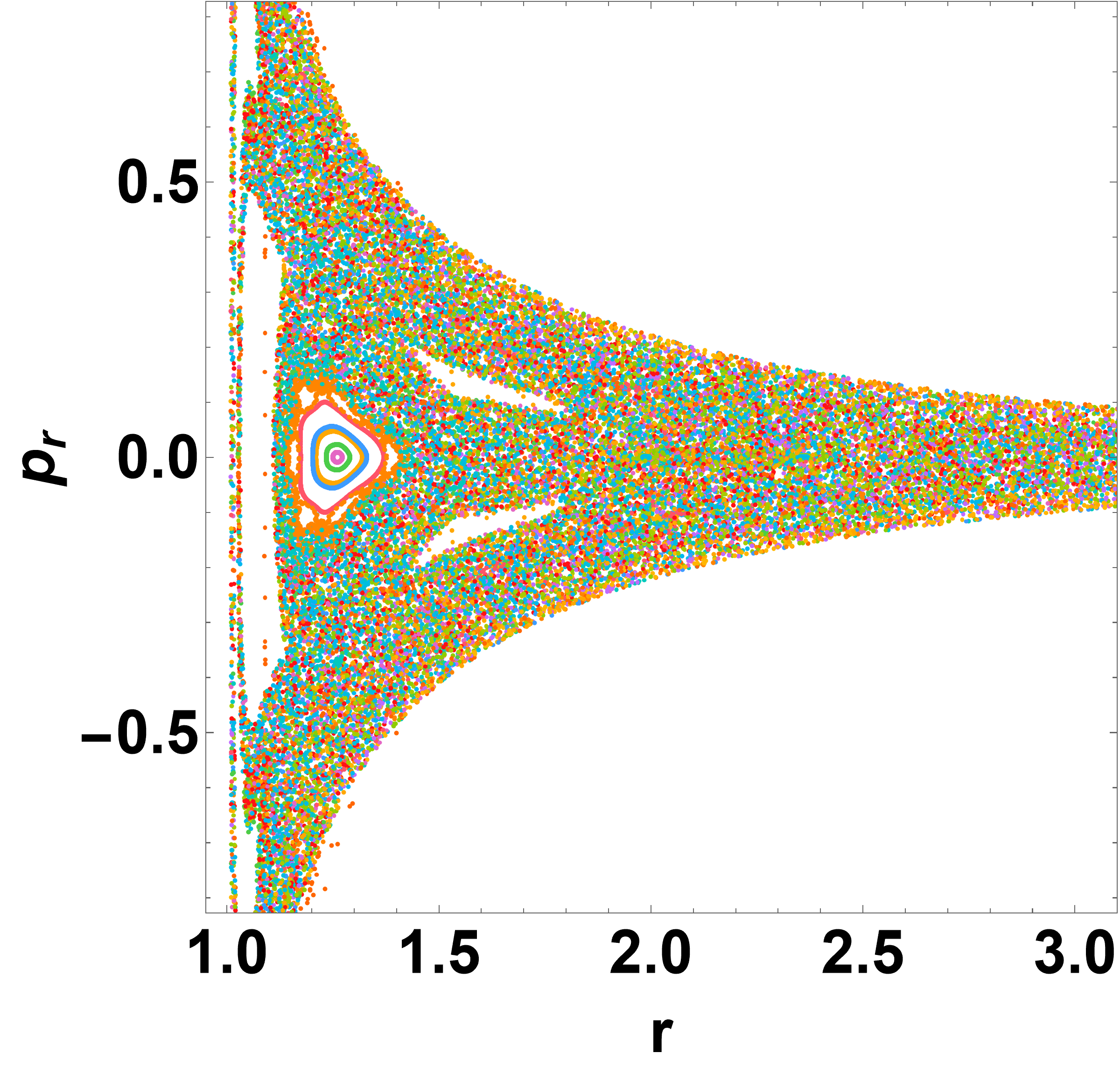}
        \end{tabular}
	\caption{Poincar\'{e} sections  for different values of the charge $q$ and energy $E$. Here $\alpha=1$, $\ell=1$, $k=0.2$, and $r_0=1$ are used. \label{fig:susy-poincare-qvary}}
\end{figure}

As predicted by the KAM theorem, the intricate chaotic phase space structure is elegantly captured through Poincar\'{e} sections. A Poincar\'{e} section records the intersections of a system’s trajectory with a selected lower-dimensional subspace, often a phase plane, providing a more accessible representation of the system’s behavior. In systems where the KAM theorem applies, the Poincar\'{e} section reveals a blend of closed, regular curves representing unbroken invariant tori alongside irregular, scattered points indicative of chaotic motion. The gradual shift from regularity to chaos as the perturbation intensifies is discernible in these sections. Poincar\'{e} sections thus offer a powerful tool for visualizing the persistence of integrable structures as well as the onset of chaos, thereby shedding light on the dynamics of the systems.

The Poincar\'{e} sections of our closed string system for a fixed charge at different energies are shown in Fig.~\ref{fig:supersymmetric-ads-soliton-poincare-E}. The phase space of our system contains four variables: $x$, $p_x$, $r$, and $p_r$. We set the initial conditions as $x(0)=0$ and $p_r(0)=0$. Next, we vary $r(0)$ and determine $p_x(0)$ using one of the Virasoro constraint's equation $\mathcal{H}=0$. With these initial conditions, we solve the Hamiltonian equations of motion numerically and construct the Poincar\'{e} sections in the $r-p_r$ plane. We observe that at small values of energy, the Poincar\'{e} sections exhibit a distinct regular structure. This is illustrated in Figs.~\ref{fig:supersymmetric-ads-soliton-poincare-E0pt2} and \ref{fig:supersymmetric-ads-soliton-poincare-E0pt4}. Here, the points coming in different colours correspond to the numerical data for the different starting conditions. As we increase the energy, some tori get distorted and eventually break into smaller tori, as shown in Figs.~\ref{fig:supersymmetric-ads-soliton-poincare-E0pt6} and \ref{fig:supersymmetric-ads-soliton-poincare-E0pt8}. Eventually, as we move to even higher energies, there is a complete collapse of the tori, and we get a collection of scattered points, also known as cantori in literature, as shown in Figs.~\ref{fig:supersymmetric-ads-soliton-poincare-E1pt0} and \ref{fig:supersymmetric-ads-soliton-poincare-E1pt2}. Thus, the Poincar\'{e} sections gradually transform from regularity to chaos as we increase the energy of the system.

We similarly analyze how the structure of the Poincar\'{e} sections changes with the charge. This is shown in Fig.~\ref{fig:susy-poincare-qvary}. We find that the Poincar\'{e} section is more structured and regular, with less scattered points for small values of $q$. As we gradually increase the charge, the regular paths transform more and more into scattered points, showing strong dependence on initial conditions. It implies that the effect of turning on the charge is to destabilize the system and enhance the chaotic behavior of the closed string. These results correlate well with our earlier observation from the power spectrum, where the chaotic nature of the string was found to be increasing with $q$.

\subsection{Lyapunov exponents}\label{sec:susylyapunov}
Lyapunov exponents are pivotal tools in the analysis of chaotic behavior in dynamical systems, capturing the rate at which nearby trajectories in phase space diverge exponentially. They offer a measure of a system's sensitivity to initial conditions, with positive values signifying the onset of chaos. For a system possessing \(n\) degrees of freedom, the Lyapunov exponents $( \lambda_1, \lambda_2, \dots, \lambda_n)$ describe the growth or decay of perturbations along different directions, each exponent representing the stability of the system in that dimension. The largest Lyapunov exponent, $\lambda_{\text{max}}$, is of particular importance, whose positive value serves as a definitive marker of chaos.

The mathematical expression for a Lyapunov exponent is given by:
\begin{equation}
\lambda = \lim_{\tau \to \infty} \lim_{\Delta Z(0) \to 0} \frac{1}{\tau} \ln \frac{|\Delta Z(\tau)|}{|\Delta Z(0)|},
\end{equation}
where $\Delta Z(0)$ denotes an infinitesimal deviation from the system's initial state, and $\Delta Z(\tau)$ represents the evolved deviation at a later time $\tau$. This formulation elegantly encapsulates the long-term dynamic evolution, and the relation $|\Delta Z(\tau)| \sim |\Delta Z(0)| e^{\lambda \tau}$, with $\lambda$ being positive, reflects the exponential growth of instability, a characteristic of chaotic systems.

\begin{figure}[htbp!]
{\def\arraystretch{2}\tabcolsep=4pt
	\begin{tabular}{|c| c c c|}
        \hline
		\textbf{Energy} & \textbf{$q=0.1$} & \textbf{$q=0.5$} & \textbf{$q=0.9$} \\
        \hline
        \rule{0pt}{11ex}
		\textbf{$E=0.7$} & \includegraphics[scale=0.16,valign=c]{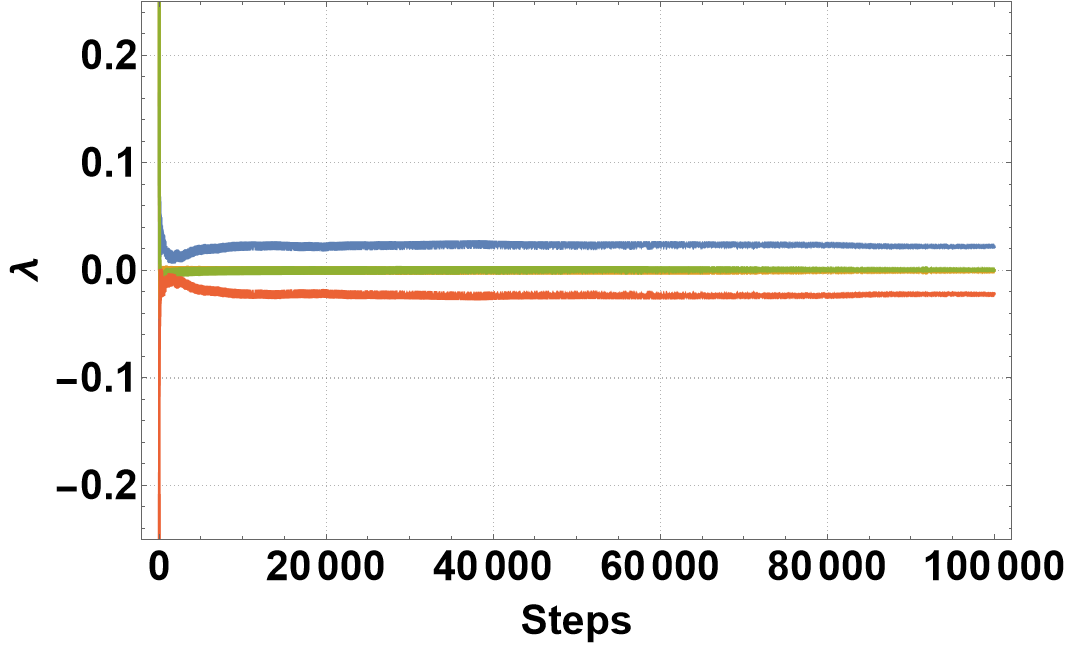} & \includegraphics[scale=0.16,valign=c]{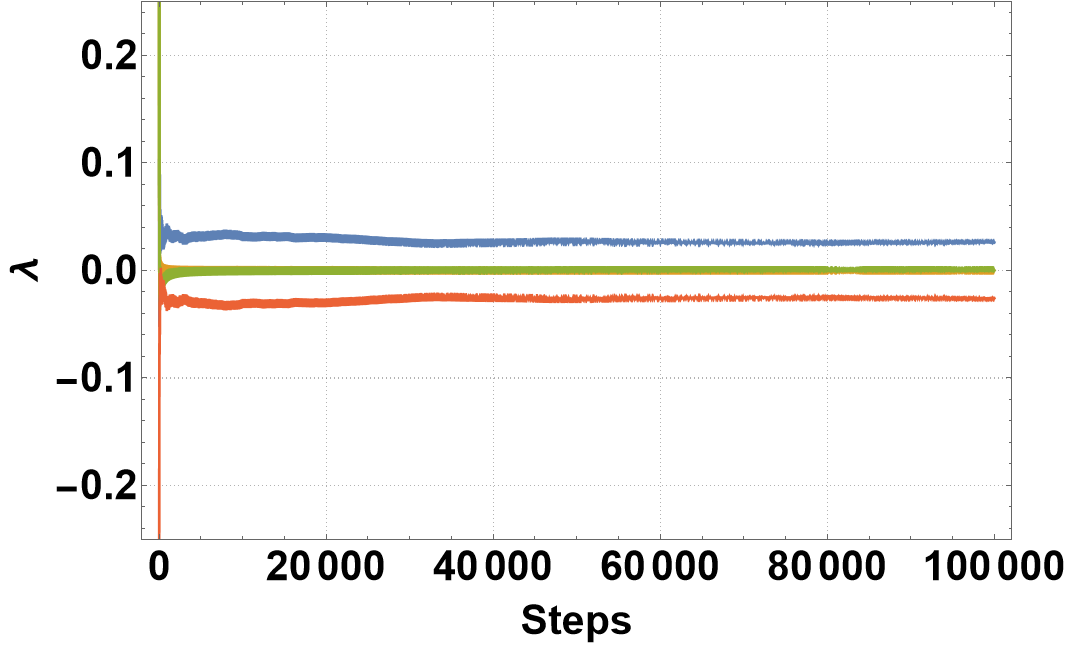} & \includegraphics[scale=0.16,valign=c]{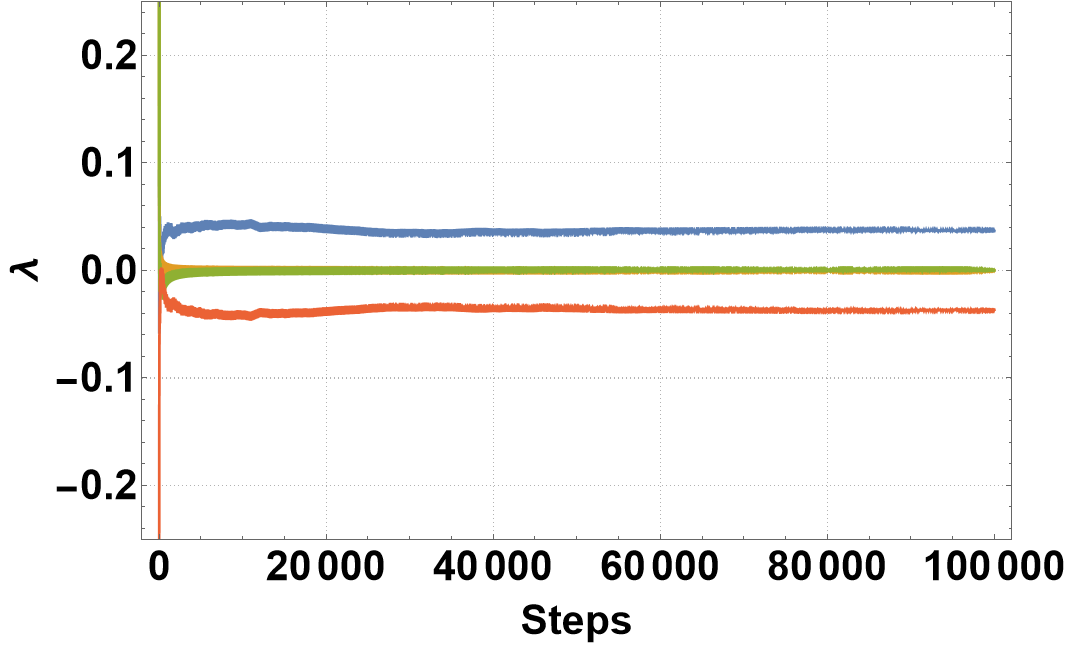} \\
        \rule{0pt}{11ex}
		\textbf{$E=0.9$} & \includegraphics[scale=0.16,valign=c]{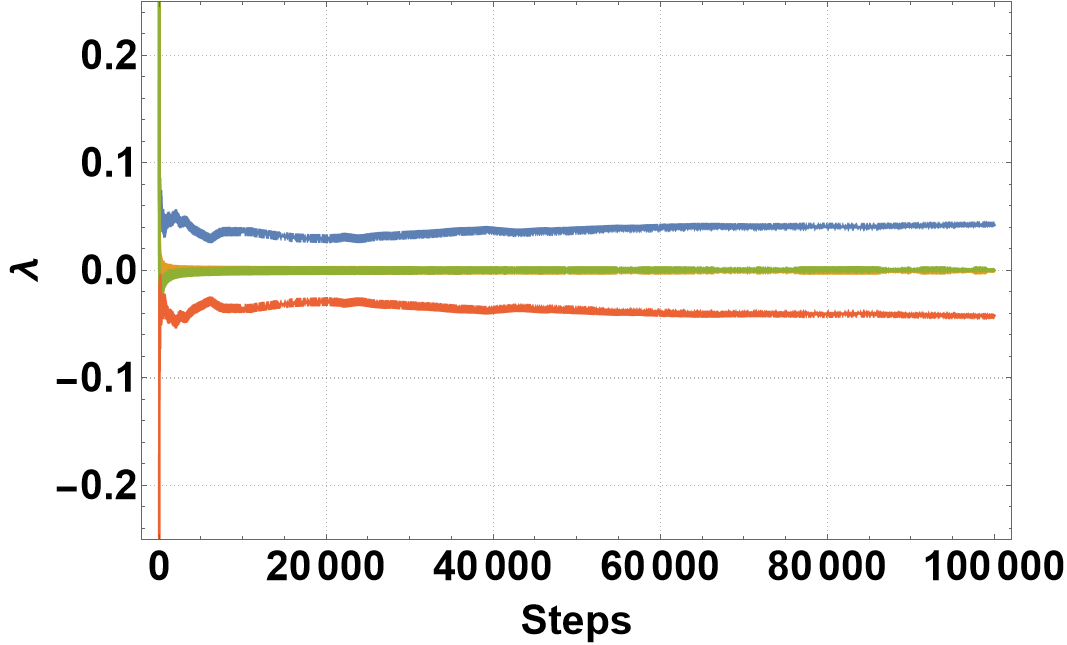} & \includegraphics[scale=0.16,valign=c]{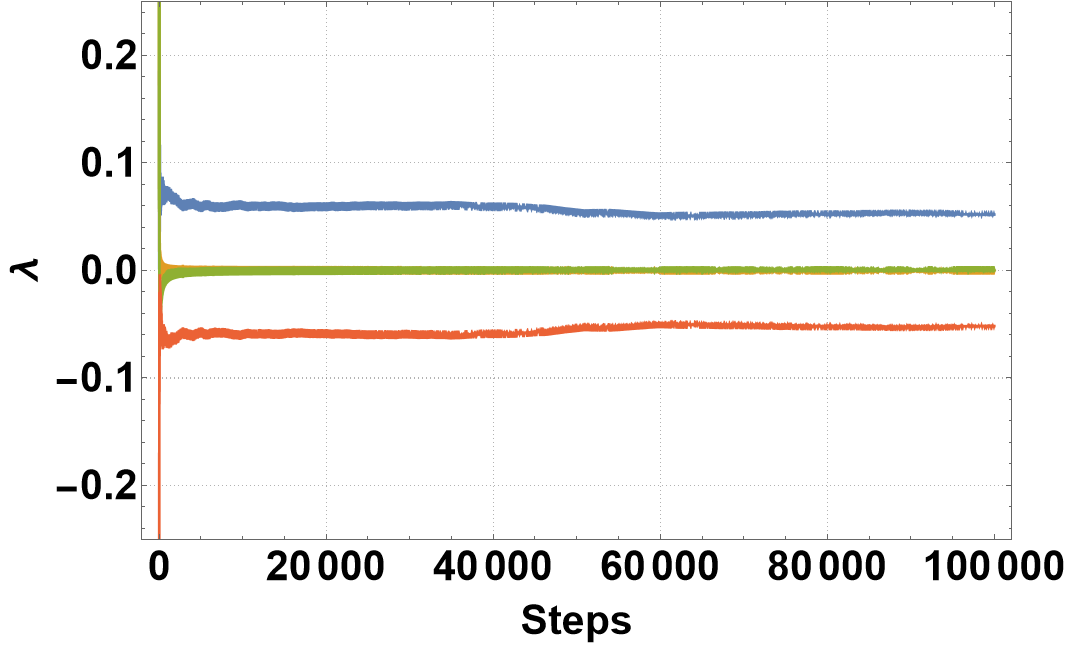} & \includegraphics[scale=0.16,valign=c]{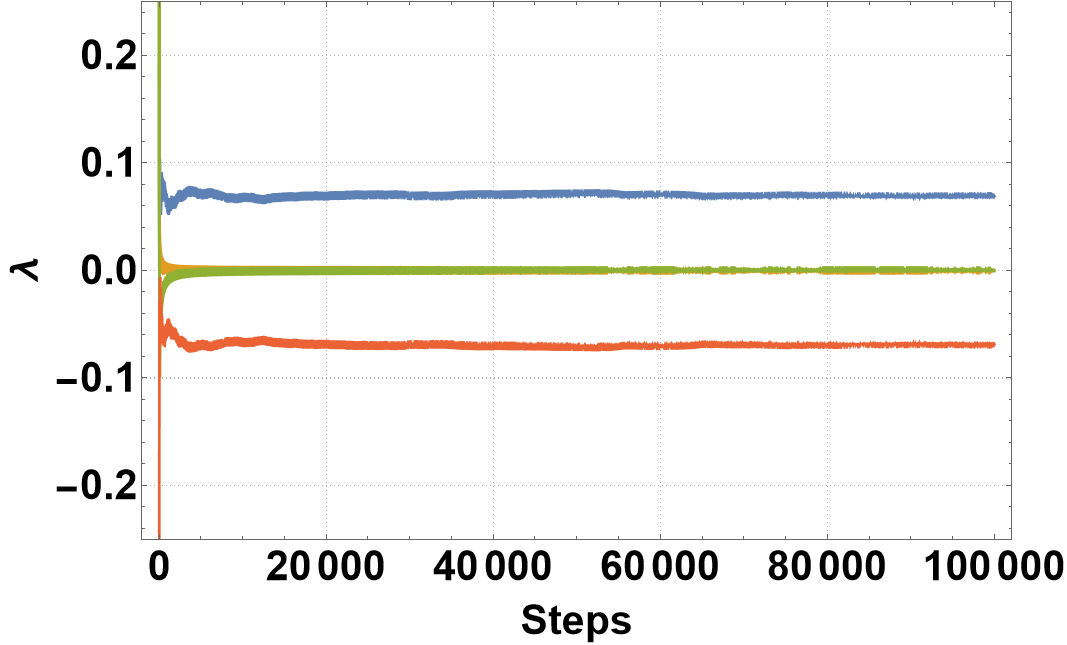} \\
        \rule{0pt}{11ex}
		\textbf{$E=1.2$} & \includegraphics[scale=0.16,valign=c]{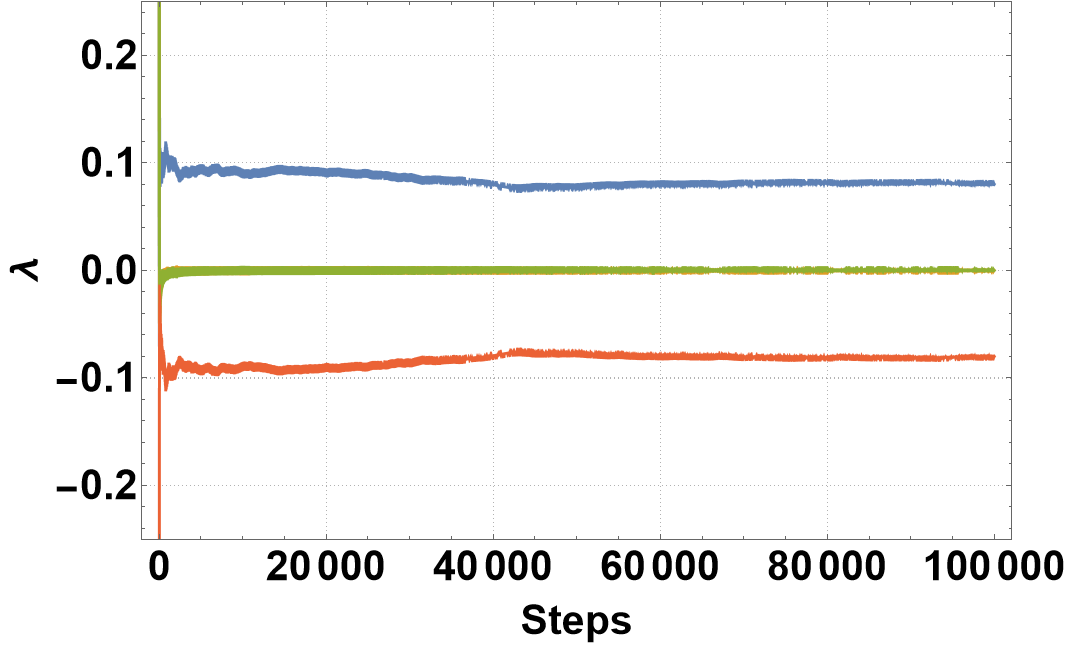} & \includegraphics[scale=0.16,valign=c]{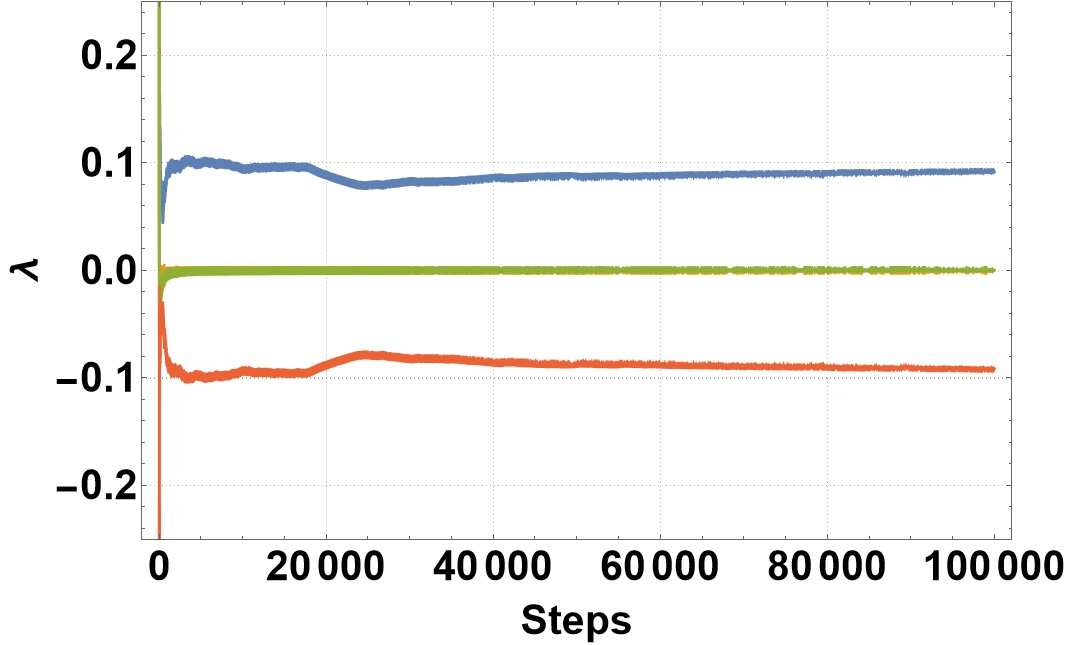} & \includegraphics[scale=0.16,valign=c]{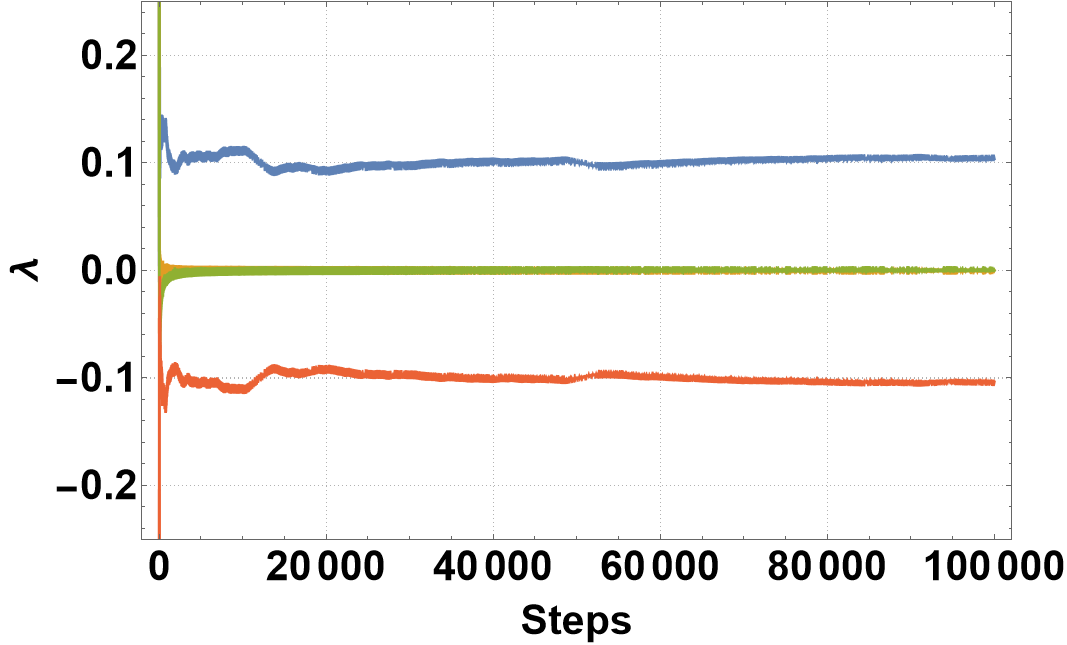} \\
        \rule{0pt}{11ex}
        \textbf{$E=1.5$} & \includegraphics[scale=0.16,valign=c]{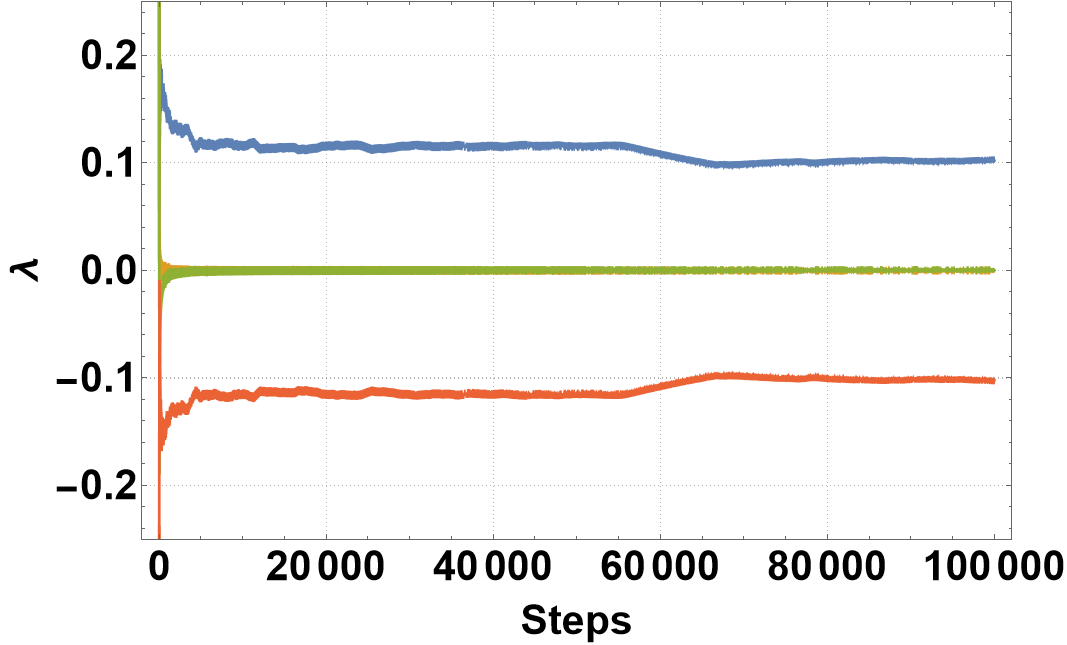} & \includegraphics[scale=0.16,valign=c]{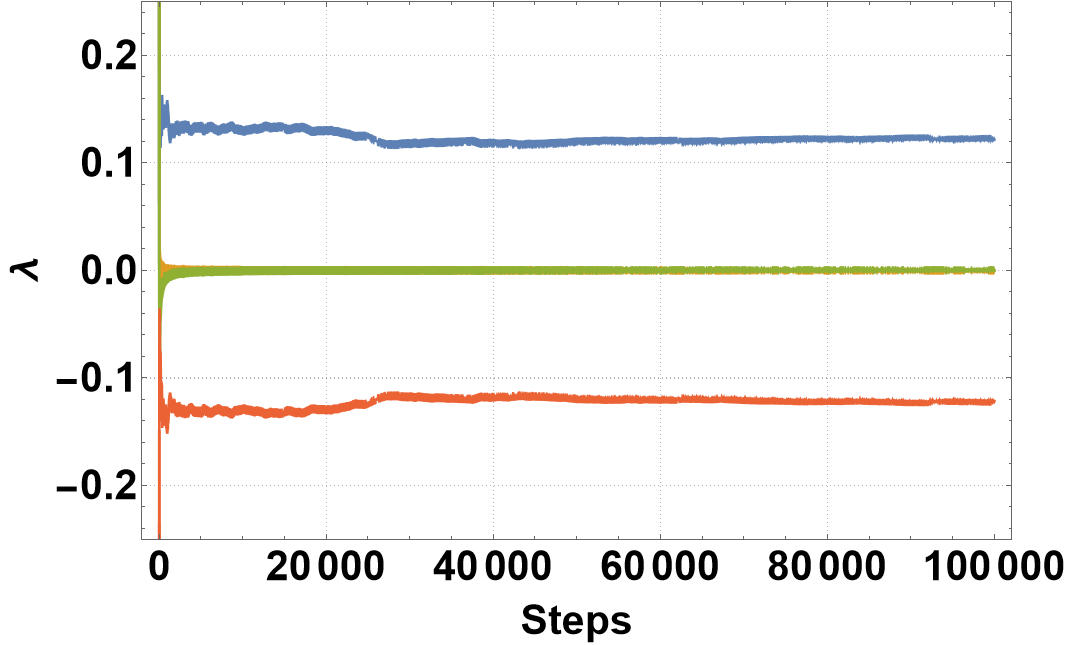} & \includegraphics[scale=0.16,valign=c]{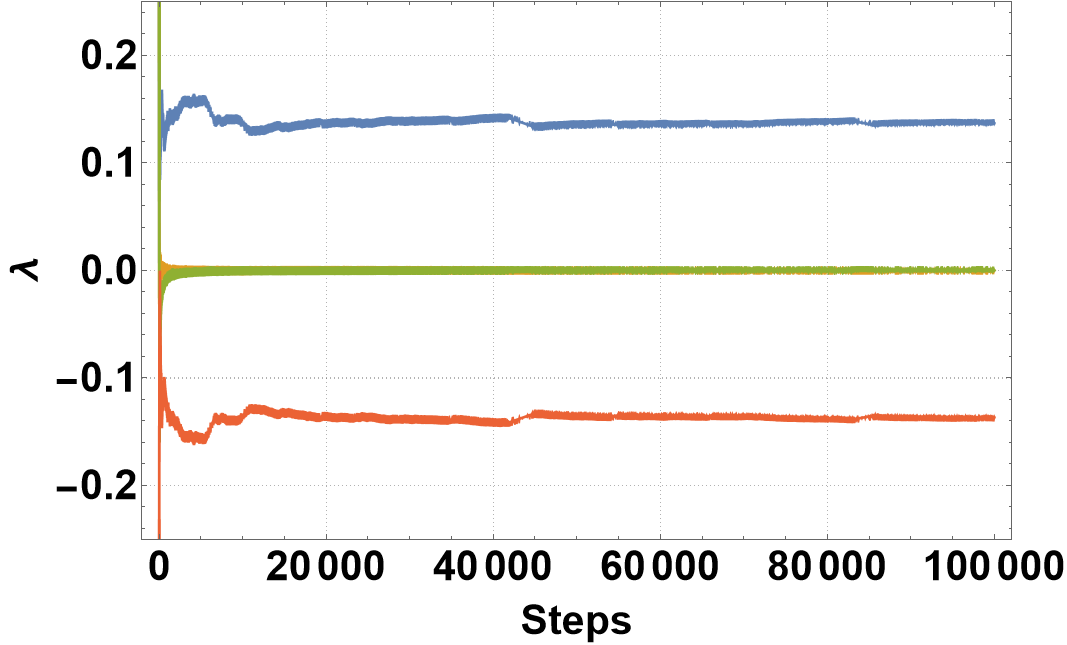} \\
        \rule{0pt}{11ex}
        \textbf{$E=1.8$} & \includegraphics[scale=0.16,valign=c]{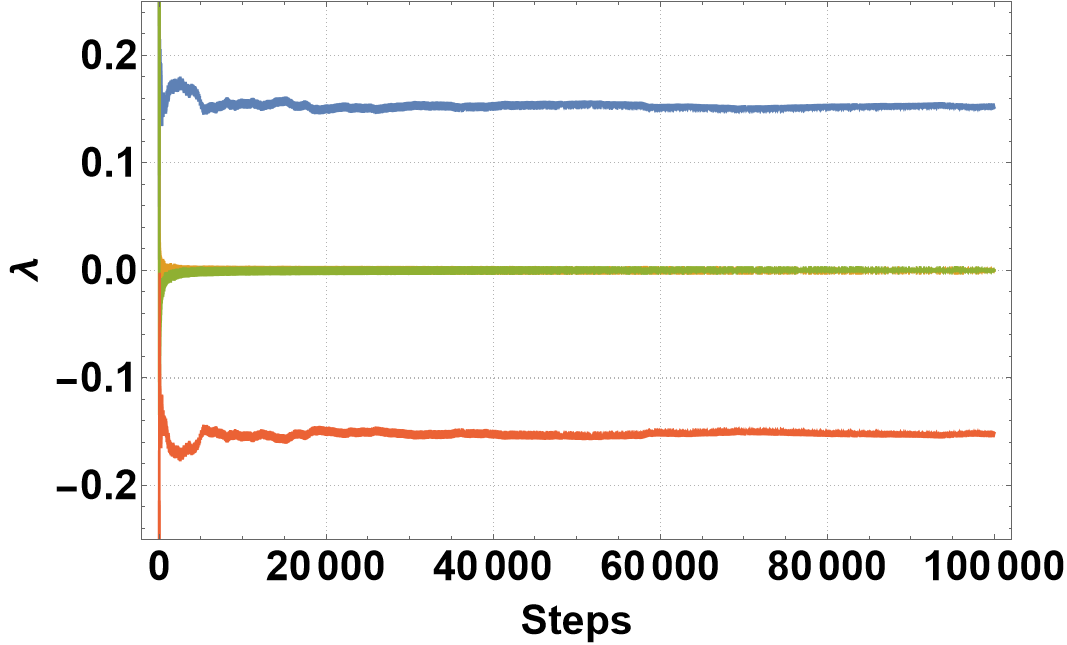} & \includegraphics[scale=0.16,valign=c]{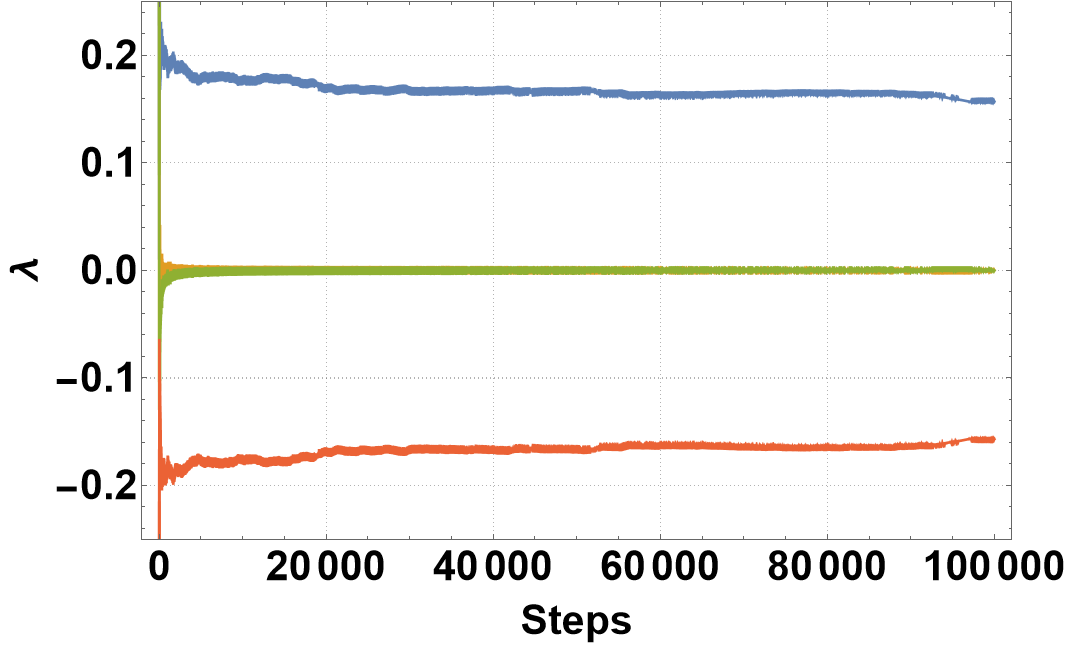} & \includegraphics[scale=0.16,valign=c]{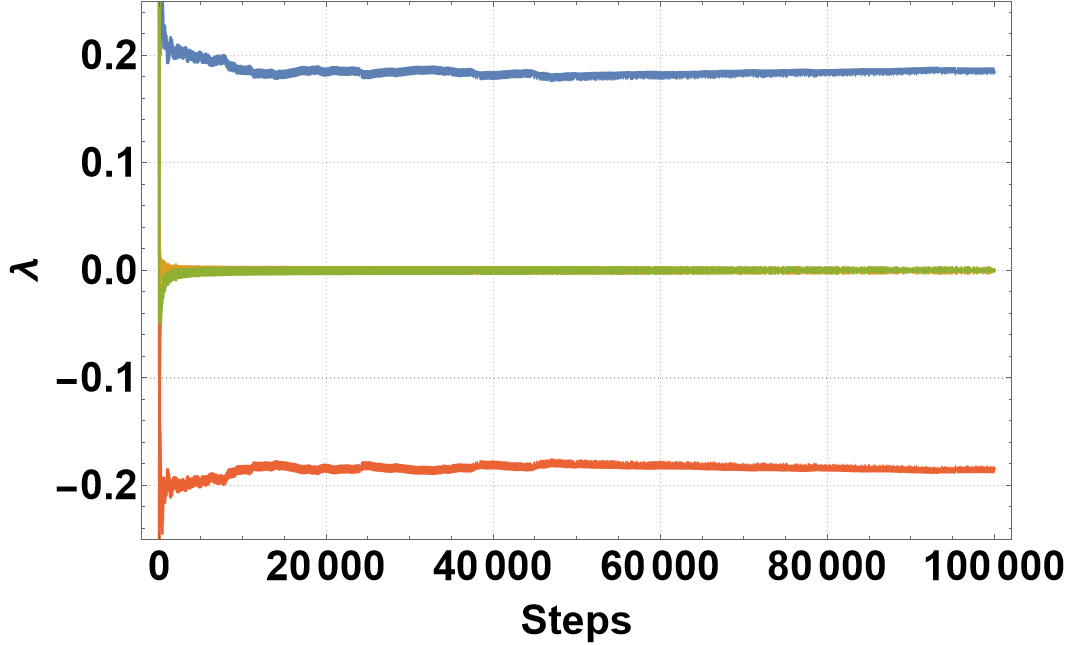} \\
        \hline\hline
        \rule{0pt}{6ex}
        \textbf{Sum of $\lambda$} & \includegraphics[scale=0.16,valign=c]{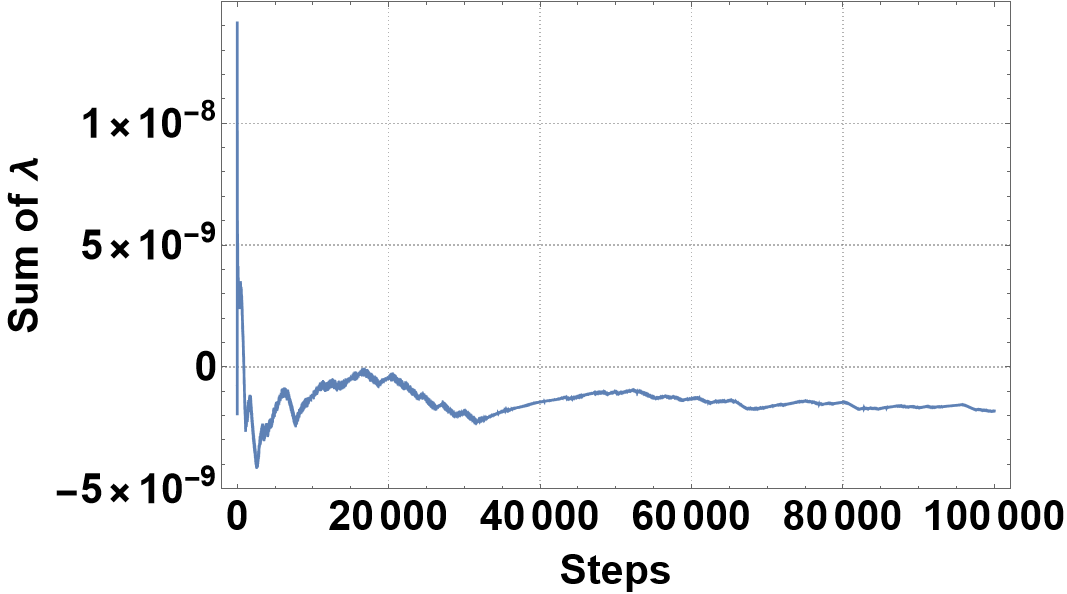} & \includegraphics[scale=0.16,valign=c]{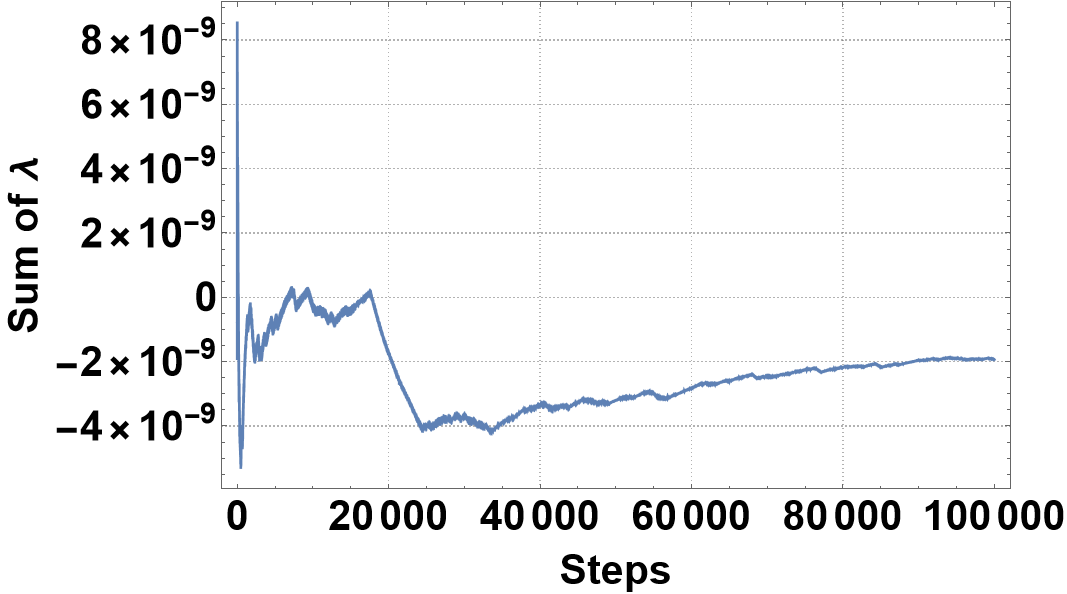} & \includegraphics[scale=0.16,valign=c]{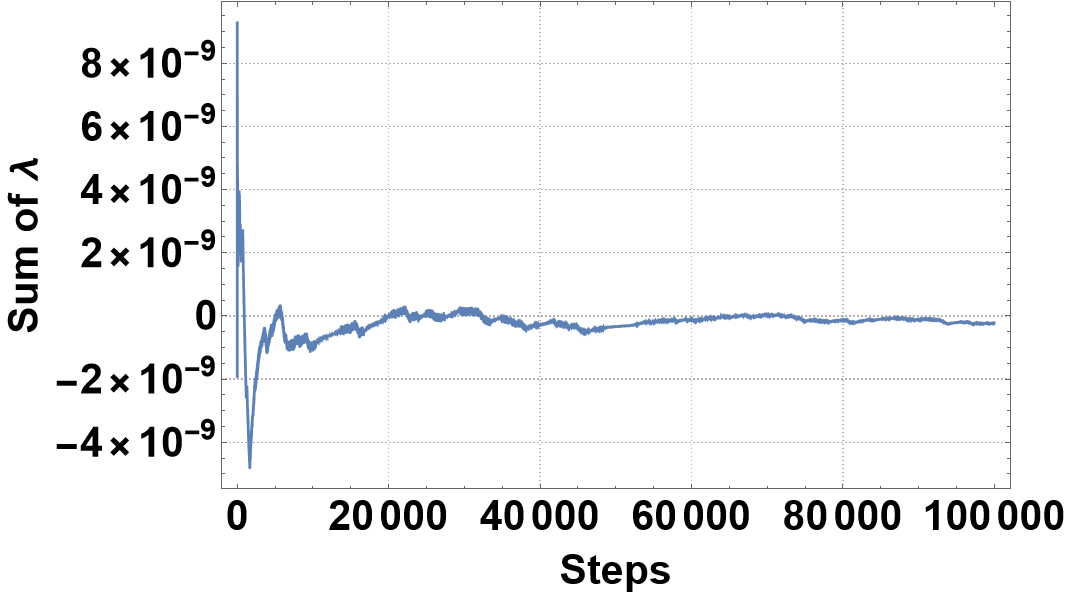}\\
        \hline
        \end{tabular}
    }
    \caption{Convergence plots of the four Lyapunov exponents for different values of $E$ and $q$. The parameter values are $\alpha=1$, $\ell=1$, $k=0.2$, and $r_0=1$. The initial conditions are $r(0)=3.3$, $p_r(0)=0$, $x(0)=0$. Here, four different colors correspond to four Lyapunov exponents. The sum of Lyapunov exponents is shown in the last row for $E=1.2$ and different values of $q$.  \label{fig:susy-lyapunov}}
\end{figure}

\begin{figure}[htbp!]
	\centering
    \subfigure[$E=0.4$]{\label{fig:susy-lyapunov-qfixed-low-energy}\includegraphics[width=0.42\linewidth]{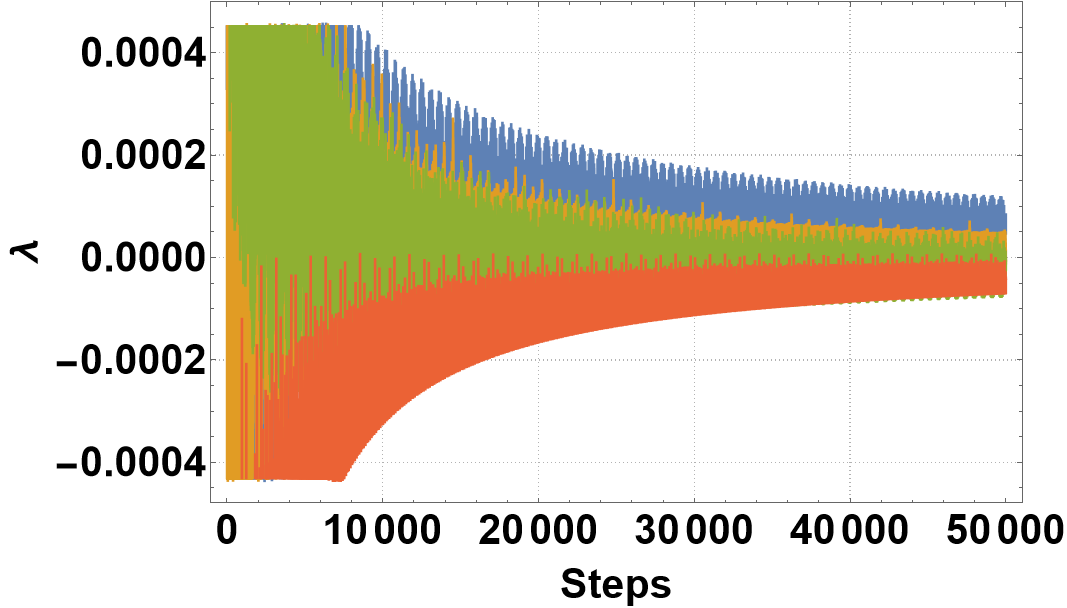}}
	\subfigure[$E=2.0$]{\label{fig:susy-lyapunov-qfixed-high-energy}\includegraphics[width=0.42\linewidth]{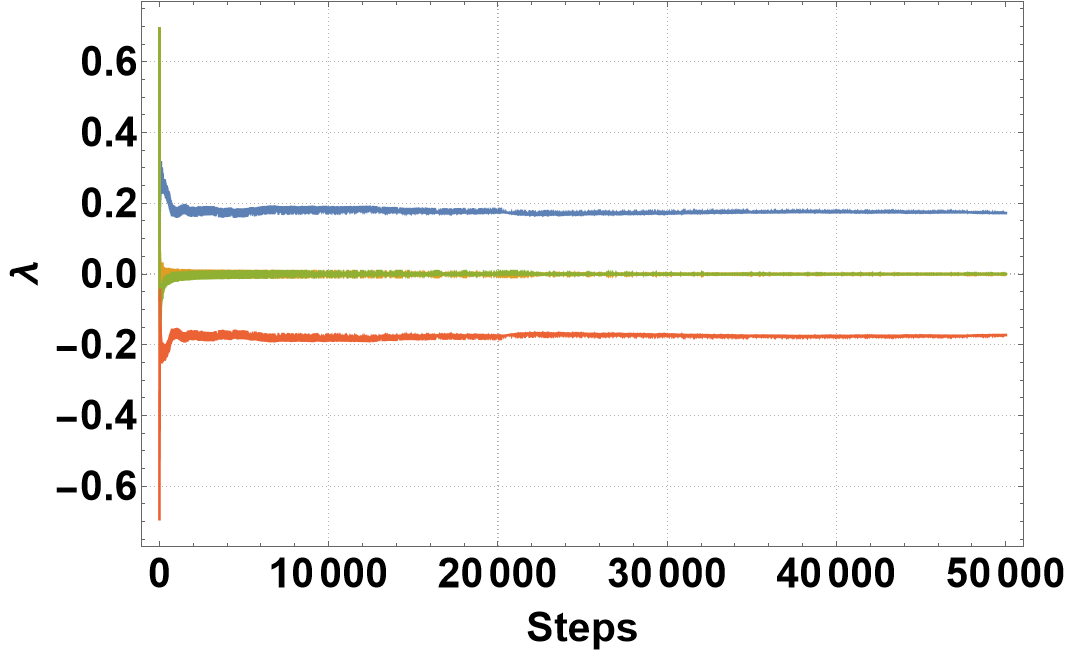}}
	\caption{Lyapunov exponents at fixed charge $q=0.2$. The parameter values are $\alpha=1$, $\ell=1$, $k=0.2$, and $r_0=1$. The initial conditions are $r(0)=2.5$, $p_r(0)=0$, $x(0)=0$. \label{fig:susy-lyapunovdamped}}
\end{figure}

Using the algorithm developed in~\cite{sandri1996numerical, wolf1985determining}, we calculate the four Lyapunov exponents associated with the four-dimensional phase space for different values of energy and charge. Our results are shown in Fig.~\ref{fig:susy-lyapunov}, where different colors correspond to different Lyapunov exponents. The convergence plots of the four Lyapunov exponents and their sum are computed. We find that the sum of Lyapunov exponents always converges to zero. It indicates the conservative nature of the system. This is shown in the last row of Fig.~\ref{fig:susy-lyapunov}. The same is true irrespective of the value of $E$ and $q$. Overall, the convergence rate of Lyapunov exponents is like a damped oscillating function. This is explicitly shown in Fig.~\ref{fig:susy-lyapunovdamped}.

\begin{figure}[htbp!]
    \centering
    \includegraphics[width=0.7\linewidth]{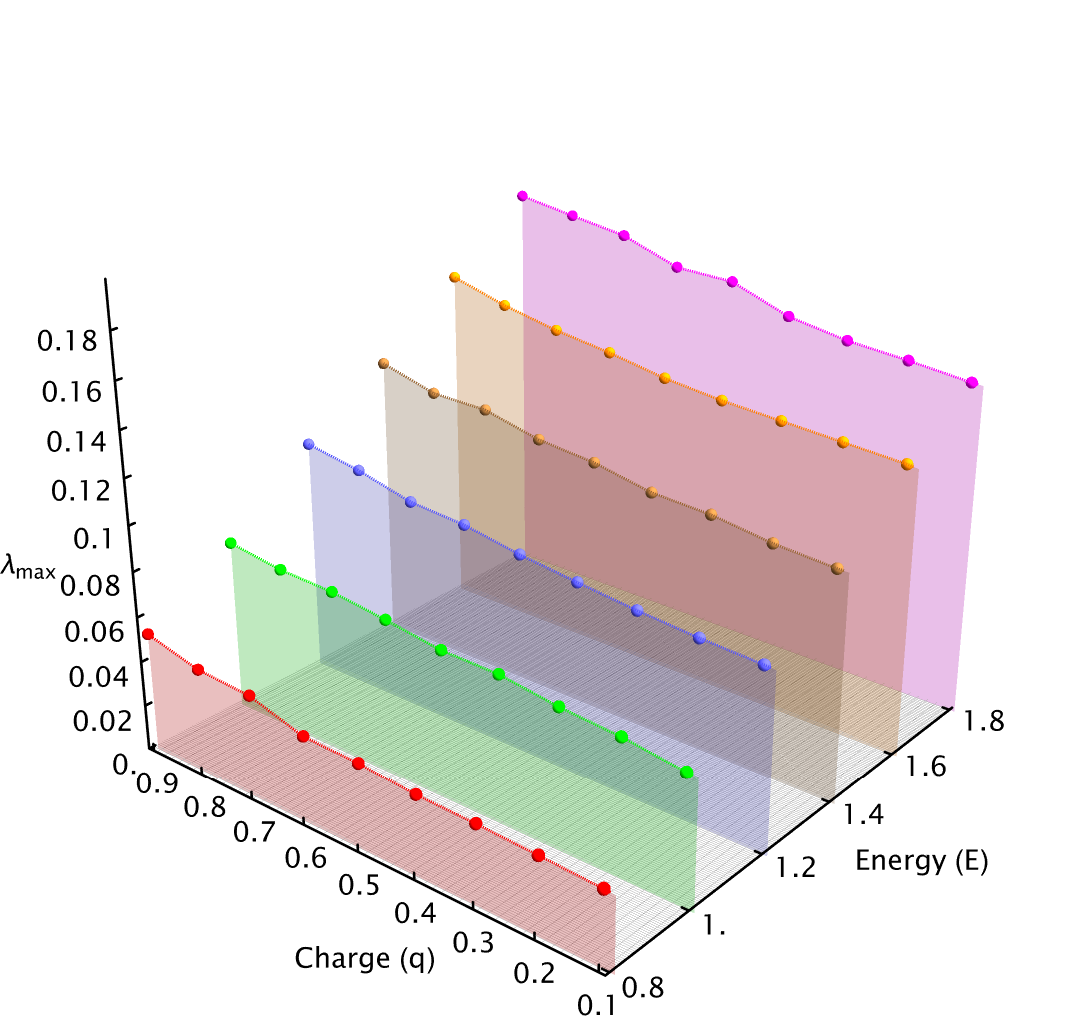}
    \caption{Comparison of the maximum Lyapunov exponent $\lambda_{\text{max}}$ for various values of energy and charge. Data points are color-coded as follows: red for $E = 0.8$, green for $E = 1.0$, blue for $E = 1.2$, brown for $E = 1.4$, orange for $E = 1.6$, and magenta for $E = 1.8$. The initial conditions are $r(0)=3.3$, $p_r(0)=0$, $x(0)=0$. }
    \label{fig:lambdamax_compare}
\end{figure}

The largest Lyapunov exponent $\lambda_\text{max}$ can be extrapolated by considering a large number of $\tau$ steps and fitting the maximum in each oscillation \cite{wolf1985determining, sandri1996numerical}. The $\lambda_\text{max}$
obtained from this fitting procedure is shown in Fig.~\ref{fig:susy-lyapunovdamped}. We find that  $\lambda_\text{max}$ is almost zero for small $E$ and a converging positive value for large $E$. For a fixed $q$, $\lambda_\text{max}$ is found to be increasing with $E$, indicating again that the chaotic nature of the closed string increases with $E$. The same conclusion was drawn in \cite{Basu:2011dg} for the AdS soliton background, and here, we have generalized this conclusion to the finite charge case. 

We also investigate the variation of $\lambda_\text{max}$ with $q$ and find that it increases with $q$ as well. This variation is similar to the variation that occurred with an increase in $E$, albeit with a relatively small change in the magnitude of $\lambda_\text{max}$ with $q$. These results confirm more qualitatively our previous discussion using the power spectrum and Poincaré sections, which suggested that chaotic dynamics of the closed string get enhanced when the energy and charge increase, with the former playing a dominant role. 

\section{Quantum chaos analysis}\label{sec:susyquantumchaos}
We now investigate the nature of quantum chaos in the closed string in the charged AdS soliton background. Quantum chaos examines the quantum properties of a system with a chaotic classical limit. As discussed in the previous section, the closed string exhibits classical chaos, with \( E \) and \( q \) serving as parameters that transition the system between chaotic and integrable behavior. It would be interesting to see how these parameters modify the quantum chaotic features, if any, of the closed string. For this purpose, we find the quantum spectrum in the framework of minisuperspace
quantization following \cite{Basu:2013uva, PandoZayas:2012ig}. 

As discussed earlier, to determine the type of motion in a system, one can examine a group of trajectories that start within a small cluster of points in phase space. In chaotic motion, the distance between any two trajectories in this group increases exponentially over time, with Lyapunov exponents quantifying the separation rate. However, when quantum effects become significant in a system, the concept of a phase-space trajectory no longer applies, and the Lyapunov exponent also loses relevance.

In the absence of the classical distinction between regular motion and chaos in the quantum domain, it becomes necessary to seek alternative, purely quantum mechanical criteria to differentiate between the two types of quantum dynamics. Ideally, this distinction should mirror the classical case, such that one group of quantum dynamics resembles regular motion while the other corresponds to chaos. There are, in fact, several quantum-specific criteria that achieve this. Some are based on the energy spectrum, others on the properties of energy eigenvectors, or on the time evolution of relevant expectation values~\cite{Haake:2010fgh}.
Here, we will discuss level-spacing distributions, Dyson-Mehta $\Delta_3$ statistics, and OTOCs for identifying quantum signatures of chaos. These methods require the quantum spectrum of the closed string, which we will obtain using the minisuperspace quantization technique.

\subsection{Level-spacing distribution}\label{sec:susylevelspacing}
Level-spacing distributions, which describe the probability distribution of the spacings between consecutive energy levels in a quantum system, have been identified as a key indicator revealing patterns related to the system's underlying dynamics. As demonstrated in~\cite{Victor1977Sep}, for quantum systems whose Hamiltonian is classically integrable, the distribution of adjacent energy level-spacing obeys quite universally the same distribution as that of a sequence of uncorrelated levels, which is a Poisson distribution: 
\begin{equation}
P(w) \simeq \exp(-w)\,.
\end{equation}
Energy levels for integrable systems show level-clustering, meaning they are spaced close to each other, as is evident from the form of the distribution function. Building on foundational insights from~\cite{McDonald:1979zz, Casati1980Jun, Berry1981Jan}, the work of \cite{PhysRevLett.52.1} further established a connection between integrable and chaotic systems and their distinct universal spectral fluctuations and demonstrated that the level-spacing distribution of eigenvalues for classically chaotic systems agreed with the predictions of random matrix theory (RMT).

\begin{table}[htbp!]
    \centering
    \begin{tabular}{|c|c|c|}
        \hline
        \begin{tabular}{c} 
        Time-reversal \\
        symmetry
        \end{tabular} & 
        \begin{tabular}{c} 
        Canonical \\
        transformations
        \end{tabular} & 
        Random matrix ensemble \\
        \hline \hline
        Yes, $T^2=1$ & $O(N)$ & Gaussian orthogonal ensemble (GOE) \\
        \hline
         No & $U(N)$ & Gaussian unitary ensemble (GUE) \\
        \hline
        Yes, $T^2=-1$ & $Sp(2N)$ & Gaussian symplectic ensemble (GSE) \\
        \hline
    \end{tabular}
    \caption{Wigner-Dyson classification of Hamiltonians based on their time-reversal symmetry and the corresponding random matrix ensemble.}
    \label{tab:wigner-dyson}
\end{table}

Depending on the universality class, Wigner~\cite{Haake:2010fgh} classified the distributions as given in Table~\ref{tab:wigner-dyson}. Here, the Hamiltonian governing the dynamics of the system either obeys or does not obey time reversal symmetry under the action of time reversal operator $T$. Depending on the ensemble of random matrices, Wigner-Dyson distribution functions for these classes take the following form:
\begin{equation}
P(w) = 
\begin{cases} 
\frac{\pi}{2} w e^{-\frac{\pi}{4} w^2} & \text{(GOE)} \\
\frac{32}{\pi^2} w^2 e^{-\frac{4}{\pi} w^2} & \text{(GUE)} \\
\frac{2^{18}}{3^6 \pi^3} w^4 e^{-\frac{64}{9\pi} w^2} & \text{(GSE)}
\end{cases}
\end{equation}
A reference plot for level-spacing distributions is given in~\cite{Basu:2013uva} for a particle in a box with a deformed potential in the nonintegrable case. The distribution functions agree with Poisson and Wigner-Dyson level statistics for the integrable and chaotic cases, respectively. The peculiar structure of spacing between adjacent levels in non-integrable systems arises from the fact that the corresponding energies show quantum mechanical level repulsion, meaning they are spaced apart. Therefore, due to level repulsion, small energy differences are suppressed. In the integrable case, eigenvalues from distinct sectors are independent, leading to an absence of level repulsion.

\subsection{Dyson-Mehta statistics}\label{sec:susydysonmehta}
The Dyson-Mehta $\Delta_3$ statistic is a measure used in quantum chaos to quantify fluctuations in the spectral rigidity of energy levels, helping to distinguish chaotic from integrable systems. Deviation from the equal spacing of the spectrum is measured using the Dyson-Mehta $\Delta_3$ statistic \cite{Dyson:1962es, Dyson1963May}:
\begin{equation}
\label{eq:dyson-mehta}
    \Delta_3(L; \epsilon) \equiv \frac{1}{L} \min_{A,B} \int_{\epsilon}^{\epsilon + L} [N(E) - A E - B]^2 \, dE\,.
\end{equation}
Here, $N(E)$ represents the count of levels with normalized energy below $E$, forming a staircase-like function with an approximate slope of one. The constants $A$ and $B$ provide the best straight-line fit to $N(E)$ within the interval $\epsilon \leq E < \epsilon + L$. The average $\Delta_3$ statistic, $\bar{\Delta}_3(L) \equiv \langle \Delta_3(L; \epsilon) \rangle_\epsilon$, is obtained by averaging over multiple windows of length $L$.

To compare our results, we will use the Poisson and Wigner GOE $\Delta_3$ statistics as benchmarks. For a random spectrum with Poisson spacing, $\Delta_3 = L/15$, independent of $\epsilon$. In the case of a GOE, $\bar{\Delta}_3(L) = (\ln L - 0.0687)/\pi^2$.

\subsection{Out-of-time-ordered correlators}\label{sec:otoc}
The origin of out-of-time-ordered correlators (OTOCs) lies in the context of quasi-classical methods in superconductors~\cite{Larkin1969}. Subsequently, they were used in the context of black hole physics, especially in the famous MSS bound~\cite{Maldacena:2015waa}. OTOCs are now readily used as a probe of quantum chaos in a plethora of works across major disciplines of physics~\cite{Swingle:2016var, Campisi:2016qlj, Kukuljan:2017xag, Li:2016xhw, Rammensee:2018pyk}. In contrast to classical chaotic systems, where disturbances can grow unboundedly, the growth of OTOC in quantum systems exhibits a fundamental limit, saturating at the Ehrenfest time $t_E$. This saturation marks a significant departure from classical chaos, as it reflects the unique interplay between quantum coherence and chaotic dynamics. The Ehrenfest time represents the timescale after which the wave function fully spreads across the system. It serves as a boundary between regimes of particle-like and wave-like behavior.

The OTOC is defined as~\cite{Hashimoto:2017oit}
\begin{equation}\label{eq:otoc-original}
    C_T(t)=-\braket{[x(t),p(0)]^2}\,,
\end{equation}
where $\braket{\mathcal{O}}\equiv \text{tr}\left[e^{-\beta H}\mathcal{O}\right]/\text{tr}e^{-\beta H}$ and $\beta=1/T$ is the inverse of the temperature. If we take energy eigenstates as the basis of the Hilbert space, the OTOC can be rewritten as
\begin{equation}\label{eq:thermalandmicrotoc}
    C_T(t)=\frac{1}{Z}\sum_n e^{-\beta E_n} C_n(t),\qquad C_n(t)=-\left<n|[x(t),p(0)]^2|n\right>\,,
\end{equation}
with Hamiltonian $H(x_1,...,x_n,p_1,...,p_n)$ and $H\ket{n}=E_n\ket{n}$. We refer to $C_T(t)$ and $C_n(t)$ as the thermal and microcanonical OTOC, respectively. For numerical purposes, it would be convenient if we express the microcanonical OTOC in a matrix form using the matrix representation of $x$ and $p$. With the help of the completeness relation $1=\sum_m\ket{m}\bra{m}$, microcanonical OTOC becomes
\begin{equation}
\label{eq:microtoc}
    C_n(t)=\sum_m b_{n m}(t) b_{n m}^{*}(t), \qquad b_{n m}(t)\equiv-i\braket{n|[x(t),p(0)]|m}\,,
\end{equation}
where $b_{n m}(t)=b_{n m}^{*}(t)$ is hermitian. Now using $x(t)=e^{i H t} x e^{-i H t}$ and inserting the completeness relation again, we get
\begin{equation}\label{eq:b-nm}
    b_{n m}(t)=-i\sum_k \left(e^{i E_{n k} t}x_{n k}p_{k m}-e^{i E_{k m} t}p_{n k}x_{k m}\right)\,,
\end{equation}
with $p_{n m}\equiv\braket{n|p|m}$, $x_{n m}\equiv\braket{n|x|m}$, and $E_{n m}=E_n - E_m$. For the Hamiltonian given by,
\begin{equation}
    H=\sum_{i=1}^{N} p_i^2+U(x_1,...,x_N)\,,
\end{equation}
we have $[H,x]=-2 i p$. Now, applying $\braket{m|...|n}$ to both sides of this equation, one gets the following expression
\begin{equation}
    p_{m n}=\frac{i}{2}E_{m n}x_{m n}\,.
\end{equation}
When we substitute the above expression in Eq.~(\ref{eq:b-nm}), we get the final formula for $b_{n m}$ as:
\begin{equation}\label{eq:bnm-expression}
    b_{n m}(t)=\frac{1}{2}\sum_k x_{n k}x_{k m}\left(E_{k m} e^{i E_{n k}t}-E_{n k} e^{iE_{km}t}\right)\,.
\end{equation}
The microcanonical OTOCs can be computed using Eqs.~(\ref{eq:microtoc}) and~(\ref{eq:bnm-expression}), with energy difference $E_{mn}$ obtained from Eq.~(\ref{eq:eig-val-eq}) following the minisuperspace quantization. Since we are working with the soliton background, the temperature does not appear in our discussion. Accordingly, we solely focus on the microcanonical OTOC.

\subsection{Minisuperspace quantization}
\label{sec:mini}
\begin{figure}[htbp!]
	\centering
    \subfigure[k = 4]{\label{fig:effective-potential-susy1}\includegraphics[width=0.45\linewidth]{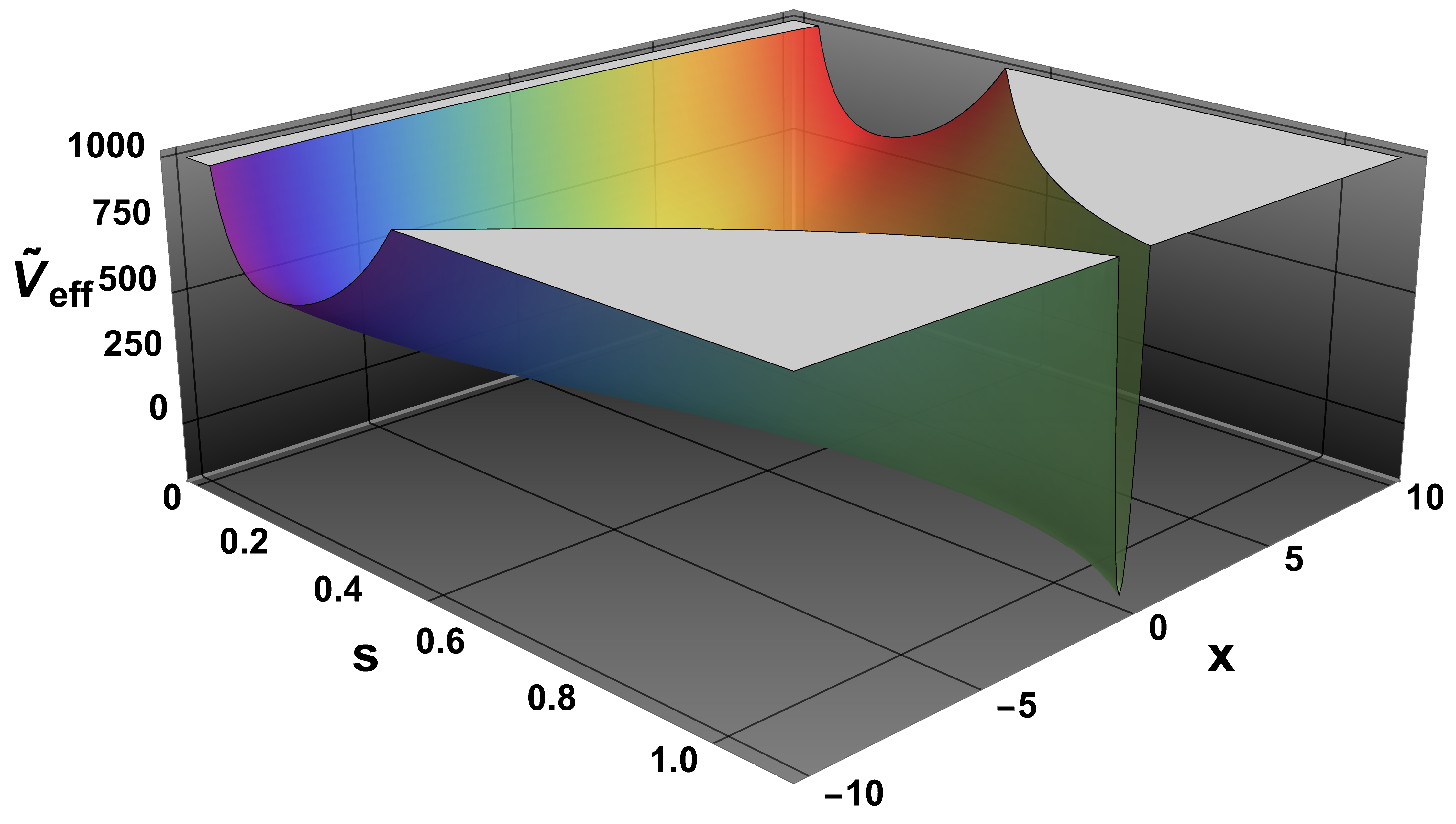}}
    \hfill
	\subfigure[k = 12]{\label{fig:effective-potential-susy2}\includegraphics[width=0.45\linewidth]{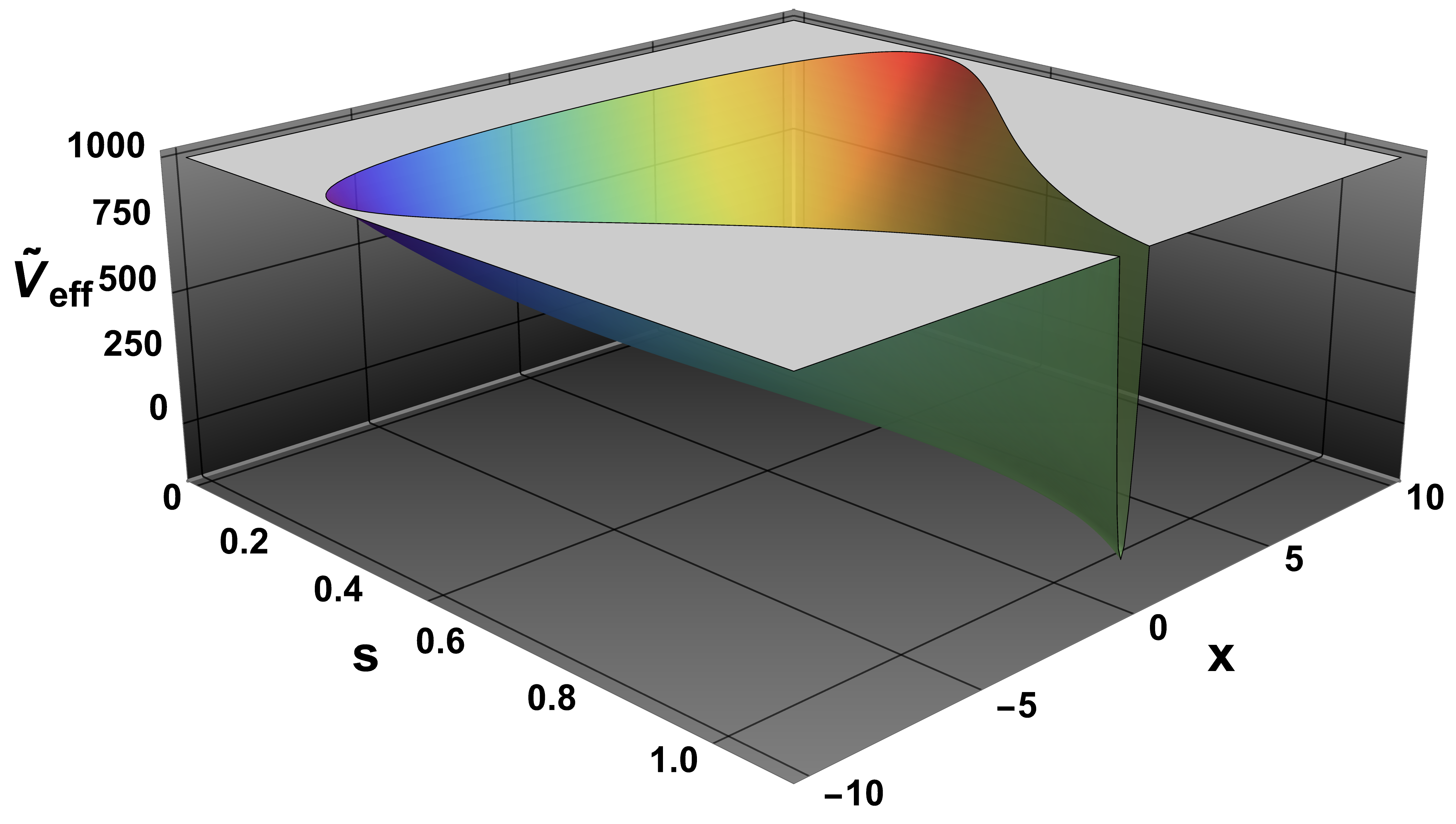}}
	\caption{\label{fig:effective-potential-susy} \( \tilde{V}_{\text{eff}} \) for different $k$ values.}
\end{figure}

To obtain the quantum energy levels of the closed string, we perform a minisuperspace quantization. The process reduces the relative degrees of freedom and makes it feasible to study the quantum aspects of our model. The minisuperspace quantization requires the following substitution in our system's Hamiltonian (\ref{eq:hamil}):
\begin{equation}
p_R^2 \rightarrow -\nabla_R^2, \quad p_x^2 \rightarrow -\nabla_x^2\,,
\end{equation}
where Laplacian is obtained from the effective metric as seen in the Lagrangian (\ref{eq:lagrangian}): 
\begin{equation}
-g_{tt} = g_{xx} = \frac{r^2}{\ell^2}, \quad g_{\theta\theta} = f(r)\frac{r^2}{\ell^2}, \quad g_{rr} = \frac{\ell^2}{r^2 f(r)}.
\end{equation}
Hence, our minisuperspace Hamiltonian becomes:
\begin{equation}
\mathcal{H} = \frac{1}{4} \left[ \left(-E^2 + \frac{k^2}{f(r)} - \partial_x^2 \right)\frac{\ell^2}{r^2} - \frac{r^2}{\ell^2}f(r) \partial_r^2 - \frac{r^2}{\ell^2}f(r)\left( \frac{f'(r)}{f(r)} - \frac{4}{r} \right)\partial_r + \frac{4\alpha^2 r^2 x^2}{\ell^2} \right].
\end{equation}
Due to the imposed Virasoro constraint, we will look for the eigenvalues of  \( \mathcal{H}\psi = 0 \). Our eigenvalue equation takes the following form:
\begin{equation}
E^2 \psi(x, r) = -\partial_x^2 \psi(x, r) - g(r) \partial_r^2 \psi(x, r) - h(r) \partial_r \psi(x, r) + V_{\text{eff}}(x, r) \psi(x, r),
\end{equation}
where we define:
\begin{eqnarray}
g(r) \equiv \frac{r^4}{\ell^4}f(r), \quad h(r) \equiv \frac{r^4}{\ell^4}f(r)\left( \frac{f'(r)}{f(r)} - \frac{4}{r} \right), \nonumber\\
V_{\text{eff}}(x, r) \equiv \frac{k^2}{f(r)} + \frac{4\alpha^2 x^2 r^4}{\ell^4}.
\end{eqnarray}
Next, we employ a coordinate transformation:
\begin{equation}
ds = \frac{dr}{\sqrt{g(r)}}\,,
\end{equation}
which maps our system to a finite domain, making it easier to do numerical analysis. We also redefine our field as \( \psi = e^{\beta} \tilde{\psi} \) such that 
\begin{equation}
\partial_s \beta = \frac{g' - 2h}{4 \sqrt{g}} 
\end{equation}
where prime denotes a derivative with respect to $r$. The form of \( \partial_s \beta\) is obtained by demanding that our transformed eigenvalue equation has no first-order partial derivatives of \(\psi\). Essentially, we have
\begin{equation}
g \partial_r^2 \psi + h \partial_r \psi = \partial_s^2 \psi - \left( \frac{g' - 2h}{2 \sqrt{g}} \right) \partial_s \psi = e^{\beta} \left\{ \partial_s^2 \tilde{\psi} + \left[ \partial_s^2 \beta - (\partial_s \beta)^2 \right] \tilde{\psi} \right\}\,,
\end{equation}
which simplifies the eigenvalue equation into the following form:
\begin{equation}
E^2\tilde{\psi}(x,s) = -\partial_x^2 \tilde{\psi}(x, s) - \partial_s^2 \tilde{\psi}(x, s) + \tilde{V}_{\text{eff}}(x, s) \tilde{\psi}(x,s),
\label{eq:eig-val-eq}
\end{equation}
where,
\begin{equation}
\tilde{V}_{\text{eff}}(x, s) \equiv \frac{k^2}{f(r)} + \frac{4\alpha^2 x^2 r^4}{\ell^4} - \left[ \sqrt{g} (\partial_s \beta)' - (\partial_s \beta)^2 \right].
\end{equation}
Due to the coordinate transformation, the domain of our problem now gets mapped from \( r \in (1, \infty) \) to \( s \in (0, s_{\infty}) \) with \( s_{\infty} \approx 1 \), making the numerical analysis convenient. In Fig.~\ref{fig:effective-potential-susy}, we plot the effective potential $\tilde{V}_{\text{eff}}(x, s)$ for $k=4$ and $k=12$ within our finite domain. In the rest of the discussion, we set $\alpha = 1$, $\ell=1$, and $r_0 = 1$. In the next subsection, we compute the quantum spectrum of the closed string and numerically analyze the corresponding level-spacing distribution, Dyson-Mehta statistic, and OTOC to study the chaotic nature of the string. 

\subsection{Numerical analysis, discussion,  and results}
Now, we numerically analyze the quantum spectrum of the closed string. Analogous to the case of \cite{Basu:2013uva}, at high \( E \) values, the system is expected to become momentum-dominated, with limited dependence on the specific form of the effective potential \( \tilde{V}_{\text{eff}} \). Consequently, one can anticipate that the energy spectrum at these higher energies will resemble that of integrable systems. 
Since we define quantum chaos through level spacing distributions~\cite{Gutzwiller1990}, a substantial number of level differences is required to closely approach a perfect GOE distribution. This general behavior applies to the complete quantum theory; however, focusing on specific sub-sectors through minisuperspace quantization, we expect to achieve only an approximate GOE distribution at both low and high energies.

To obtain eigenvalues $E^2$ numerically, we restrict our study to a finite region \( s_{\text{min}} < s < s_{\text{max}}, \, x_{\text{min}} < x < x_{\text{max}} \), and impose hard-wall Dirichlet boundary conditions at the boundaries. To satisfy these conditions, the eigenfunctions must approach zero near the boundary. This is achieved by considering only eigenvalues \( E^2 \) significantly lower than the boundary potential. Hence, this sets an upper limit on the system's energy.

\begin{figure}[htbp!]
	\centering
    \subfigure[3d plot]{\label{fig:eigenfunction-plot-susy}\includegraphics[width=0.4\linewidth]{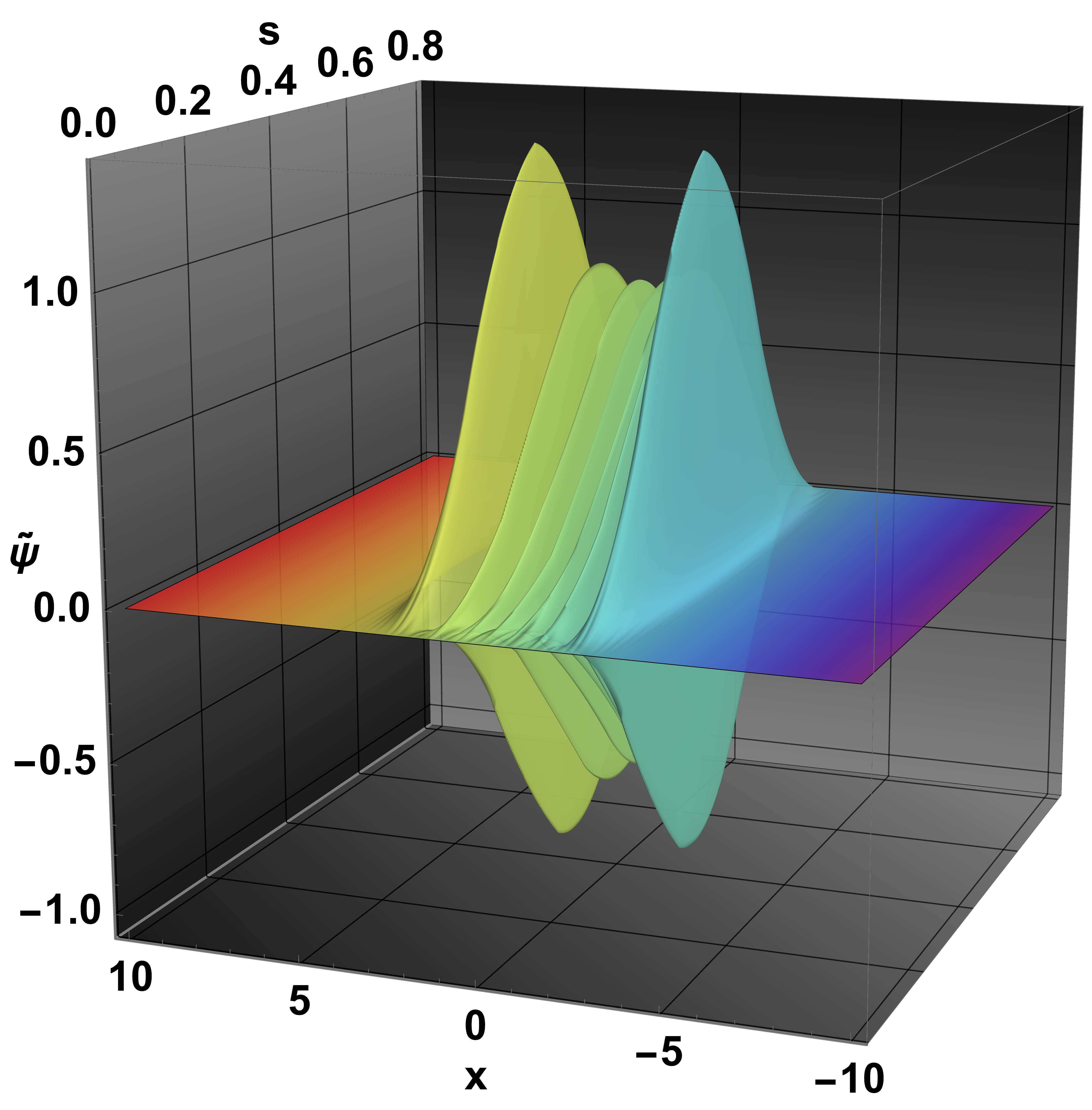}}
    \hfill
	\subfigure[Contour plot]{\label{fig:eigenfunction-contour-susy}\includegraphics[width=0.4\linewidth]{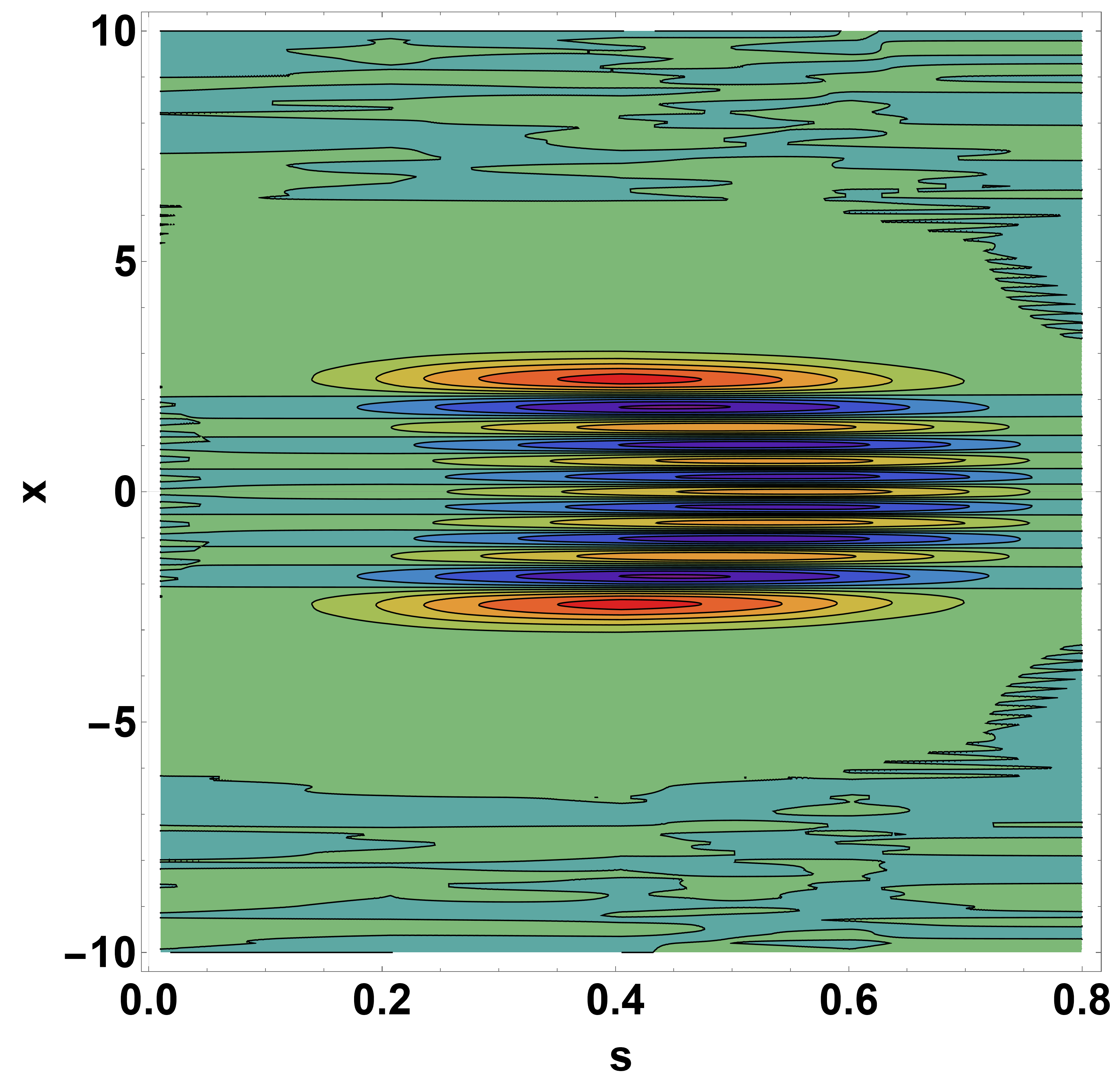}}
	\caption{\label{fig:eigenfunction-susy}The 3d plot (left) and contour plot (right) of the eigenfunction corresponding to the $15^{\text{th}}$ energy level with $k=5$ having eigenvalue $E^2=155.65$.}
\end{figure}

\begin{figure}[htbp!]
	\centering
    \subfigure[Small $E$ gives nearly Wigner GOE]{\label{fig:susy-soliton-level-spacing-low-energy}\includegraphics[width=0.42\linewidth]{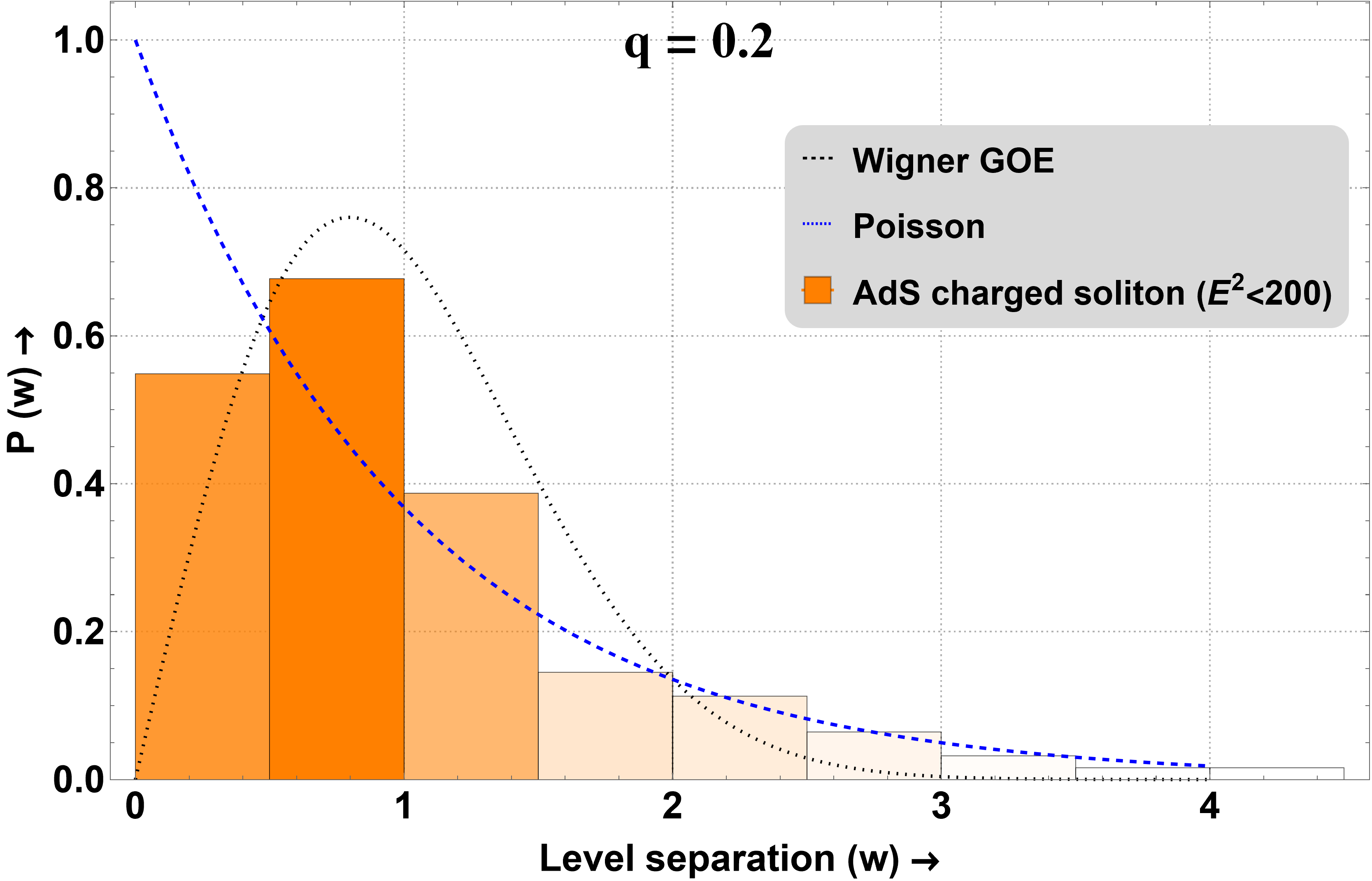}}
	\subfigure[Large $E$ gives nearly Poisson]{\label{fig:susy-soliton-level-spacing-high-energy}\includegraphics[width=0.42\linewidth]{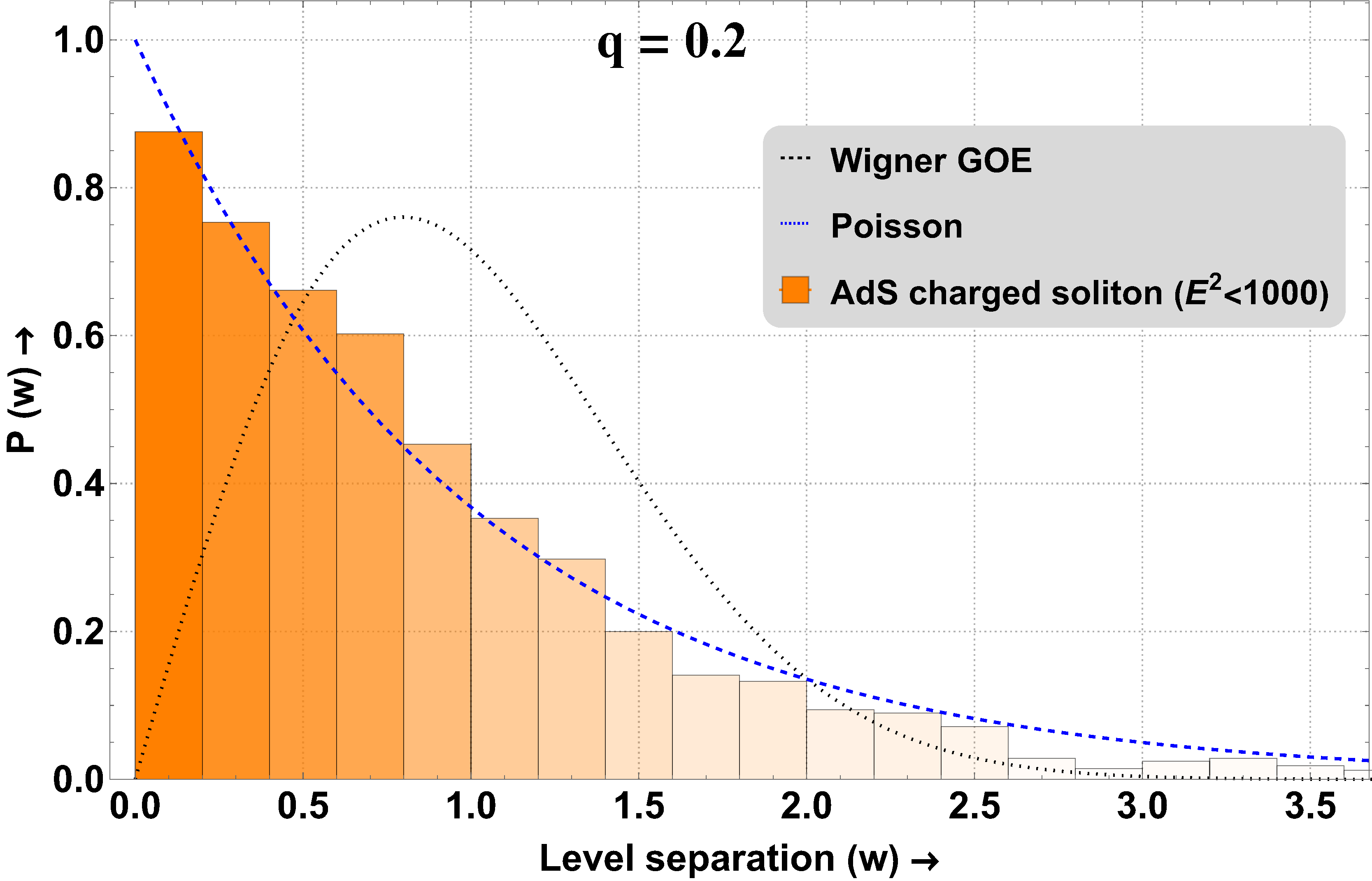}}
	\caption{\label{fig:susy-soliton-level-spacing-constantcharge}Level-spacings distribution for different energies. Here, $q=0.2$ is fixed.}
\end{figure}

For our numerical analysis, we select \( x_{\text{max}} = 10 = -x_{\text{min}}, \, s_{\text{min}} = 0, \, \text{and} \, s_{\text{max}} = 1 \). From the potential plots [Fig. \ref{fig:effective-potential-susy}], we observe that \( \tilde{V}(x, s) \gtrsim 1000 \) at the boundary. Consequently, we examine eigenvalues with \( E^2 \lesssim 1000 \) within the range \( 4 \leq k \leq 12 \) to ensure that the Dirichlet boundary conditions are well respected. A sample eigenfunction is depicted in Fig.~\ref{fig:eigenfunction-plot-susy}, with a contour plot provided in Fig.~\ref{fig:eigenfunction-contour-susy}. After calculating the energies from the eigenvalues \( E^2 \) for each value of \( k \), we normalize these energies, compute the nearest-neighbor differences, and then plot a histogram of the level-spacing distribution as shown in Fig.~\ref{fig:susy-soliton-level-spacing-constantcharge}.

For small energy values, \( E^2 < 200 \), we obtain a distribution similar to the Wigner GOE with a clear signature of level repulsion, indicative of quantum chaos as the energy levels are spaced apart. However, going up to higher energies \( E^2 < 1000 \), we obtain a distribution agreeing with the Poisson distribution of spacing between random and uncorrelated levels. This shows that the system is asymptotically integrable at higher energies and responds to the change in energy as a parameter that dials the system between chaos and integrability for a certain range of energies. 

\begin{figure}[htbp!]
	\centering
    \subfigure[$q=0.1$]{\label{fig:susy-soliton-level-spacing-q0pt1}\includegraphics[width=0.32\linewidth]{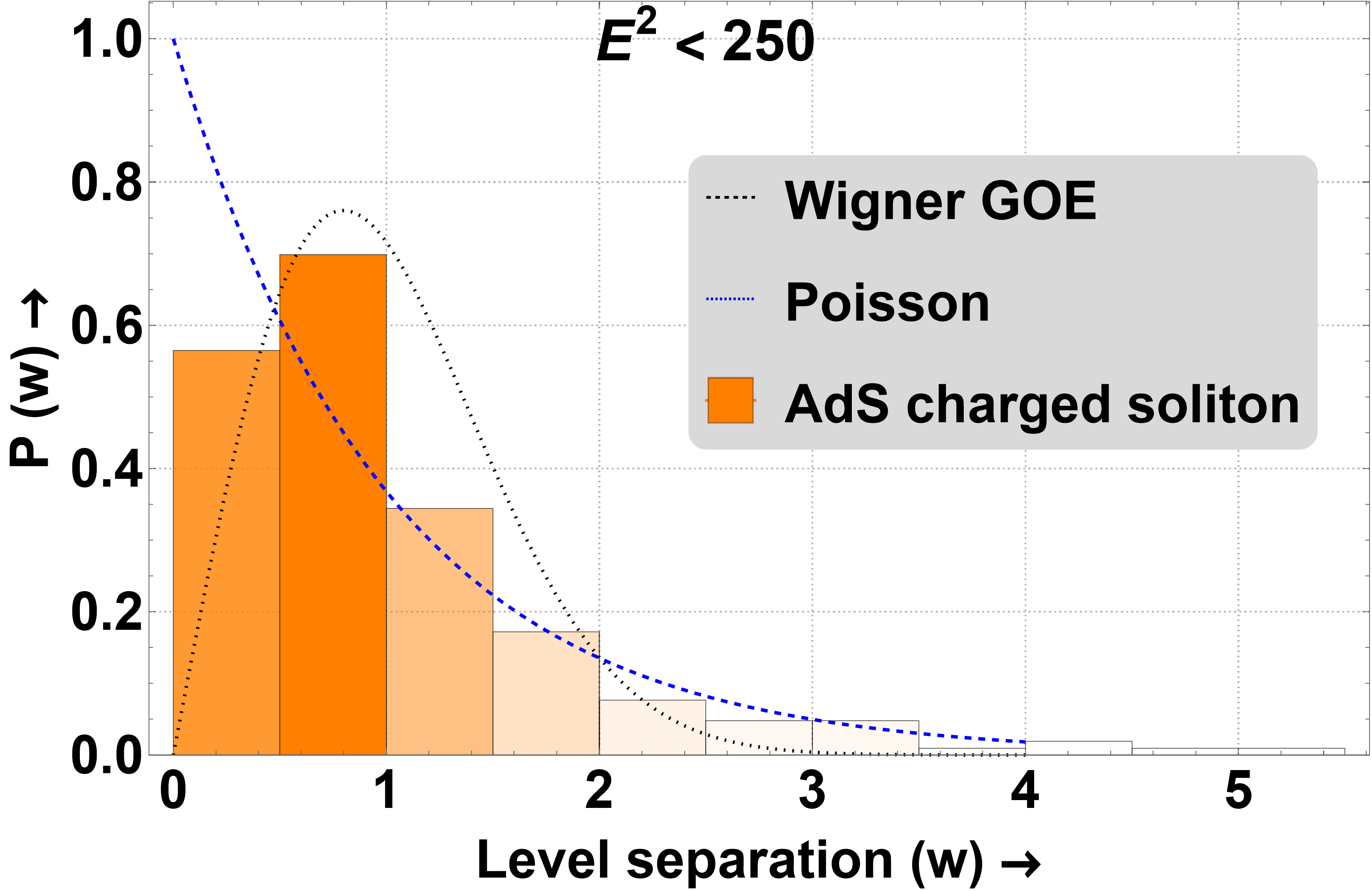}}
    \subfigure[$q=0.5$]{\label{fig:susy-soliton-level-spacing-q0pt5}\includegraphics[width=0.32\linewidth]{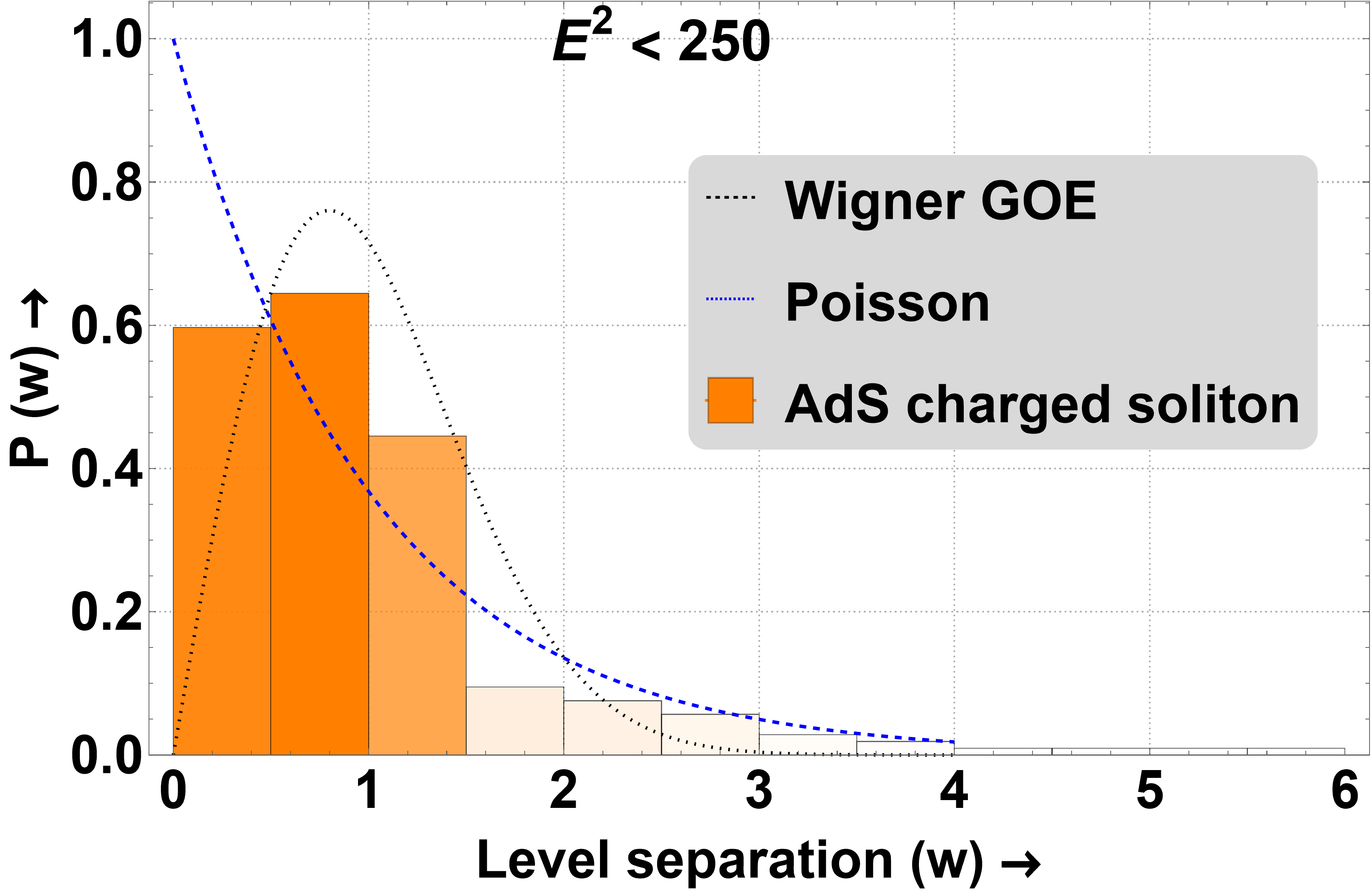}}
    \subfigure[$q=0.9$]{\label{fig:susy-soliton-level-spacing-q0pt9}\includegraphics[width=0.32\linewidth]{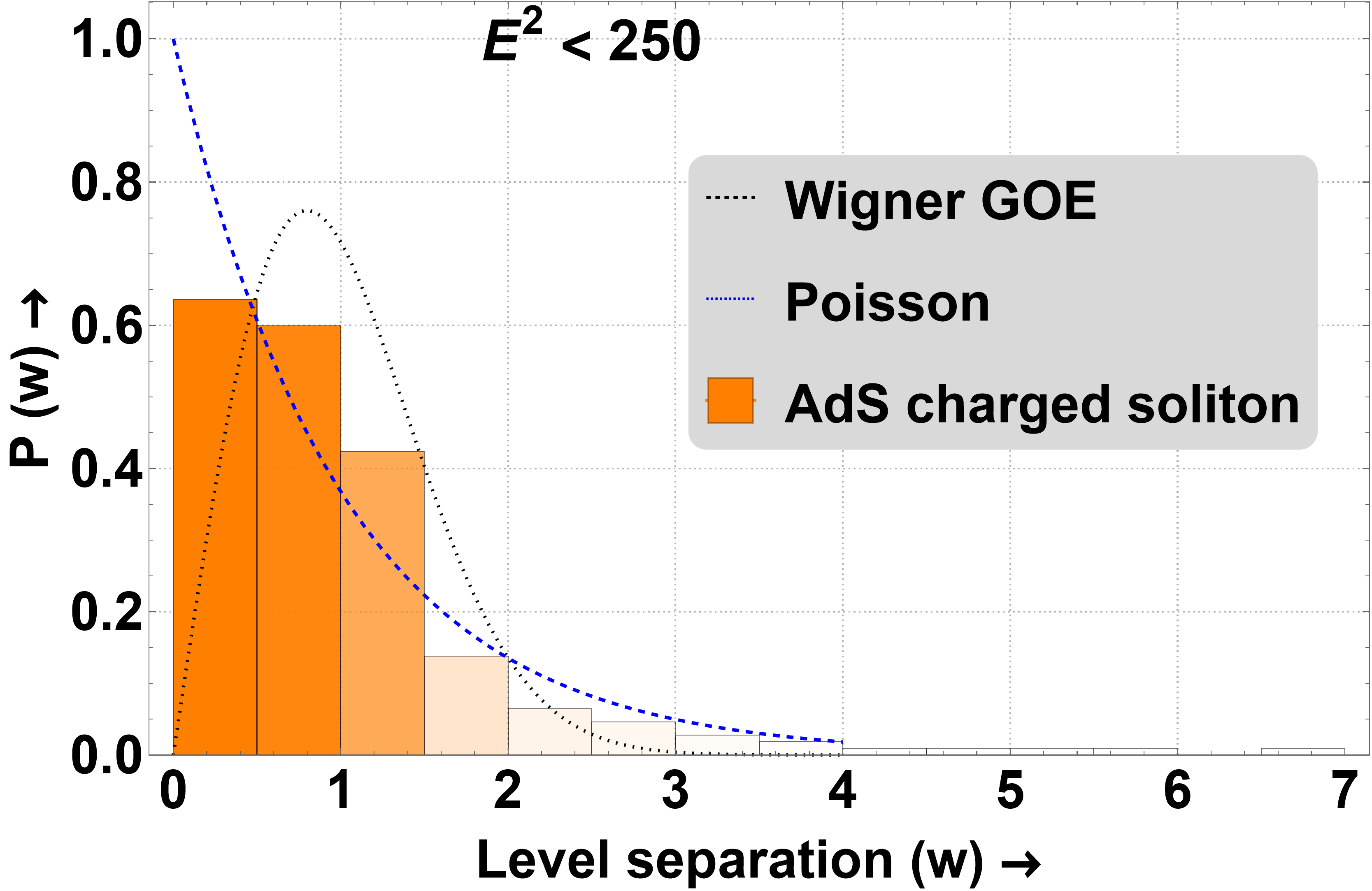}}
	\caption{\label{fig:susy-soliton-level-spacing-constantenergy}Level-spacings distribution for different charge values. Here $E^2<250$ is considered.}
\end{figure}

We further analyze the energy level distribution as we dial the charge parameter. This is shown in Fig. \ref{fig:susy-soliton-level-spacing-constantenergy}, where we observe the effect of varying \( q \) on the distribution of level spacing, keeping the bound on maximum energy \( E^2 \) constant. We find that, like in the case of classical chaos, the effect of $q$ is similar to the effect of $E$ here as well. In particular, at low $q$, the distribution is of Wigner GOE type, whereas at high $q$, the distribution approaches the Poisson type. The distribution shifts as the charge \( q \) increases, suggesting that the system is becoming integrable for higher values of $q$ and that it can also serve as a parameter for asymptotic integrability.

\begin{figure}[htbp!]
	\centering
    \subfigure[Keeping the charge constant at $q=0.2$, we see an approximate agreement of \( \overline{\Delta}_3 \) with Wigner GOE at low energies (blue triangles, $E^2 < 200$) and agreement with Poisson for higher energies (orange dots, $200 < E^2< 1000$). ]{\label{fig:dysonmehta-fixedcharge}\includegraphics[width=0.45\linewidth]{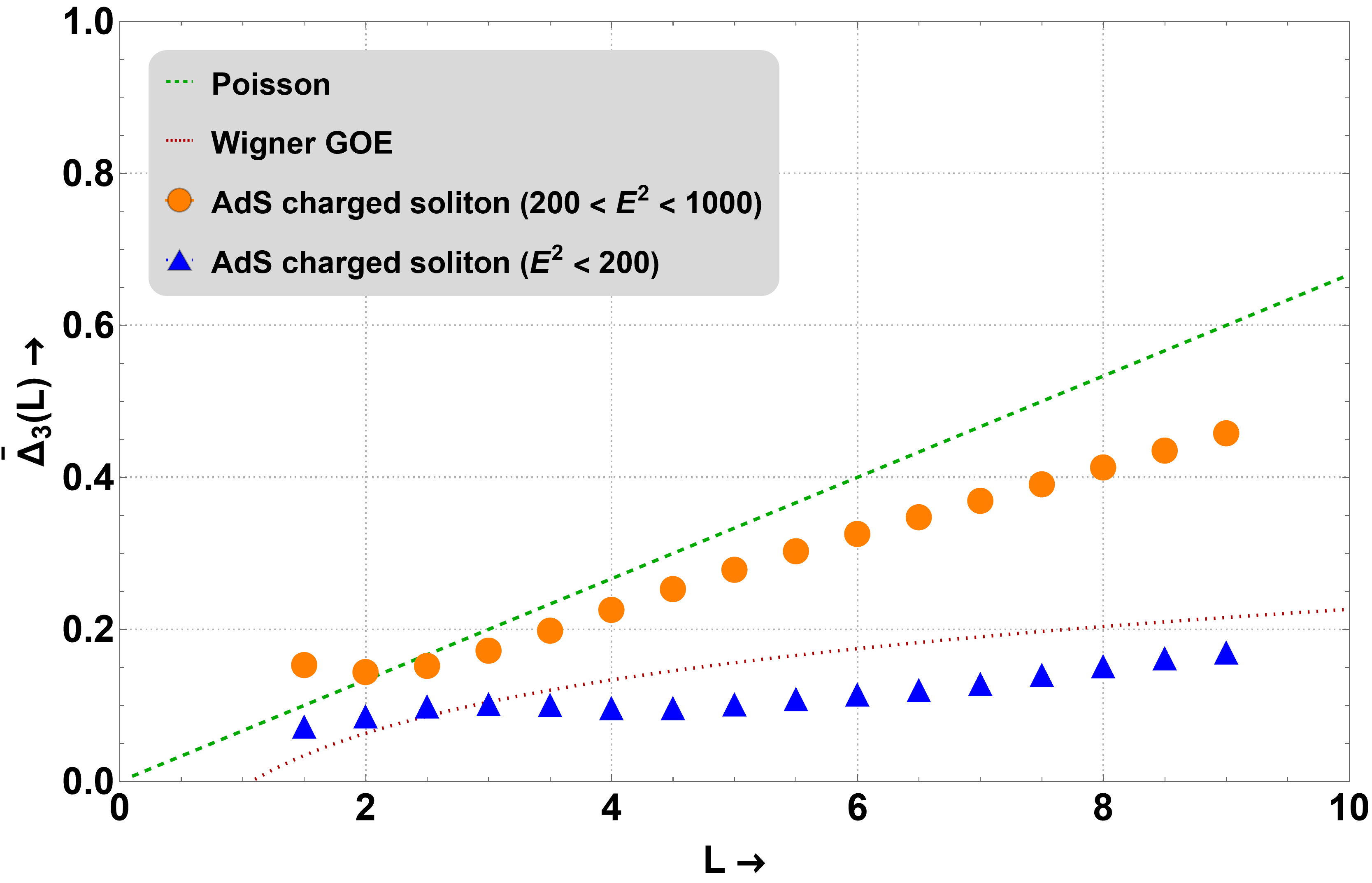}}
    \hspace{0.5cm}
	\subfigure[Keeping the energy constant at $E^2 < 350$, we increase the charge from $q=0$ to $q=0.99$ and observe a splitting between the two lines indicating the role of charge as a stabilizer lifting the spectrum towards the integrable regime. ]{\label{fig:dysonmehta-fixedenergy}\includegraphics[width=0.45\linewidth]{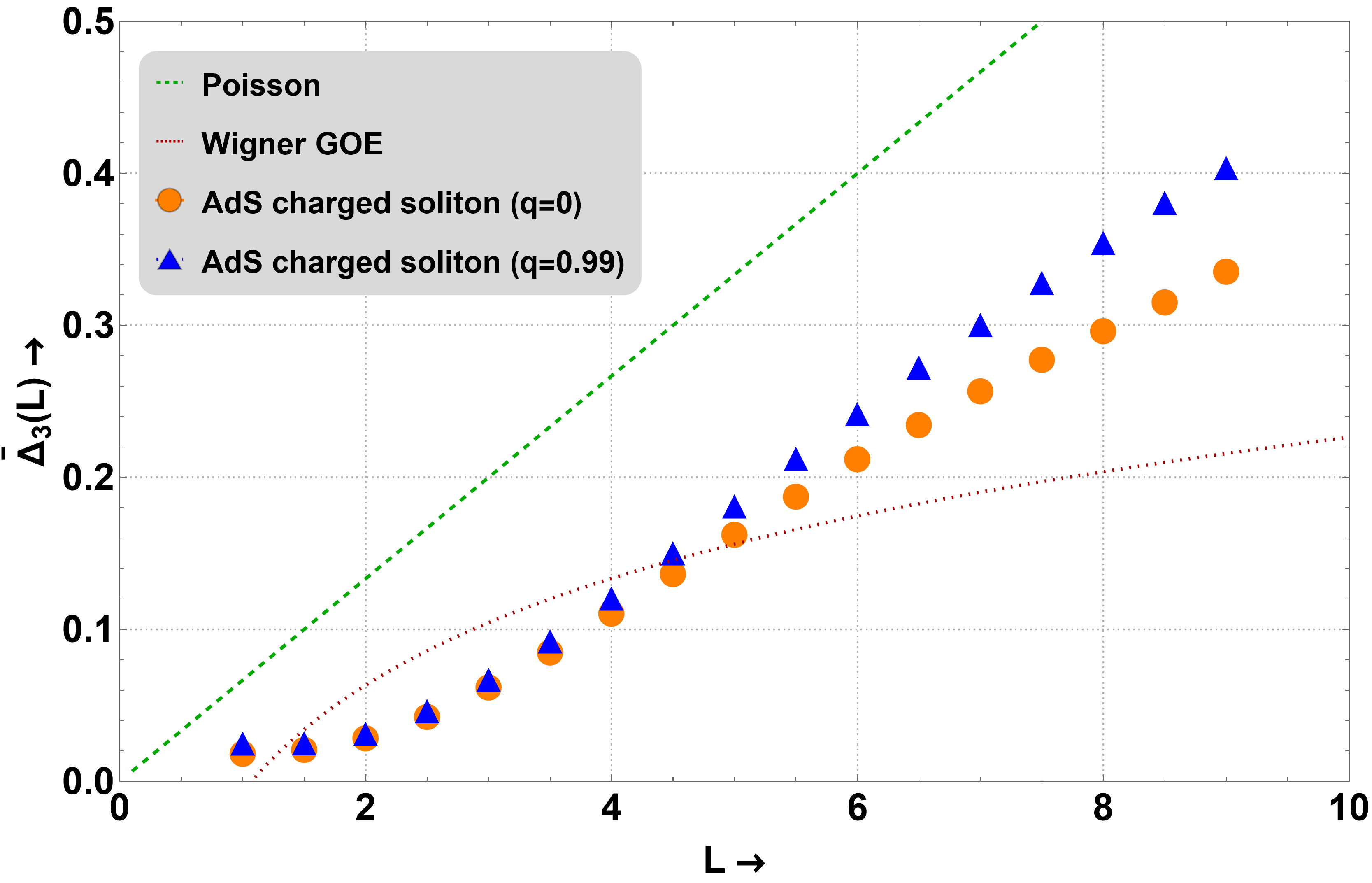}}
	\caption{\label{fig:dysonmehta-statistics}Dyson-Mehta $\Delta_3$  statistic and spectral rigidity for the closed string in the AdS charged soliton background. The dashed green and dotted red lines correspond to Poisson and Wigner GOE distributions.}
\end{figure}

\begin{figure}[htbp!]
	\centering
    \subfigure[$k=6$, \quad $q=0.0$]{\label{fig:otock6opt010to80}\includegraphics[width=0.42\linewidth]{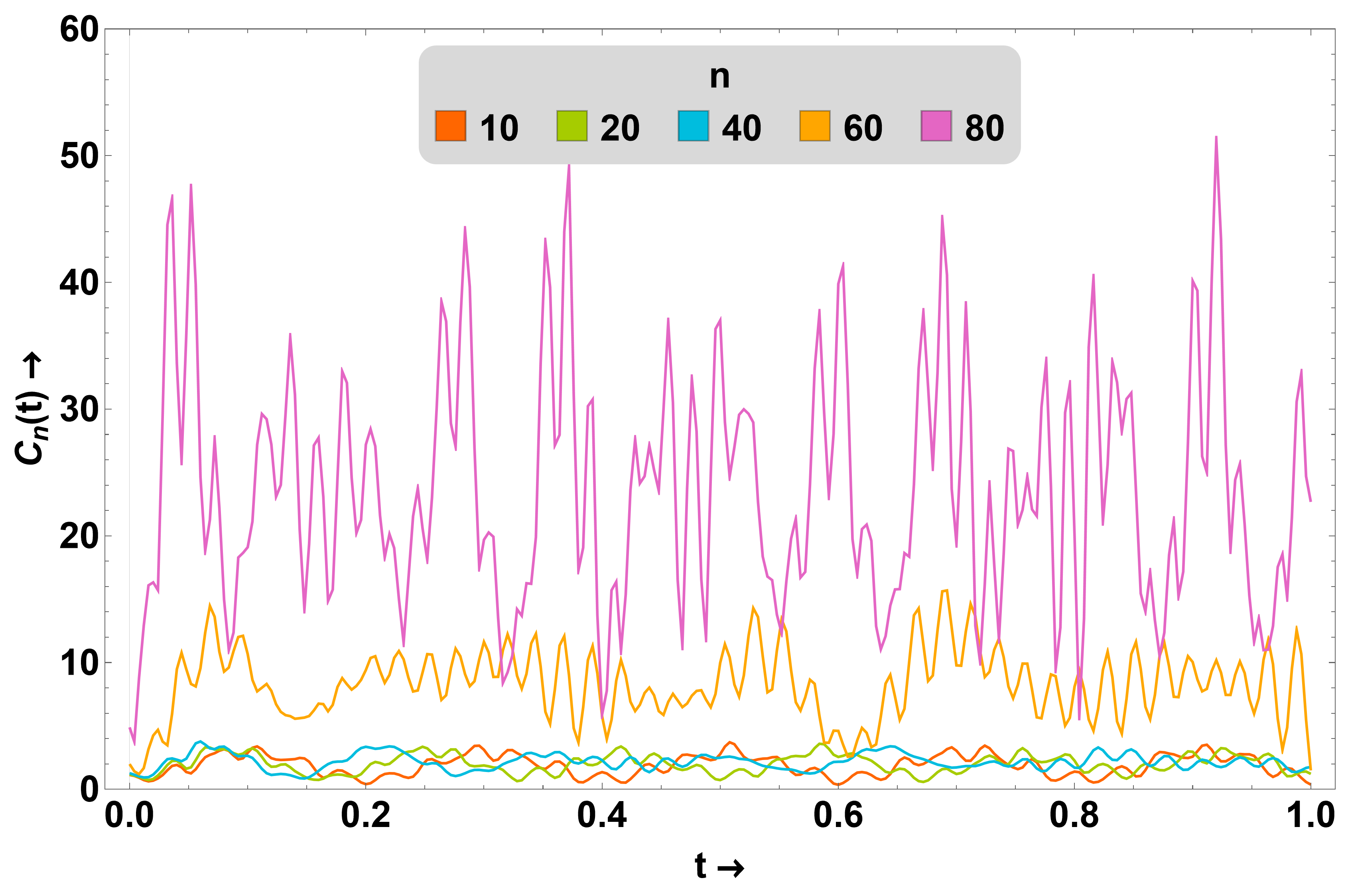}}
	\subfigure[$k=6$, \quad $q=0.9$]{\label{fig:otock6opt910to80}\includegraphics[width=0.42\linewidth]{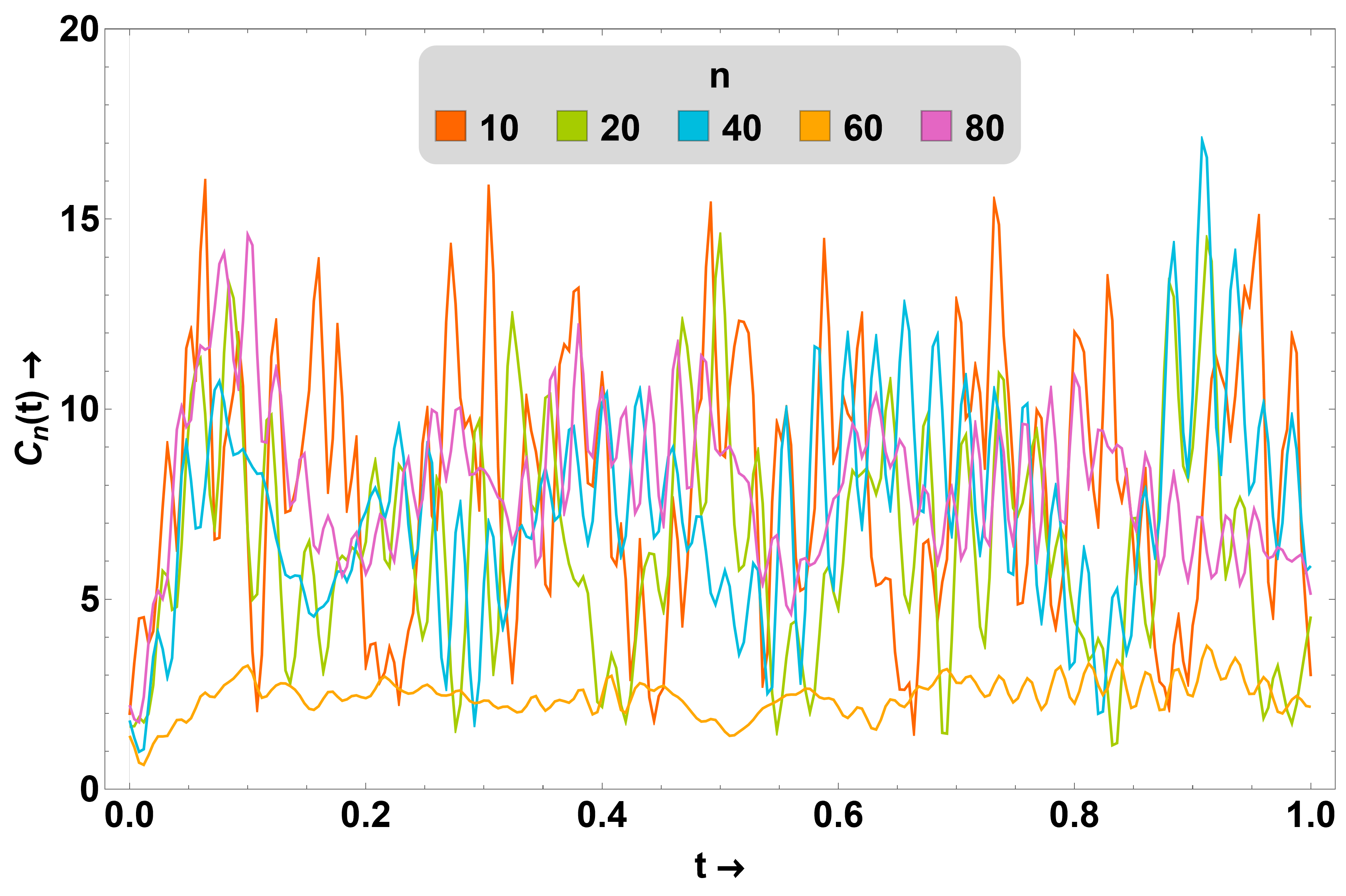}}\\
    \subfigure[$k=6$, \quad $q=0.0$]{\label{fig:otock6opt0100to180}\includegraphics[width=0.42\linewidth]{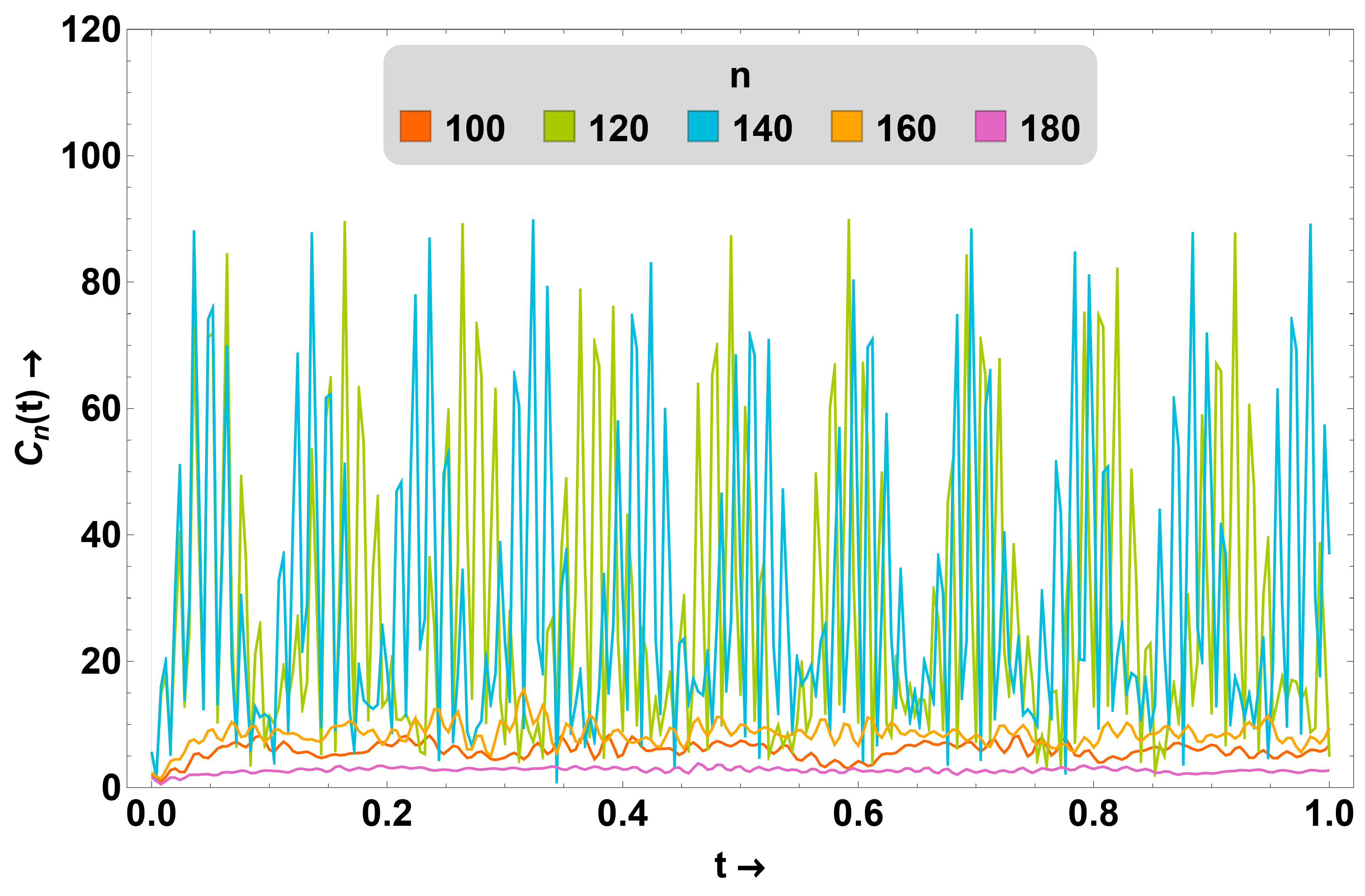}}
	\subfigure[$k=6$, \quad $q=0.9$]{\label{fig:otock6opt9100to180}\includegraphics[width=0.42\linewidth]{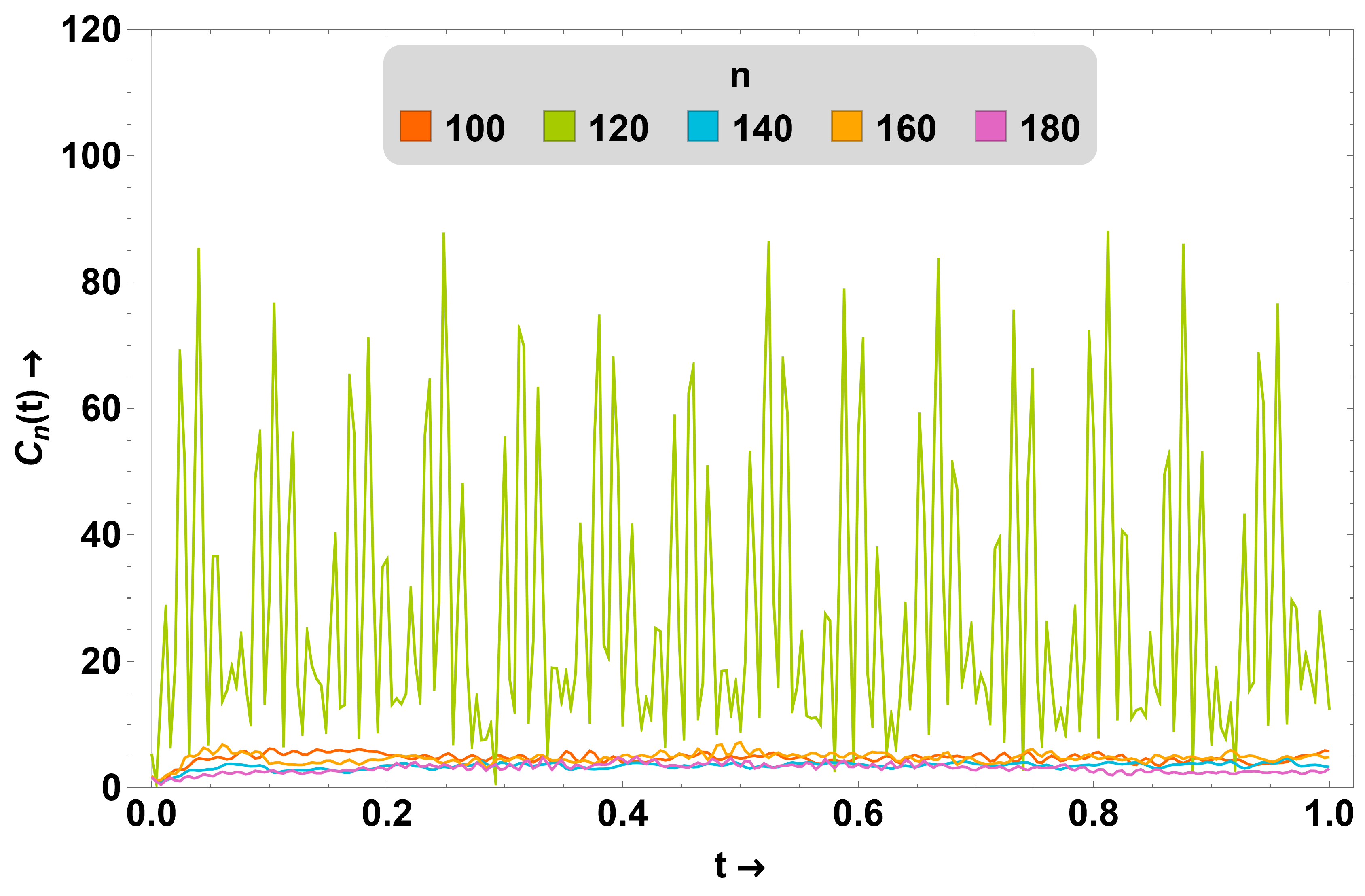}}
	\caption{\label{fig:otoc-k6-qvary}Microcanonical OTOCs for different eigenvalues at $k=6$ and $q=0$, $q=0.9$.}
\end{figure}

To further examine the chaotic nature of the system, we compute the \( \Delta_3 \) statistic, which measures spectral rigidity and is defined in (\ref{eq:dyson-mehta}). The results are illustrated in Fig.~\ref{fig:dysonmehta-statistics}. At a fixed charge $q=0.2$, we analyze our system for lower energy levels \( E^2 < 200 \) and observe that the values of \( \overline{\Delta}_3(L) \) align with the curve for quantum chaotic behavior. As the energy increases to \( 200 < E^2 < 1000 \), \( \overline{\Delta}_3(L) \) values rise, approaching the \( L/15 \) line that characterizes the spectral rigidity of random, uncorrelated levels, which reflects the larger spacing deviations typical of integrable systems as shown in  Fig.~\ref{fig:dysonmehta-fixedcharge}.

\begin{figure}[htbp!]
{\def\arraystretch{2}\tabcolsep=4pt
	\begin{tabular}{|c|c|c|c|}
        \hline
		\textbf{Energy level} & \textbf{$k=1$} & \textbf{$k=2$} & \textbf{$k=3$} \\
        \hline
        \rule{0pt}{10ex}
		\textbf{$n=5$} & \includegraphics[scale=0.14,valign=c]{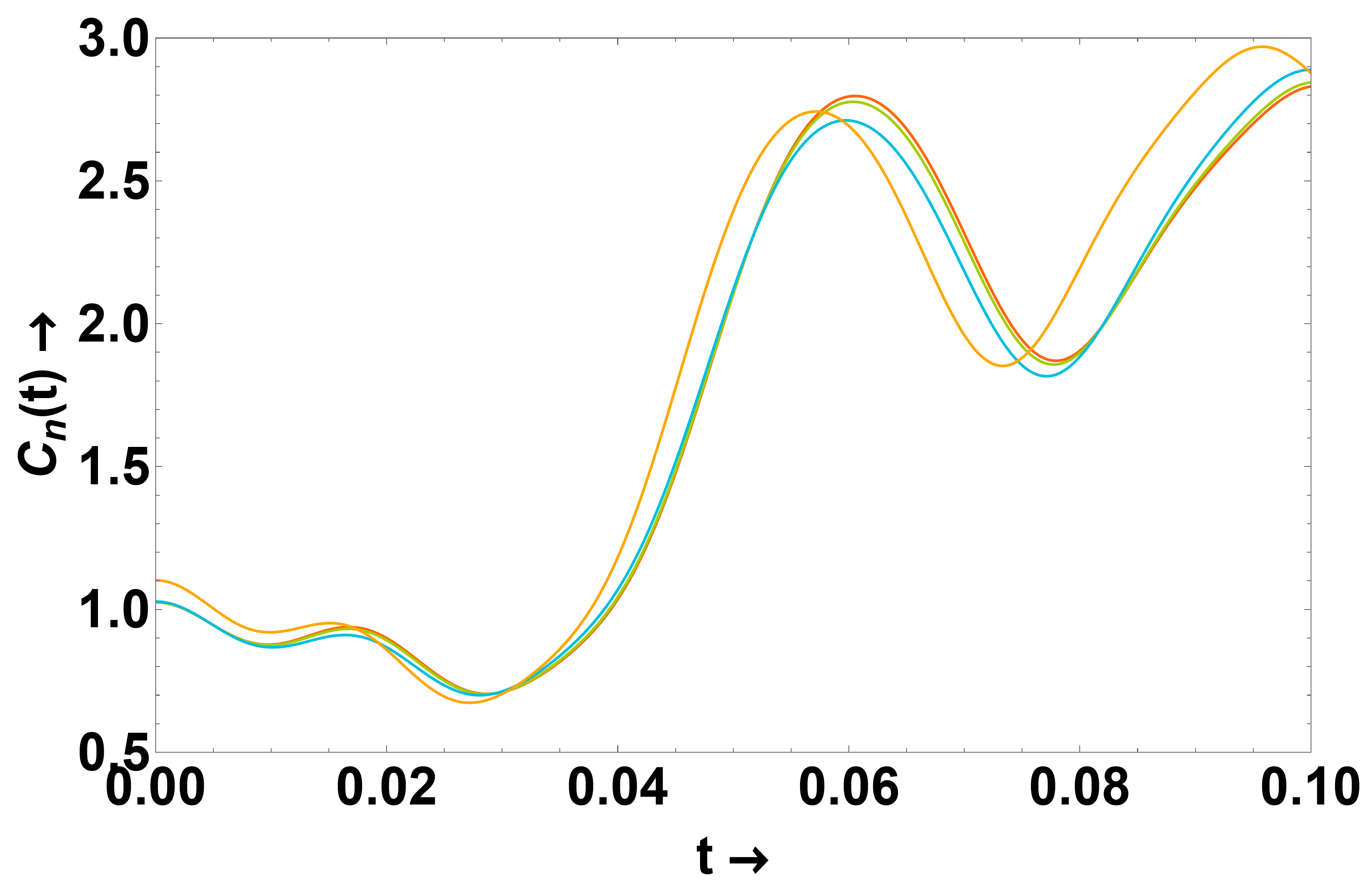} & \includegraphics[scale=0.14,valign=c]{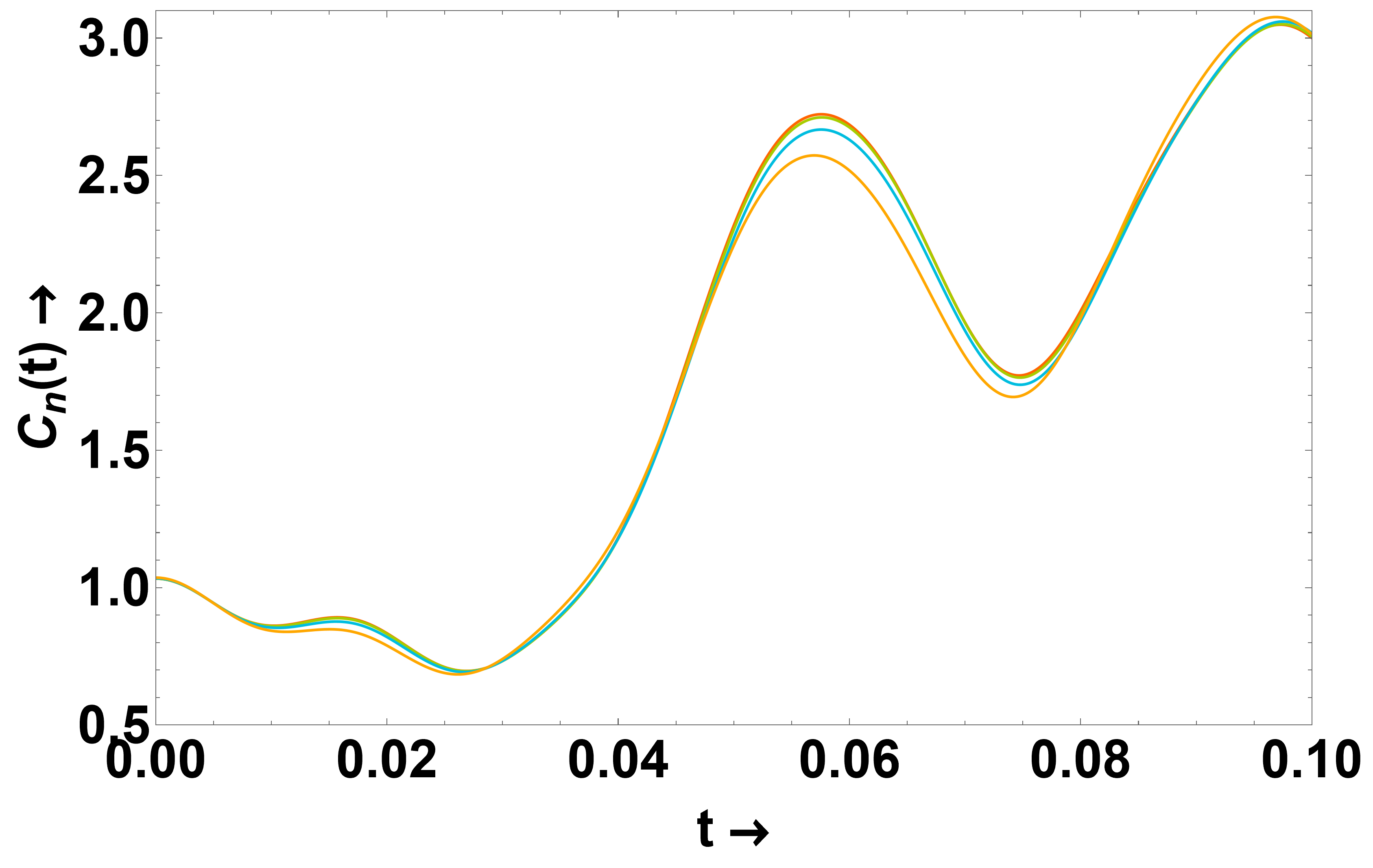} & \includegraphics[scale=0.14,valign=c]{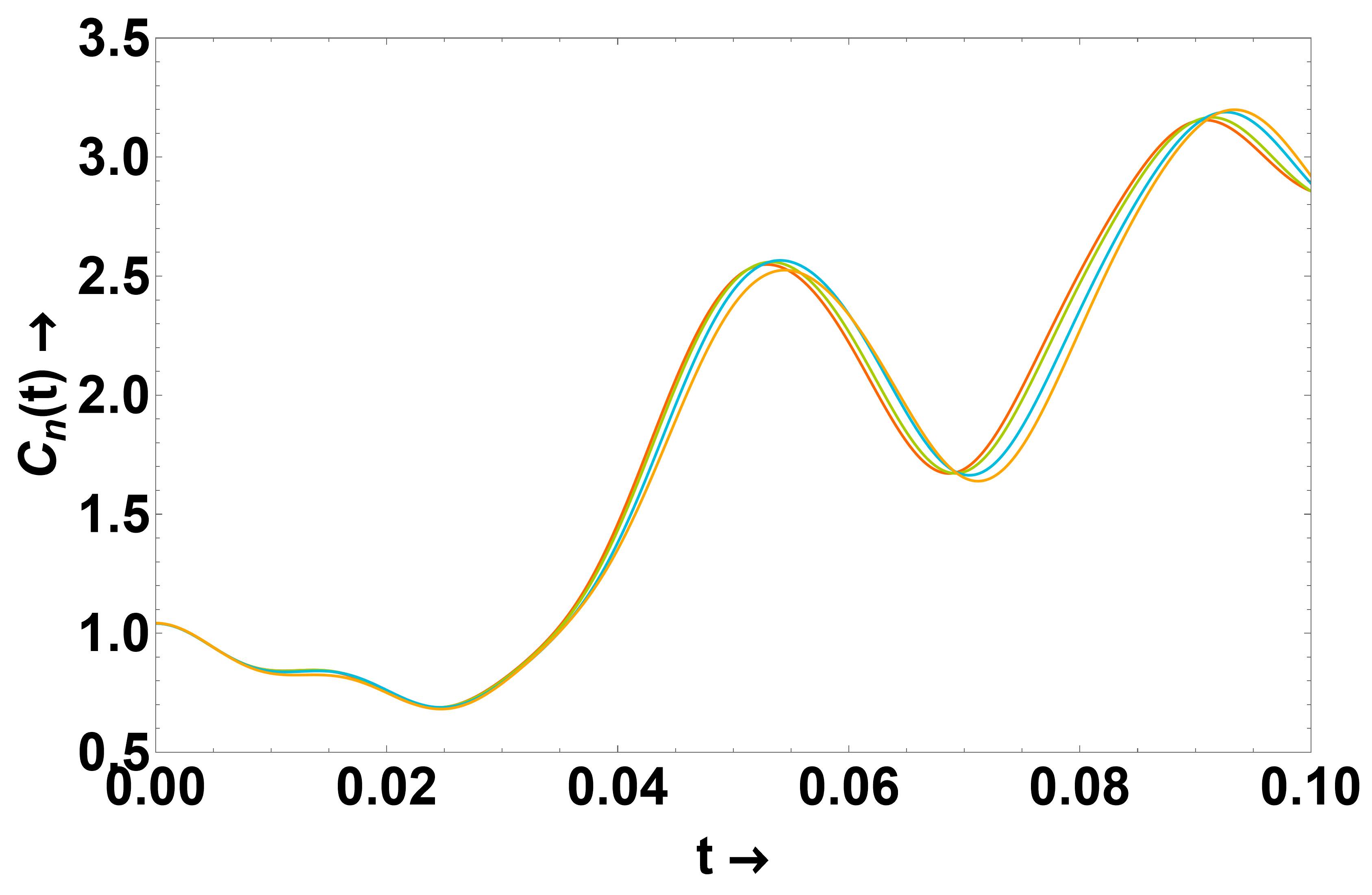} \\
        \rule{0pt}{10ex}
		\textbf{$n=10$} & \includegraphics[scale=0.14,valign=c]{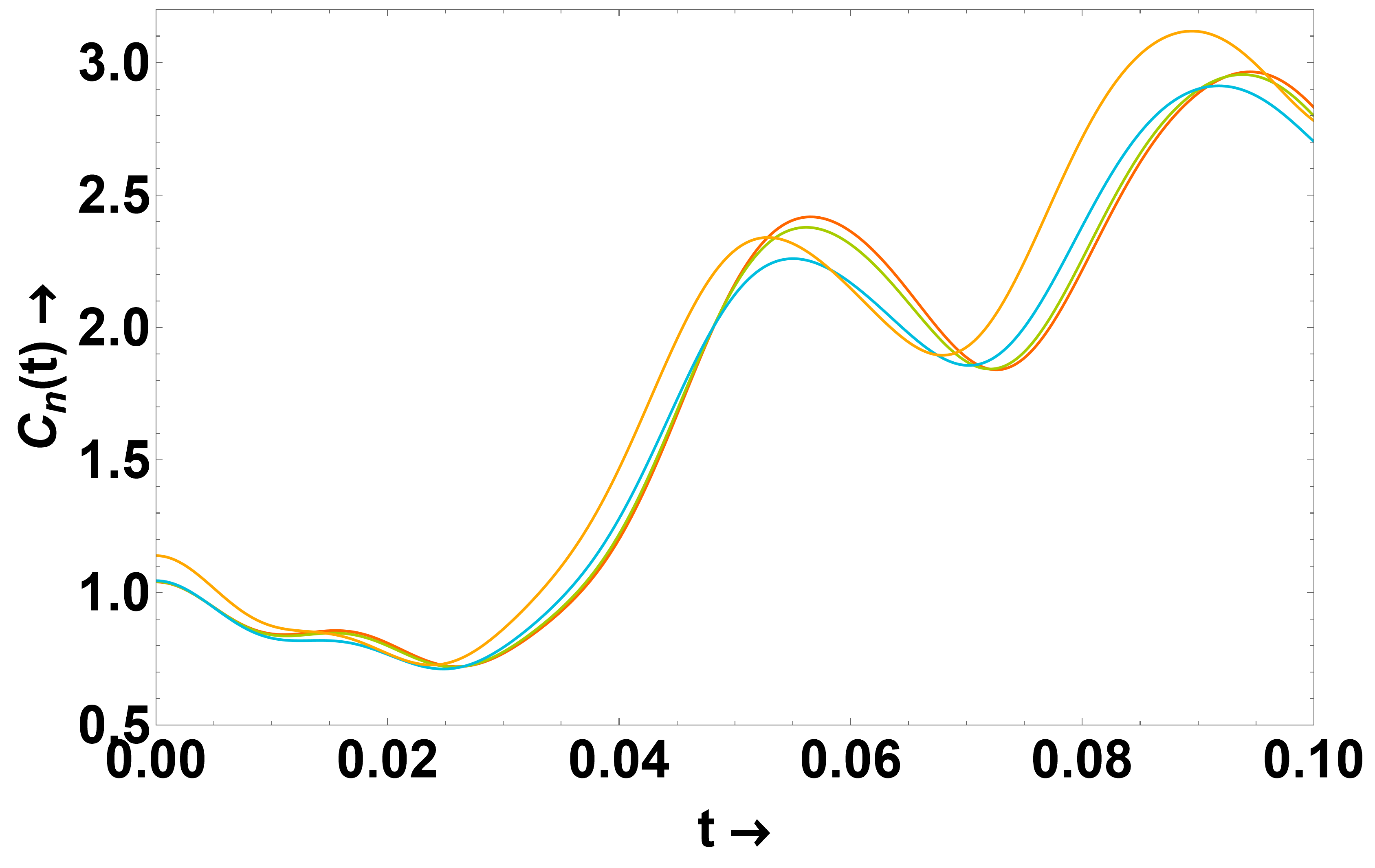} & \includegraphics[scale=0.14,valign=c]{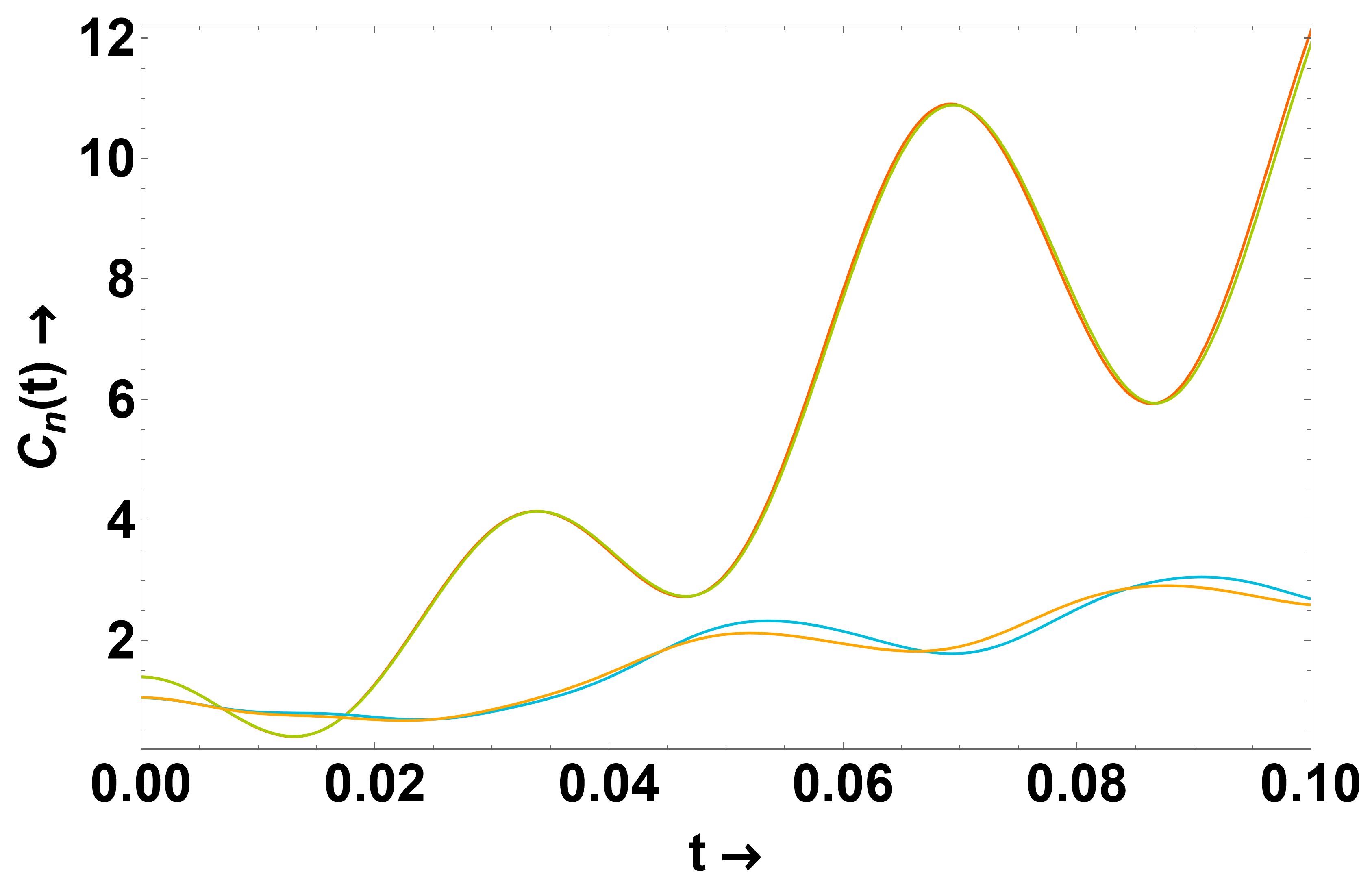} & \includegraphics[scale=0.14,valign=c]{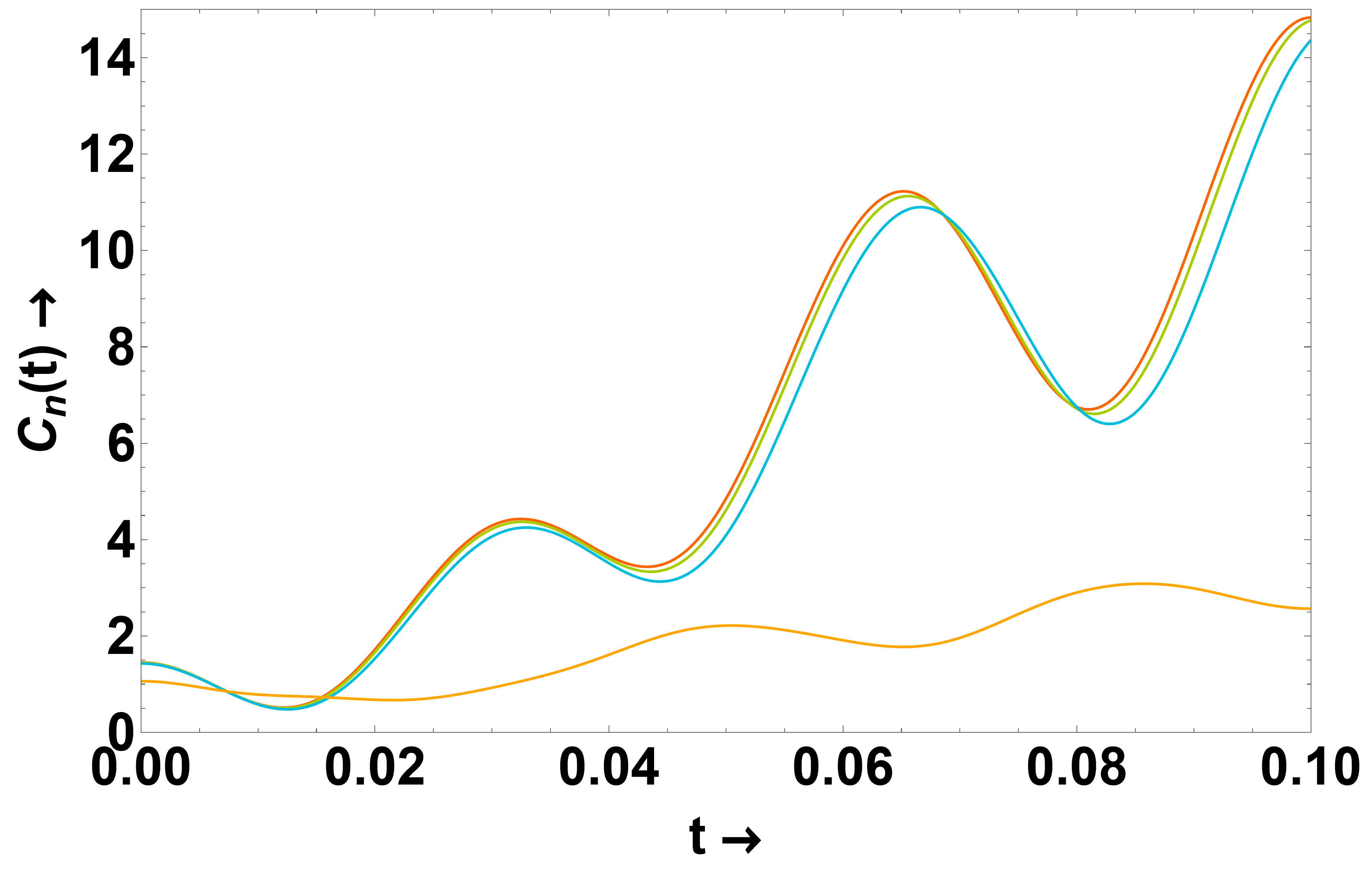} \\
        \rule{0pt}{10ex}
		\textbf{$n=20$} & \includegraphics[scale=0.14,valign=c]{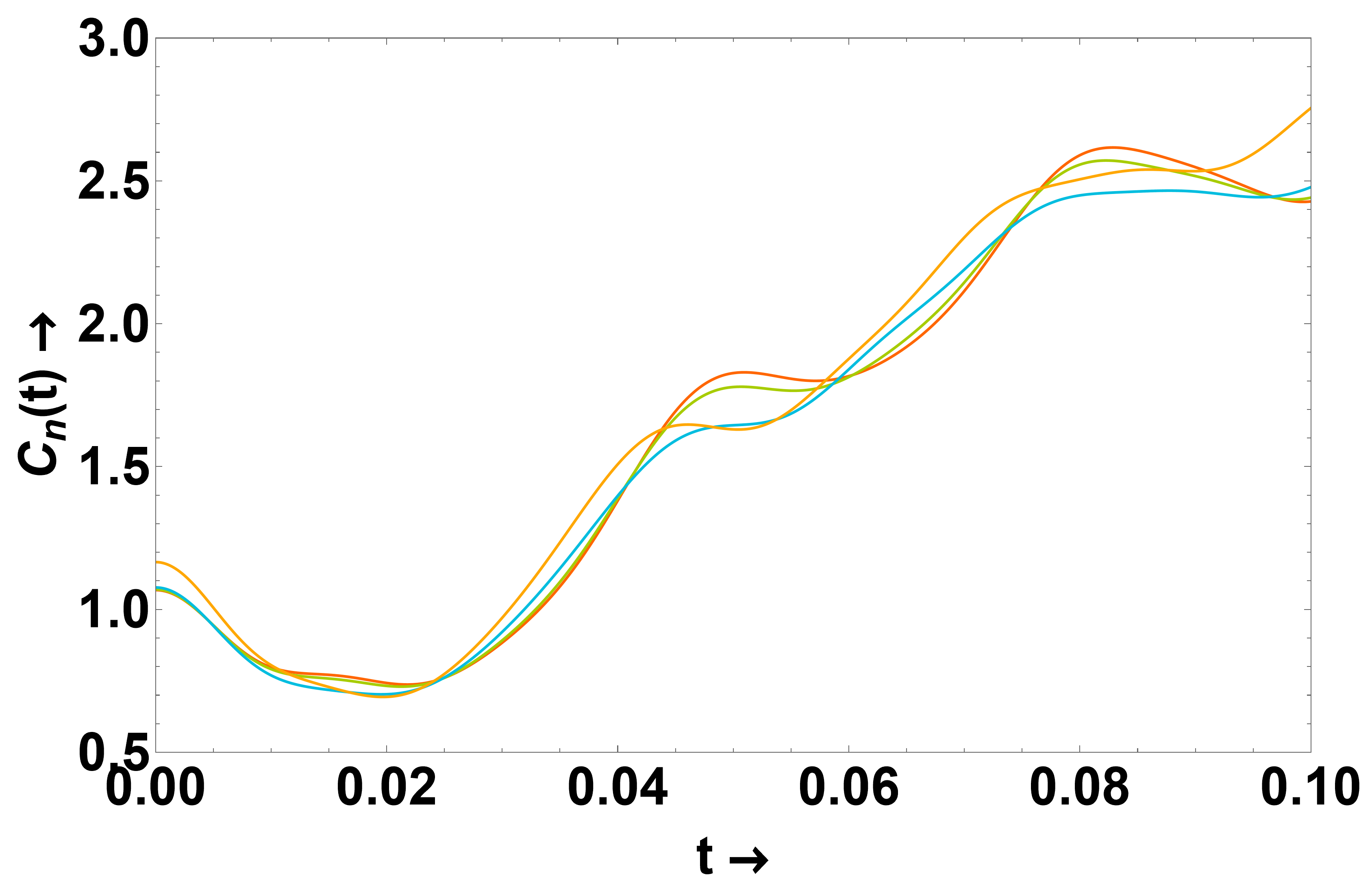} & \includegraphics[scale=0.14,valign=c]{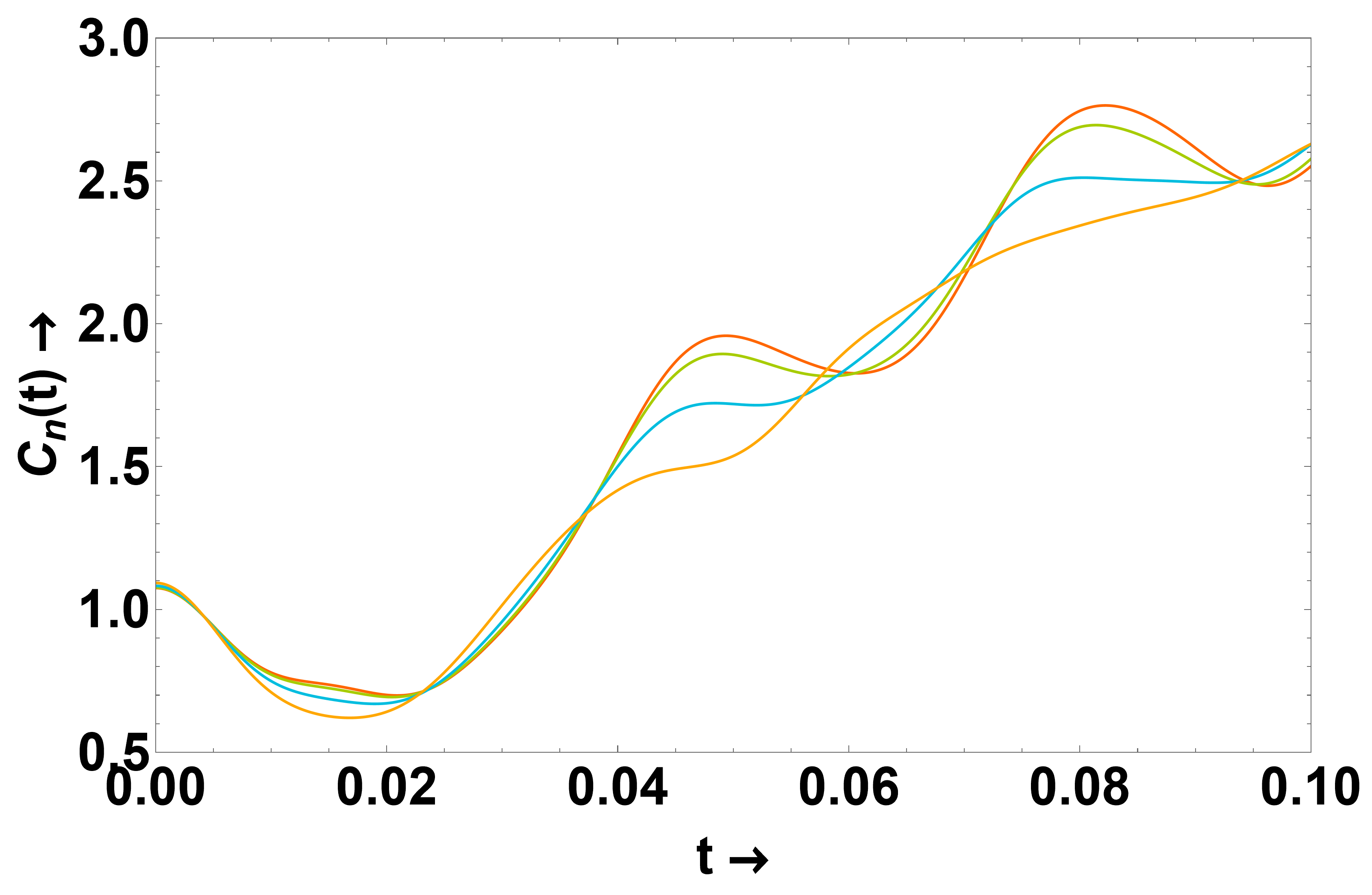} & \includegraphics[scale=0.14,valign=c]{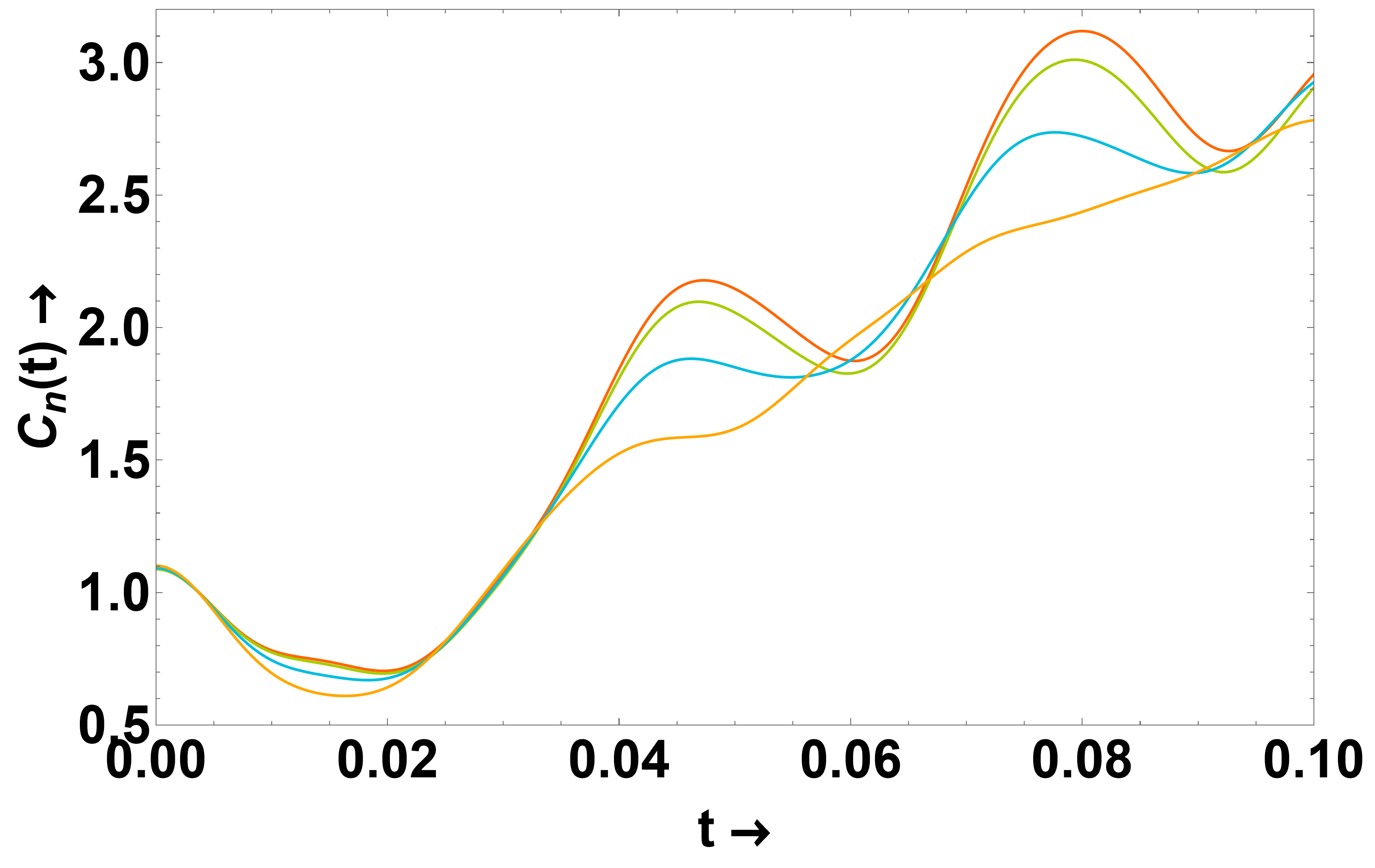} \\
        \hline\hline
        \rule{0pt}{10ex}
        \textbf{$n=110$} & \includegraphics[scale=0.14,valign=c]{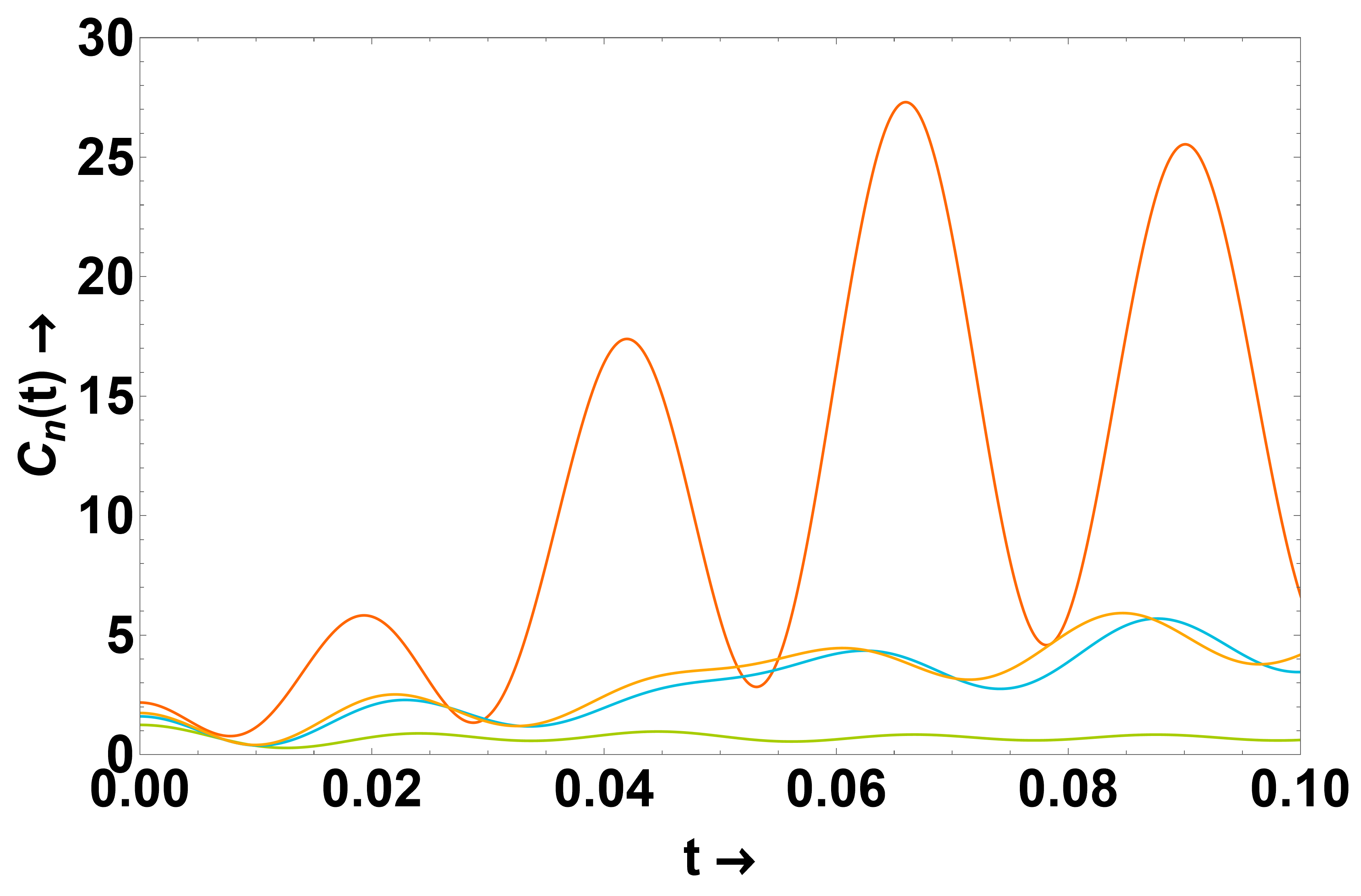} & \includegraphics[scale=0.14,valign=c]{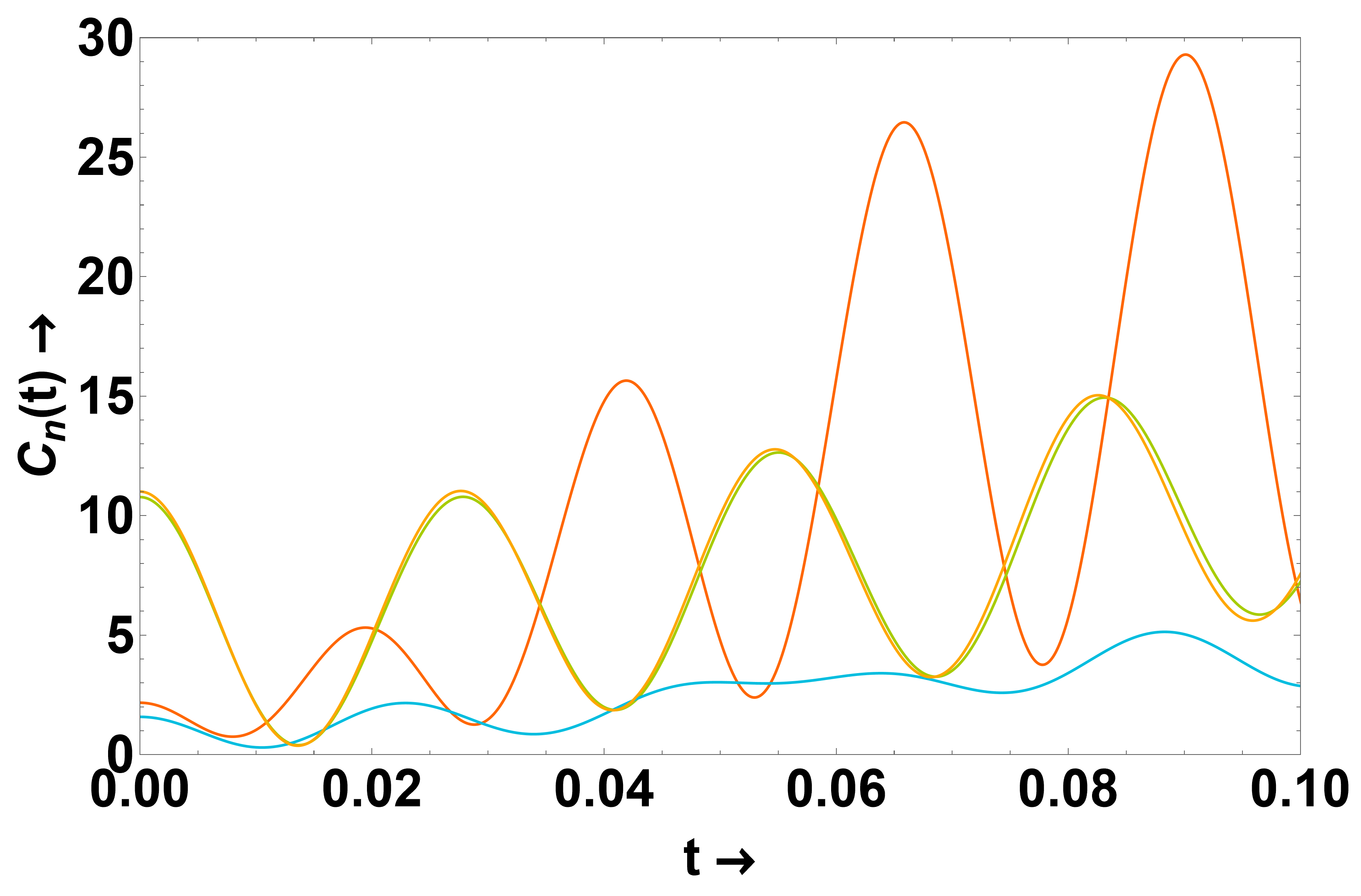} & \includegraphics[scale=0.14,valign=c]{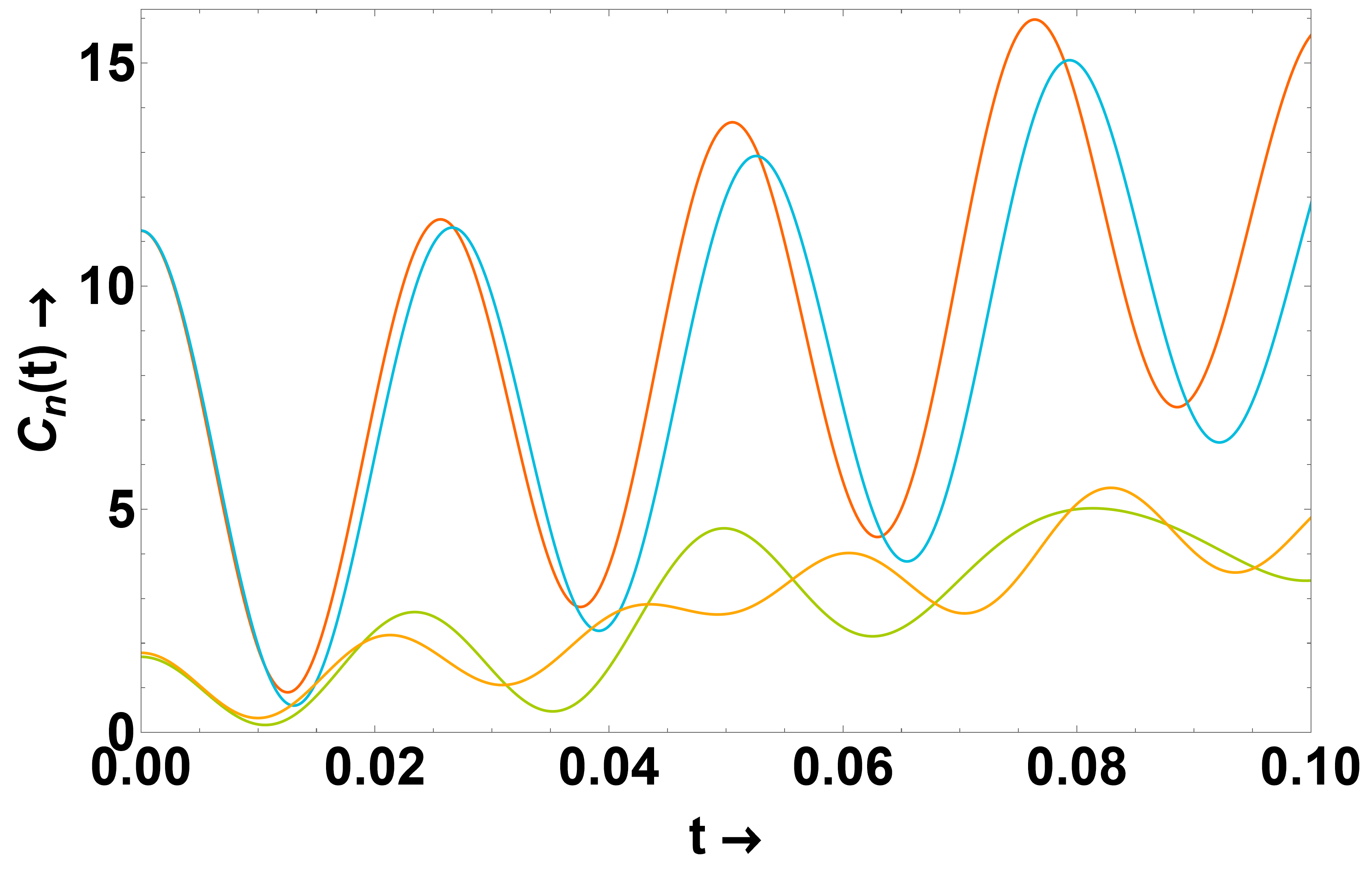} \\
        \rule{0pt}{10ex}
        \textbf{$n=120$} & \includegraphics[scale=0.14,valign=c]{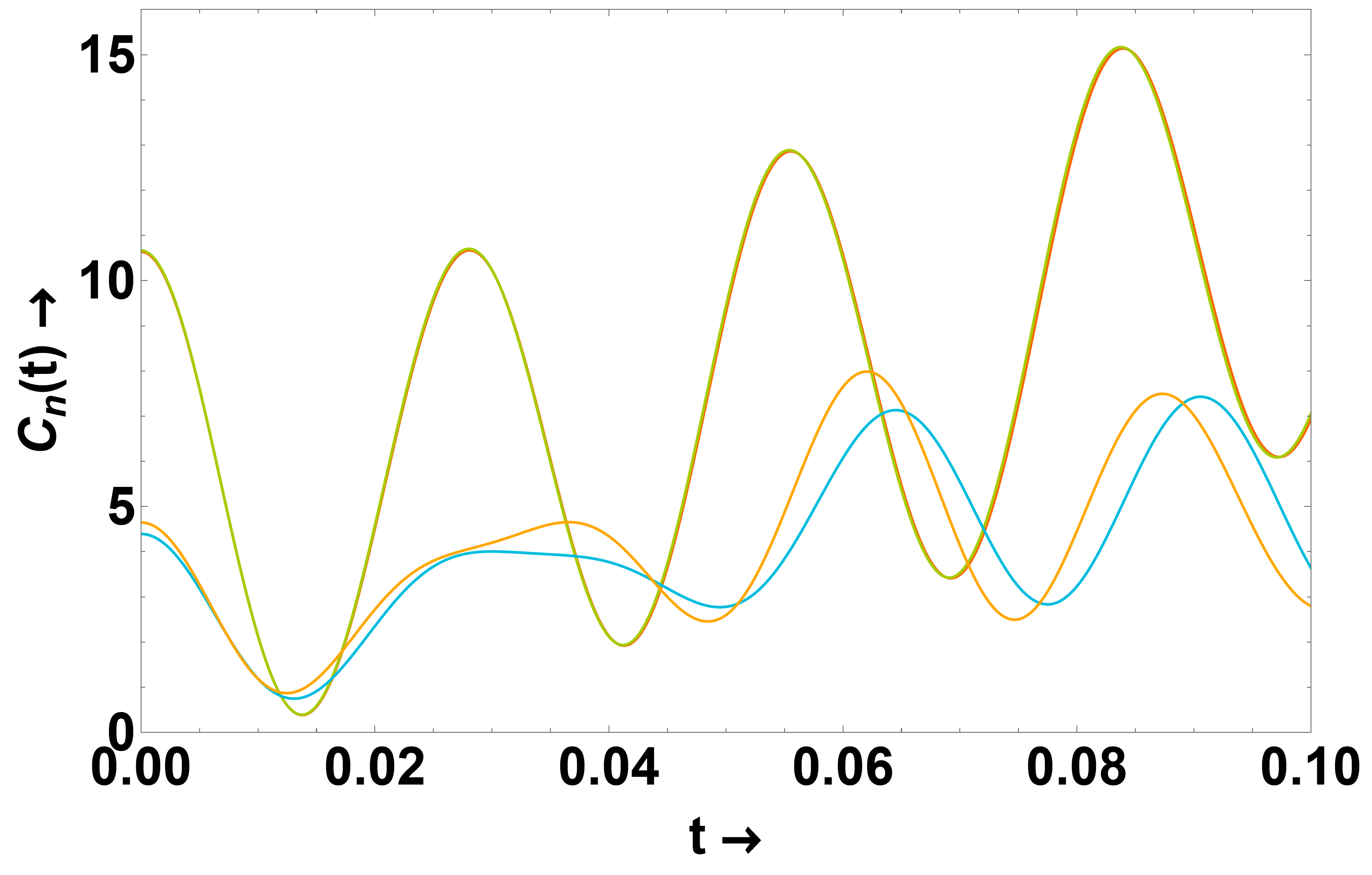} & \includegraphics[scale=0.14,valign=c]{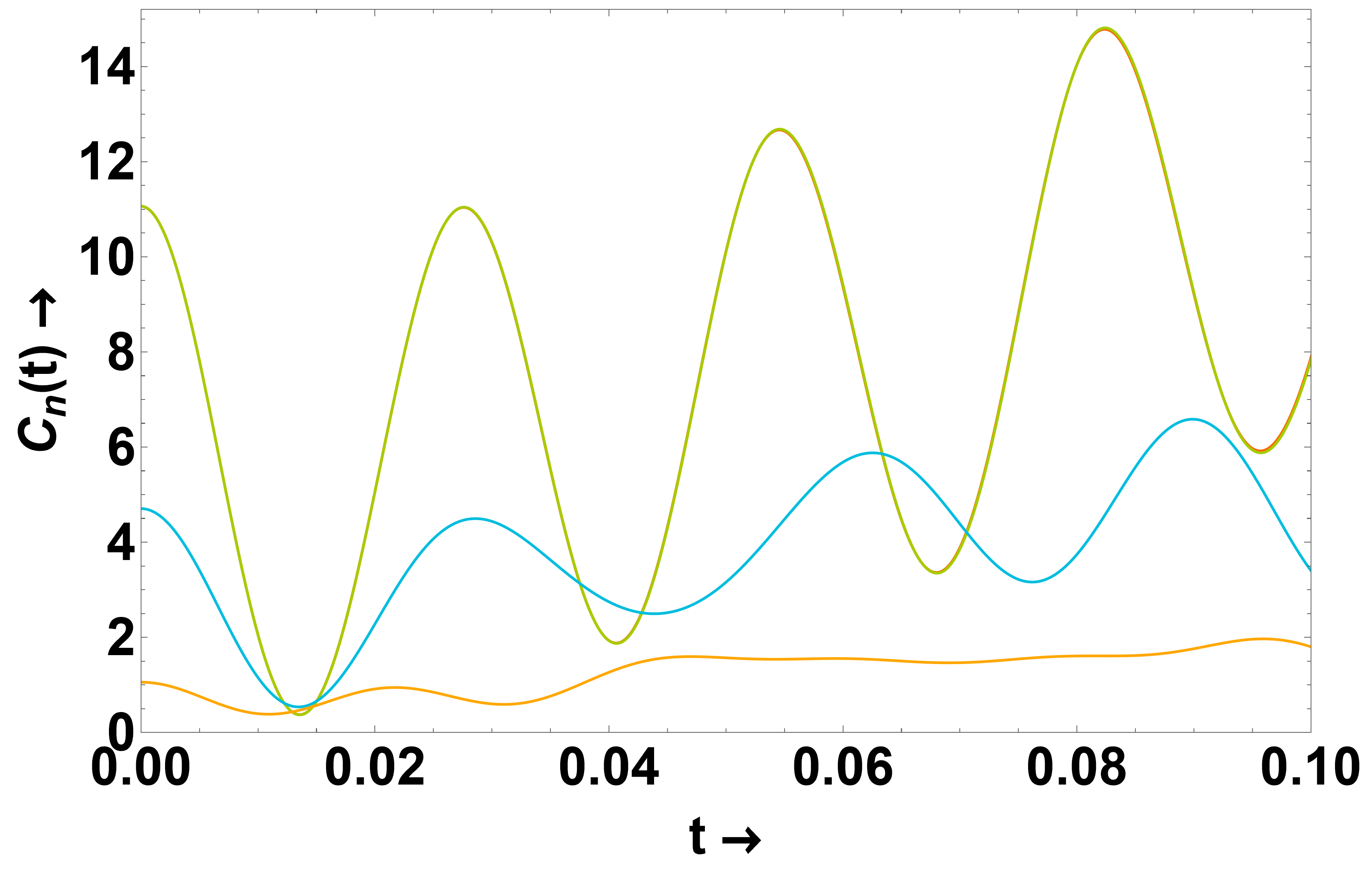} & \includegraphics[scale=0.14,valign=c]{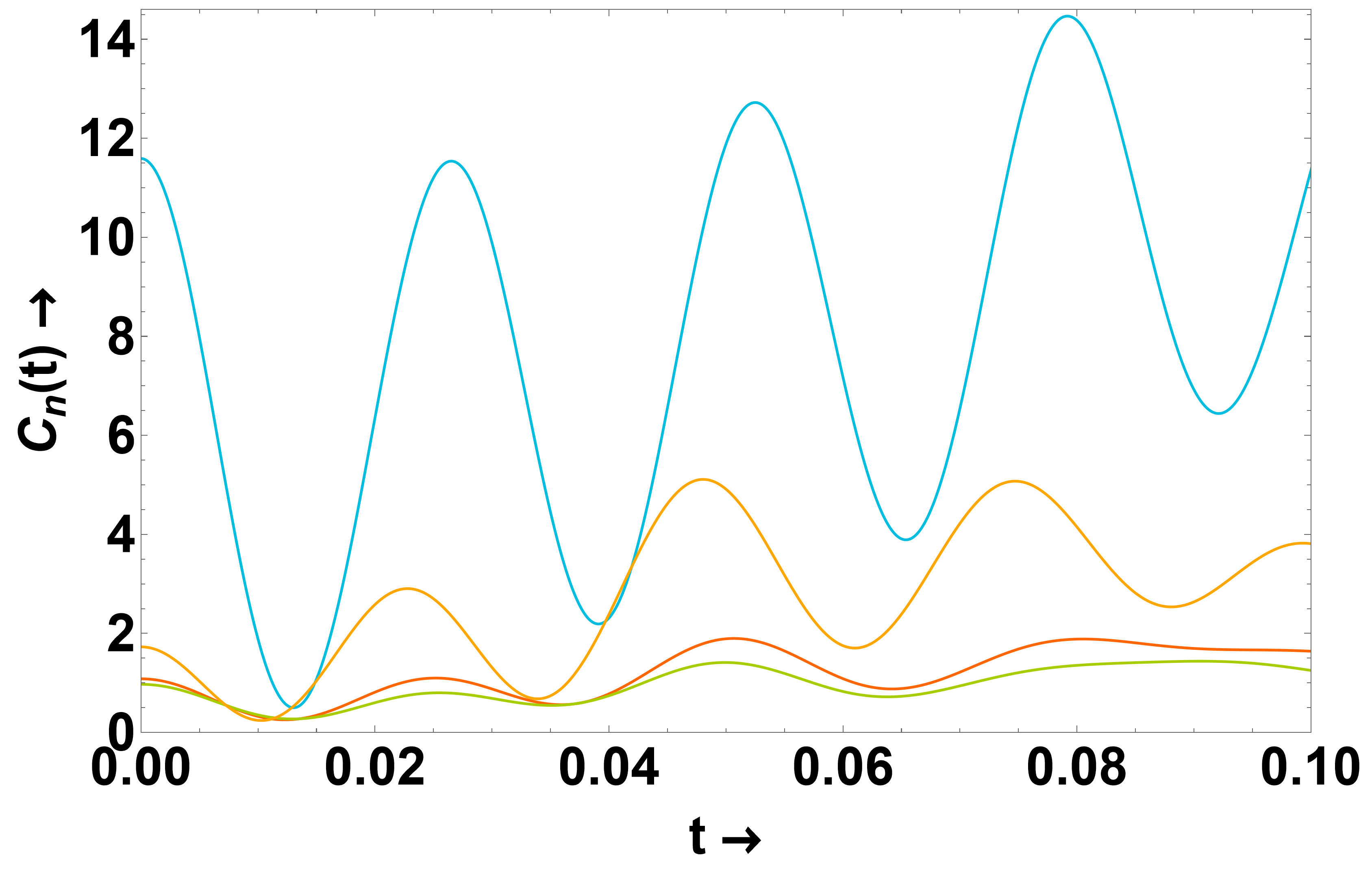} \\
        \rule{0pt}{10ex}
        \textbf{$n=150$} & \includegraphics[scale=0.14,valign=c]{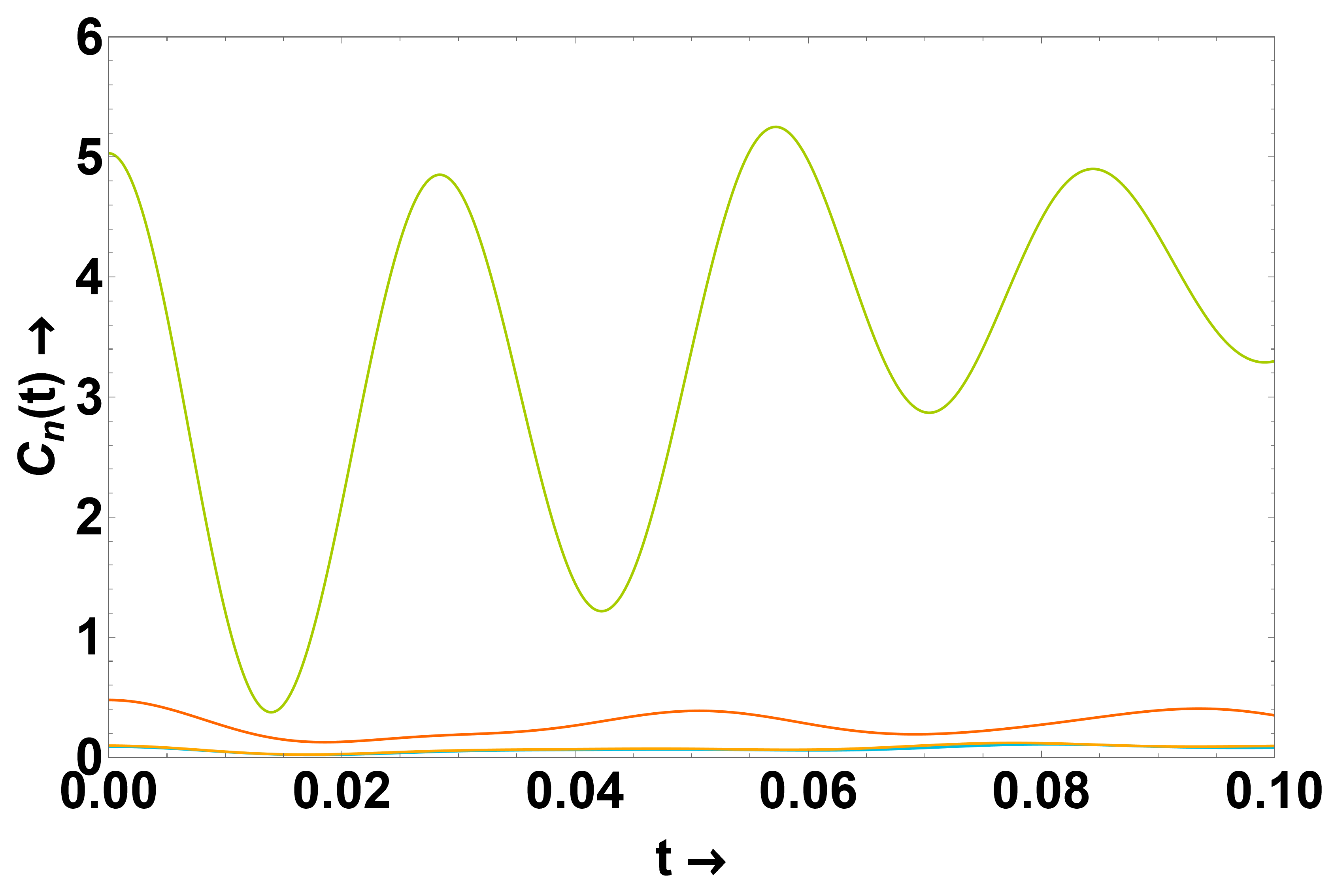} & \includegraphics[scale=0.14,valign=c]{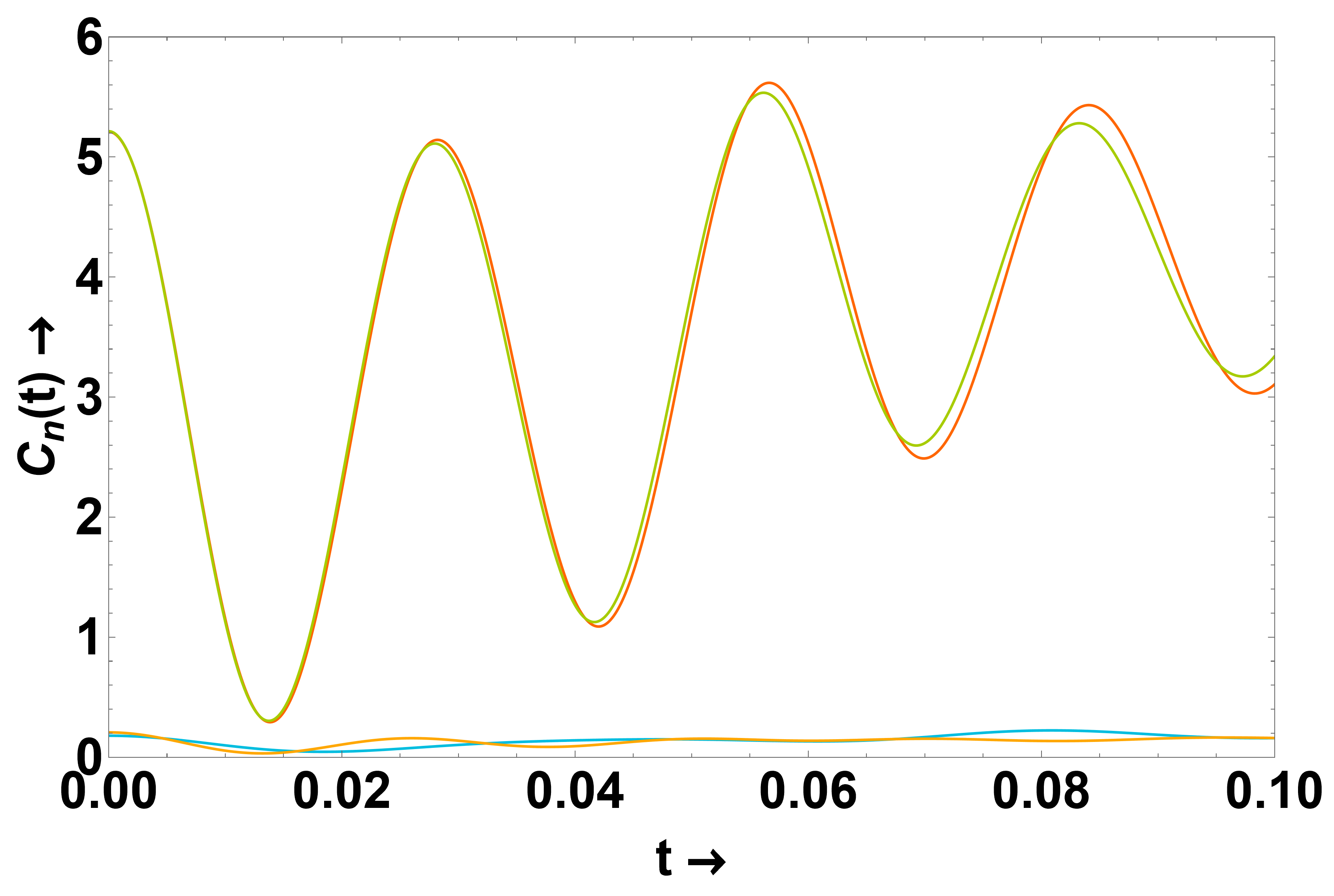} & \includegraphics[scale=0.14,valign=c]{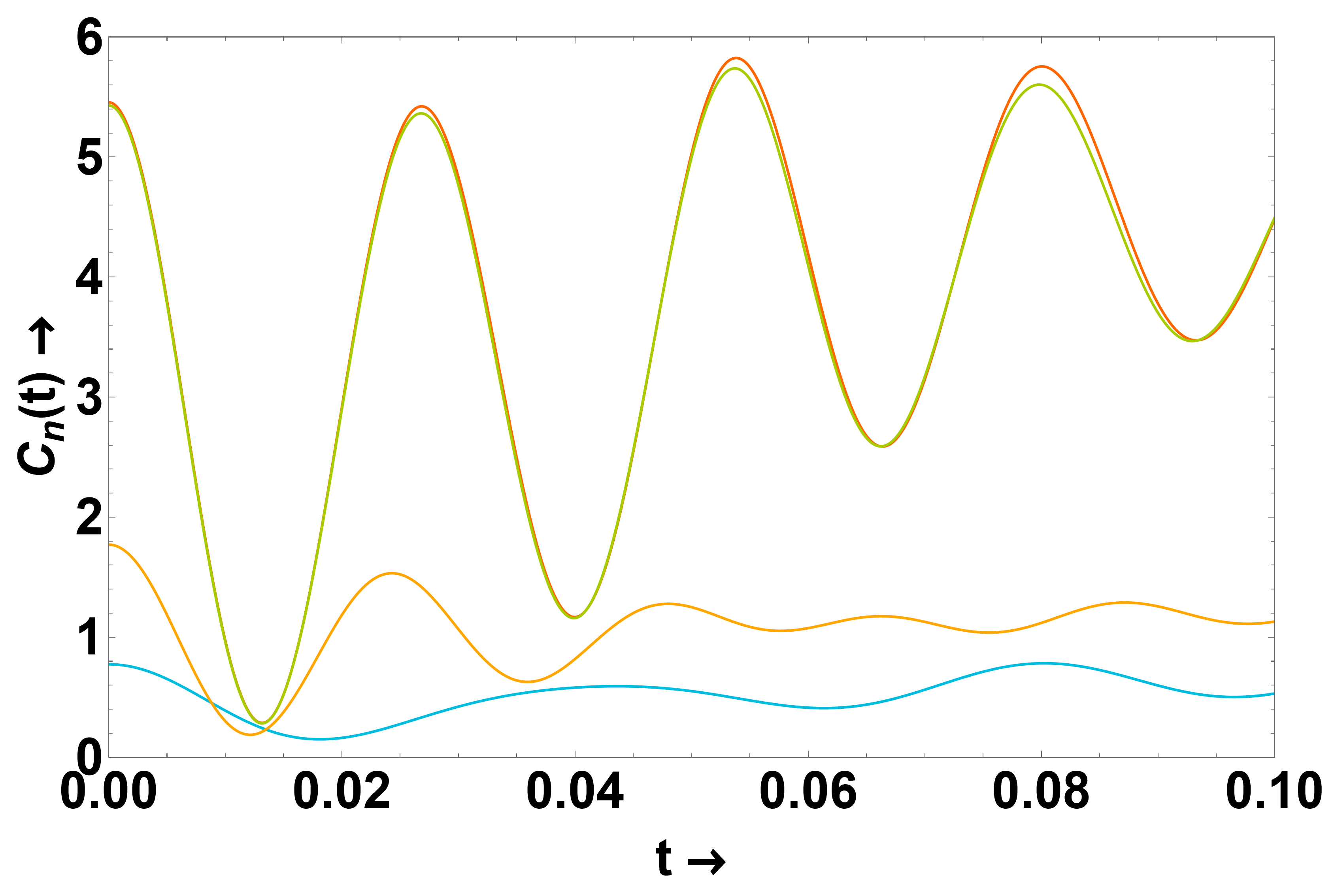} \\
        \hline
        \end{tabular}
    }
    \caption{Early time microcanonical OTOCs for different values of $n$ and $k$. Here first 200 eigenvalues and eigenfunctions were taken during the calculations. The results are divided into two categories: low energy ($n=5, 10, 20$) and high energy ($n=110, 120, 150$). The color coding is done as: red for $q = 0.0$, green for $q = 0.3$, cyan for $q = 0.6$ and orange for $q = 0.9$. There is a visible effect of charge for low-energy microcanonical OTOCs in the view of lowering the early exponential growth rate, while for the high-energy case, we only get aperiodic oscillations and no exponential growth.\label{fig:early-otoc}}
\end{figure}

We also examine our system's dynamics by dialing the charge between low and high values at a fixed energy $E^2 < 350$ as shown in Fig.~\ref{fig:dysonmehta-fixedenergy}. On increasing the charge, we observe a lifting, albeit small, of \( \overline{\Delta}_3(L) \) values towards that of uncorrelated levels, hinting that the charge may act as a stabilizer, lifting the spectrum towards the integrable regime. Our overall analysis with the \( \Delta_3 \) statistic suggests that the charge and energy parameters tend to lift the spectrum towards the integrable regime, with the latter playing a more prominent role. 

We also computed the microcanonical OTOCs for our system. The numerical results are shown in Fig.~\ref{fig:otoc-k6-qvary}. Here, we calculated $C_n(t)$ at $k=6$ for $n^\text{th}$ eigenvalues using Eqs.~(\ref{eq:microtoc}) and (\ref{eq:bnm-expression}). We observe two distinct effects of $q$ for low and high $n$ values. The early time growth rate for $n=\{10-80\}$ is suppressed when we increase $q$ from $q=0.0$ to $q=0.9$. This can be easily seen from Figures~\ref{fig:otock6opt010to80} and~\ref{fig:otock6opt910to80}. Since the early growth of microcanonical OTOC signifies chaos, this means the effect of the charge is to decrease the chaos in the system for low energies. In contrast, when we analyze the Figures~\ref{fig:otock6opt0100to180} and~\ref{fig:otock6opt9100to180}, we find that for $n=\{100-180\}$, there is no significant difference when we increase $q$ from $q=0.0$ to $q=0.9$. This implies for large energies, the effect of charge on the system is not very prominent, which seems obvious since we have found from our previous analysis that energy is a much more dominant factor than charge when it comes to the effect on chaos.

To make our point clear, we also show the early time evolution of microcanonical OTOCs in Fig.~\ref{fig:early-otoc}. We present this for three different $k$ values: $k=1, 2, 3$ and also for different $n^\text{th}$ eigenvalues. Again, for low energies, i.e., for $n=5, 10, 20$, we observe that the effect of $q$ is to suppress the initial microcanonical OTOC growth of the system. Since the early growth of the microcanonical OTOC plays a big role in the computation of the quantum Lyapunov exponent, we can indirectly say that increasing the charge decreases the quantum Lyapunov exponent of the system to some extent. For the high energies, i.e., for $n=110, 120, 150$, we observe that there is no growth in the microcanonical OTOCs even at early times. Instead, we get aperiodic oscillations. This suggests that at high energies, the system is already approaching integrability, as reflected by oscillatory behavior for the early-time microcanonical OTOCs and $q$ has no effect on the system. Thus, we can conclude that the effect of charge $q$ is to make the system less chaotic in the quantum domain for low energies. This can be contrasted with our results in the classical domain, where larger $q$ tends to make the system more chaotic.

\section{Conclusion}\label{sec:conclusion}
In this paper, we analyze the chaotic dynamics of a closed string in the charged AdS soliton background using classical and quantum chaos diagnostic tools. The charged soliton is a solution of minimal $d=5$ gauged supergravity and corresponds to a confining phase in the dual field theory. In the classical domain, we used the power spectrum, Poincar\'{e} sections, and Lyapunov exponents to analyze the effect of energy and charge on the chaotic dynamics of the closed strings. Our findings suggest that increases in charge and energy destabilize the system, thereby making the closed string dynamics more chaotic.

We then analyzed the distribution of energy level-spacings, Dyson-Mehta $\Delta_3$  statistic, and microcanonical OTOCs to investigate the effect of charge and energy on the quantum chaos of closed string in the confined phase. For this purpose, we utilized the minisuperspace quantization prescription and obtained the quantum energy spectrum of the closed string. We analyzed the distribution of energy level-spacings for a fixed charge and a fixed energy. In the fixed-charged case, we observed distribution changing from nearly Wigner GOE to nearly Poisson as we increased the set from smaller energy values ($E^2<200$) to larger energy values ($E^2<1000$), indicating that at higher energies, the system asymptotes to integrability. Similarly, for the fixed energy case, we found the effect of increasing the charge is to shift the system from a nearly Wigner GOE distribution to a nearly Poisson one, which means the role of charge is to stabilize the system in the quantum domain.

Our overall investigation here not only demonstrates the interplay of charge and energy in shaping the chaotic behavior of the closed string across both classical and quantum domains in the confined phase but also provides a new window for further studies. For instance, the transition from chaotic dynamics, marked by level repulsion at lower energies, to an integrable regime as energy increases underscores a complex energy dependence that merits further in-depth investigation of the intermediate energy regimes, which could potentially clarify the subtle transitions between chaos and integrability. 

However, the full string spectrum remains an open question, especially in the context of a complete quantum theory without the simplifying assumptions of minisuperspace quantization. The minisuperspace approximation provides a viable framework for probing quantum chaos in the closed string spectrum, despite inherent truncations of the full string degrees of freedom. This approach retains only the center-of-mass dynamics and a subset of spatial modes, neglecting fermionic fields, background fluxes, and oscillator sectors of the full theory, as noted in studies of other confining geometries~\cite{PandoZayas:2012ig}. While some level of systematic errors are expected due to this approximation, its validity in capturing universal spectral properties is supported by precedents across quantum cosmology, string theory, and lattice QCD. The minisuperspace methodology, initiated in quantum cosmology~\cite{Hartle:1983ai}, simplifies dynamics by restricting to symmetric configurations, arguing these dominate the path integral. Similarly, in Liouville theory~\cite{Seiberg:1990eb}, truncating to minisuperspace preserves critical features like critical exponents despite omitting oscillator modes. For highly symmetric systems, such as AdS\(_3\) string configurations~\cite{Maldacena:2000hw}, this approximation reproduces exact spectral results in integrable sectors. Crucially, as shown in~\cite{Teschner:1997fv, Douglas:2003up}, even truncated dynamics retain key statistical signatures (e.g., level repulsion) when the classical limit is chaotic, aligning with studies where minisuperspace results match lattice QCD observations of hadronic spectral statistics~\cite{Bittner:2004ff, Markum:2005ft}. This indicates that while absolute energies may deviate, the universality of level spacing distributions (GOE vs. Poisson) might remain robust. This resilience mirrors findings in Liouville theory and related studies, where high symmetry allows the approximation to capture dominant spectral features with reasonable accuracy~\cite{Seiberg:1990eb, Teschner:1997fv}.  


A rigorous quantification of deviations from the full quantum theory remains an open challenge. Future work could address this by comparing minisuperspace results with full-string spectra in solvable limits, such as near-integrable deformations~\cite{Douglas:2003up}. Nonetheless, the ability of the minisuperspace framework to reproduce chaotic signatures, which is supported by string-theoretic precedents and lattice QCD parallels, underscores its utility as an important and relevant tool for studying quantum chaos in holographic confining gauge theories, despite its simplified nature.  

Moreover, exploring the influence of additional parameters such as chemical potential or temperature within this framework holds promise for offering valuable insights into QCD chaos from the gauge/gravity perspective, particularly in the deconfined phase and out-of-equilibrium settings. Such extensions could enhance our understanding of how holography encodes chaotic dynamics within QCD and other strongly coupled quantum field theories. By investigating these aspects, we can advance efforts to bridge classical chaos diagnostics with quantum mechanical properties, potentially unlocking new avenues in the study of holographic QCD chaos.

\section*{Acknowledgments}
The work of S.M.~is supported by the core research grant from the Science and Engineering Research Board, a statutory body under the Department of Science and Technology, Government of India, under grant agreement number CRG/2023/007670.

\bibliography{biblio.bib}
\bibliographystyle{unsrt}

\end{document}